\documentclass[prx,twocolumn,superscriptaddress,10pt]{revtex4-2}
\usepackage[utf8]{inputenc}
\usepackage{amsmath}
\usepackage{amssymb}
\usepackage{graphicx}
\usepackage{pdfpages,wasysym}
\makeatletter
\AtBeginDocument{\let\LS@rot\@undefined}
\makeatother

\usepackage{color}
\usepackage{verbatim}
\usepackage{booktabs}
\usepackage{float} 
\usepackage{listings}
\usepackage{alltt}
\usepackage{subfigure}
\usepackage{multirow}
\usepackage{xcolor}

\usepackage{array}
\newcolumntype{C}[1]{>{\centering\let\newline\\\arraybackslash\hspace{0pt}}m{#1}}

\usepackage{hyperref}
\usepackage{cleveref}

\usepackage[normalem]{ulem}

\crefname{equation}{Eq.}{Eqs.}
\Crefname{equation}{Equation}{Equations}
\crefname{figure}{Fig.}{Figs.}
\Crefname{figure}{Figure}{Figures}
\crefname{section}{Sect.}{Sects.}
\Crefname{section}{Section}{Sections}

\newcommand{\ket}[1]{| #1 \rangle}
\newcommand{\bra}[1]{\langle #1 |}
\newcommand{\ketbra}[1]{| #1 \rangle\langle #1 |}

\newcommand{\expo}[1]{\text{e}^{ #1 }}
\newcommand{\appropto}{\mathrel{\vcenter{
  \offinterlineskip\halign{\hfil$##$\cr
    \propto\cr\noalign{\kern2pt}\sim\cr\noalign{\kern-2pt}}}}}

\newcommand{\ha}{\hat{a}}
\newcommand{\had}{\hat{a}^\dagger}
\newcommand{\hx}{\hat{q}}
\newcommand{\hp}{\hat{p}}

\newcommand{\hU}{\hat{U}}
\newcommand{\hH}{\hat{H}}
\newcommand{\hd}{\hat{d}}
\newcommand{\hq}{\hat{q}}
\newcommand{\hg}{\hat{g}}

\newcommand{\hvx}{\hat{\mathbf{x}}}
\newcommand{\hT}{\hat{T}}
\newcommand{\vlambda}{\pmb{\lambda}}
\newcommand{\vmu}{\pmb{\mu}}
\newcommand{\vupsilon}{\pmb{\upsilon}}

\DeclareFontFamily{U}{stixextrai}{}
\DeclareFontShape{U}{stixextrai}{m}{n}
 { <-> stix-mathtt }{}

\newcommand{\lltriangle}{{\usefont{U}{stixextrai}{m}{n}\symbol{153}}}

\makeatletter
\providecommand{\vrectangle}{\mkern1mu\mathpalette\v@rectangle\relax\mkern1mu}
\newcommand{\v@rectangle}[2]{%
  \hbox{
  \fboxrule=0.5\fontdimen 8
    \ifx#1\displaystyle\textfont\else
    \ifx#1\textstyle\textfont\else
    \ifx#1\scriptstyle\scriptfont\else
    \scriptscriptfont\fi\fi\fi 3
  \fboxsep=-\fboxrule
  \fbox{$\m@th#1\phantom{(}$}%
  }
}
\makeatother

\begin{document}
\title{Encoding qubits in multimode grid states}
\author{Baptiste Royer}
\affiliation{Department of Physics, Yale University, New Haven, Connecticut 06511, USA}
\affiliation{Yale Quantum Institute, Yale University, New Haven, Connecticut 06511, USA}

\author{Shraddha Singh}
\affiliation{Yale Quantum Institute, Yale University, New Haven, Connecticut 06511, USA}
\affiliation{Department of Applied Physics, Yale University, New Haven, Connecticut 06511, USA}

\author{S.M. Girvin}
\affiliation{Department of Physics, Yale University, New Haven, Connecticut 06511, USA}
\affiliation{Yale Quantum Institute, Yale University, New Haven, Connecticut 06511, USA}

\begin{abstract}
Encoding logical quantum information in harmonic oscillator modes is a promising and hardware-efficient approach to the realization of a quantum computer. In this work, we propose to encode logical qubits in grid states of an ensemble of harmonic oscillator modes. We first discuss general results about these multimode bosonic codes; how to design them, how to practically implement them in different experimental platforms and how lattice symmetries can be leveraged to perform logical non-Clifford operations. We then introduce in detail two two-mode grid codes based on the hypercubic and $D_4$ lattices, respectively, showing how to perform a universal set of logical operations. 
We demonstrate numerically that multimode grid codes have, compared to their single-mode counterpart, increased robustness against propagation of errors from ancillas used for error correction. 
Finally, we highlight some interesting links between multidimensional lattices and single-mode grid codes concatenated with qubit codes.
\end{abstract}

\maketitle

By redundantly encoding logical quantum information, quantum error correction (QEC) can protect quantum computations against the effects of decoherence, allowing the realization of quantum algorithms requiring large circuit depths.  
One promising approach to QEC is to encode logical qubits in harmonic oscillator modes, a strategy named bosonic codes from the statistics obeyed by the oscillator excitations. 
Amongst the attractive features of bosonic codes is their large (formally infinite) Hilbert space which allows a high degree of redundancy and their relatively simple error model compared to multi-qubit QEC codes. Moreover, oscillators can exhibit very long lifetimes. 
For example, microwave cavities with lifetime up to 2 seconds have been demonstrated~\cite{Romanenko20a}, which is orders of magnitude longer than state-of-the-art superconducting qubits.

Several important milestones have already been reached with bosonic codes, demonstrating the appeal of this approach. For example, using the cat encoding~\cite{Mirrahimi14a}, the lifetime of a logical qubit reached the lifetime of an unencoded qubit~\cite{Ofek16a}. Operation of a cat code and binomial code~\cite{Michael16a} in a manner transparent to ancilla errors was demonstrated~\cite{Rosenblum18a,Reinhold20n,Hu19a,Ma20a}, and an autonomous error-correction scheme was realized~\cite{Gertler21l}. 
In order to further increase the logical lifetime of bosonic qubits, one approach is to increase the size of the code, i.e.\ increase the number of excitations in the code words. For example, binomial codes can be made to correct multiple errors by increasing the distance between the Fock states constituting the code words~\cite{Michael16a}. 
However, oscillator error channels such as amplitude damping or spurious non-linearities typically scale with the number of excitations, and increasing the size of bosonic codes beyond a certain point leads to diminishing return. 
Practical implementations of bosonic codes therefore impose constraints on the number of excitations in code words. 

\begin{figure}[t]
    \centering
    \includegraphics[scale=1.]{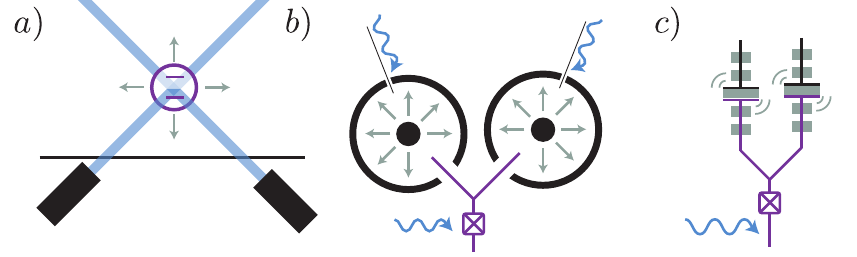}
    \caption{Cartoon representation of possible platforms for the implementation of two-mode grid codes. a) Motional modes (gray) of a trapped ion (purple), with lasers used for control (blue). b) Two single-post microwave cavities (black), coupled to a transmon qubit (purple). Electromagnetic modes (gray) are controlled through pulses (blue) applied on the cavities and transmon qubit. c) Phononic-crystal-defect resonator modes (gray) coupled to a transmon qubit (purple).}
    \label{fig:PracticalImplementation}
\end{figure}

To improve the QEC properties of bosonic codes, an alternative approach is to consider an \emph{ensemble} of harmonic oscillators. 
The first such multimode code to be introduced was the dual-rail encoding~\cite{Chuang95d}, where the logical information is encoded in the single-excitation subspace of two distinct modes.
Since then, several other multimode constructions have been made based on superposition of Fock states
~\cite{Chuang97a,Wasilewski07i,Bergmann16v,Niu18f,Albert18a,Ouyang20d}, generalization of cat states~\cite{Albert19f} and other ideas~\cite{Gottesman01a,Hayden16h,Lau16m,Niu18f}. Here, we study the grid state codes introduced by Gottesman, Kitaev and Preskill (GKP), and more precisely their multimode extension~\cite{Gottesman01a}. Clifford operations can be implemented in a fault-tolerant manner in ideal GKP codes using Gaussian operations~\footnote{Clifford operations in GKP codes are fault-tolerant in the sense that translation errors are not amplified too much by Gaussian operations.}, and in principle their single-mode version is very robust against dominant error channels such as amplitude damping~\cite{Albert18a}. As a result, GKP codes have attracted a lot of interest, but most work has been focused either on single-mode codes, or the concatenation of multiple single-mode codes with a qubit QEC code~\cite{Fukui18r,Vuillot19a,Wang19k,Noh20b,Hanggli20s,Noh22b}. 

In this paper, we are instead interested in ``genuine'' multimode GKP codes, i.e.\ the codes defined in Ref.~\cite{Gottesman01a}, and show that these codes possess properties that cannot be achieved in their single-mode counterpart.
First, we introduce new codes based on lattices with large sphere packing ratios, which translates into an increased distance between code words. Interestingly, we find that the allowed code dimensions are related to number theoretic results such as Legendre’s three-square theorem. 

We also show that logical operations in multimode codes can be simpler than single-mode codes. For example, we introduce a two-mode code where all single-qubit Clifford operations can be realized with passive linear optics, which contrasts with single-mode codes where general single-qubit Clifford gates require squeezing. Leveraging lattice symmetries, we also introduce non-Clifford gates based on Kerr-type interactions for GKP codes. In stark contrast to the non-Clifford cubic phase gate originally proposed in Ref.~\cite{Gottesman01a}, these new gates are \emph{exact} for finite-energy states. 
Some of the results we present for multimode codes can be used to improve single-mode codes. For example, the teleportation-based error correction circuit for single-mode codes introduced in Ref.~\cite{Walshe20q} can also be reframed in the multimode context. More importantly, the non-Clifford gates we introduce can also be implemented in the single-mode square GKP code. 

Bosonic codes are designed to be robust against oscillator errors, but a recurring limitation to the logical lifetime is the propagation of logical errors from the ancilla required for quantum control of the modes. To overcome this issue, we design grid codes where, with high probability, ancilla errors propagate to the oscillator modes as correctable errors instead of logical errors. In particular, the probability of a propagated logical error decreases with the size of the grid codes. In contrast, in single-mode GKP codes where a two-level system is used as ancilla~\cite{Campagne-Ibarcq20a,Royer20a,de-Neeve20a}, the probability that an ancilla error propagates as a logical error is roughly $50\%$, a constant independent of the number of excitations in the GKP code words.
General approaches such as path-independence~\cite{Ma20y,Ma22u} or biased-noised ancillas~\cite{Puri19a} have been suggested to reduce the impact of ancilla errors in bosonic codes, and these approaches could also be compatible with multimode GKP codes.

We present practical error-correction methods for multimode codes, showing that they could be implemented in a variety of experimental platforms, some of which are illustrated in \cref{fig:PracticalImplementation}. The preparation of grid states was first demonstrated in the motion of a trapped ion~\cite{Fluhmann19a,de-Neeve20a}, and multimode codes could be implemented by considering multiple modes of a single ion, as illustrated in \cref{fig:PracticalImplementation}a), or the collective motion of multiple ions. Quantum control of multiple modes has already been demonstrated in other contexts~\cite{Zhang19a}. Multimode GKP codes could also be implemented in microwave cavities, as illustrated in \cref{fig:PracticalImplementation}b). This is the first platform where error correction in a GKP code was demonstrated~\cite{Campagne-Ibarcq20a}, and the preparation of grid states has been used to benchmark cavity control methods~\cite{Sivak21w,Eickbusch21f}. Quantum control of multiple cavities has also been demonstrated in various contexts, for example the preparation of two-mode cat states~\cite{Wang16y} or the realization of an engineered exchange interaction~\cite{Gao19d}, and the fabrication of high-quality multi-cavity systems is an active research area~\cite{Lei20p,Chakram21q}. Another promising platform for GKP codes is phononic-crystal-defect resonators, which have a smaller physical footprint than microwave cavities~\cite{Arrangoiz-Arriola16a,Arrangoiz-Arriola19i}. A schematic illustration of this platform is shown in \cref{fig:PracticalImplementation}c). 
Finally, we enumerated above experimental platforms where control of oscillator modes is performed using a nonlinear ancilla, but a natural platform that could host multimode GKP codes is propagating photonic modes, where the entanglement of thousands of modes has been demonstrated~\cite{Larsen19k,Asavanant19i}. Given the ability to prepare single-mode GKP states, one could leverage precise homodyne measurements to prepare and correct errors in multimode GKP codes.
To the best of our knowledge, the required preparation of single-mode grid states has not yet been realized in an optical platform, although several promising proposals exist~\cite{Travaglione02a,Pirandola04a,Vasconcelos10a,Terhal16a,Motes17a,Shi19a,Weigand20a,Tzitrin20a,Royer20a,Hastrup21j}.

The rest of the paper is organized as follows. First, in \cref{sect:Notation}, we set the notation used throughout the rest of the paper. Then, in \cref{sect:multimodeGKP}, we review and extend the theory on multimode GKP codes. These first mathematically oriented sections are then followed by a presentation of practical error correction strategies in \cref{sect:ErrorCorrection}. We then study two concrete examples of two-mode codes, namely the tesseract code in \cref{sect:Cubiclattice} and the $D_4$ code in \cref{sect:D4lattice}. We highlight some similarities between concatenated codes and multimode lattices in \cref{sect:qubitCodeConstruction} and conclude in \cref{sect:conclusion}.

\section{Preliminaries and Notation}\label{sect:Notation}

In this paper, we consider $m$ harmonic oscillator modes, and we aim to establish correspondences between translations in a (symplectic) vector space $\mathbb R^{2m}$ and the quantum Hilbert space of these $m$ modes.
We denote the dimensionless creation and annihilation operators for the $j$th mode $\ha_j$ and $\had_j$, respectively, obeying $[\ha_j,\had_k] = \delta_{jk}$. We work in units of $\hbar = 1$ and denote the quadrature coordinates $\hx_j = (\ha_j + \had_j)/\sqrt 2$ and $\hp_j = -i(\ha_j - \had_j)/\sqrt 2$ such that $[\hx_j,\hp_k] = i \delta_{jk}$.
We arrange the quadrature coordinates in vectors such that $\hvx = (\hx_1,\,\hp_1,\,\hx_2,\,...,\,\hx_m,\,\hp_m\,)$, and points in phase space correspond to vectors $\mathbf v\in \mathbb R^{2m}$ arranged in the same order as $\hvx$.  We decorate all quantum operators with a hat and denote vectors in bold fonts.

We define a translation in phase space by $\mathbf v$ in units of $l=\sqrt{2\pi}$ as
\begin{equation}
\begin{aligned}
\hat T(\mathbf v) &\equiv \expo{-i l \hvx^T \Omega \mathbf v  }\\
 &= \bigotimes_{j=1}^m \hat D_j\left(l\left[ C\mathbf v\right]_j \right),
\end{aligned}
\end{equation}
where $\hat D_j(\alpha) = \exp\{\alpha \had_j - \alpha^* \ha_j\}$ with $\alpha \in \mathbb C$ is the standard displacement operator for the $j$th mode~\cite{C.W.-Gardiner00a} and $C$ is a $m \times 2m$ matrix that maps the real vector $\mathbf v \in \mathbb R^{2m}$ to a complex vector $C\mathbf v \in \mathbb C^{m}$,
\begin{equation}
C = \frac{1}{\sqrt 2}
\begin{pmatrix}
1 & i & 0 &0 & ... \\
0 & 0 & 1 & i & ... \\
\multicolumn{5}{c}{...} \\
\end{pmatrix}.
\end{equation}
We have also defined the anti-symmetric matrix
\begin{equation}
\Omega = \bigoplus_m
\begin{pmatrix}
0 & 1 \\
-1 & 0 \\
\end{pmatrix},
\end{equation}
from which we define the symplectic form
\begin{equation}
\omega(\mathbf u,\mathbf v) = \mathbf u^T \Omega \mathbf v.
\end{equation}
This form is alternating, $\omega(\mathbf u,\mathbf v) = - \omega(\mathbf v,\mathbf u)$, which implies that $\omega(\mathbf u,\mathbf u) = 0$.
The commutation relations of the quadrature coordinates impose that
\begin{equation}
\label{eq:commRelDisplacements}
\hat T(\mathbf u)\hat T(\mathbf v) = \hat T(\mathbf v)\hat T(\mathbf u) \expo{i2\pi \mathbf v^T \Omega \mathbf u} = \hat T(\mathbf u + \mathbf v)\expo{i\pi \mathbf v^T \Omega \mathbf u}.
\end{equation}
By defining translations in units of $l=\sqrt{2\pi}$, the translation operators associated with two vectors $\mathbf u$ and $\mathbf v$ commute if and only if their symplectic form is an integer, $[\hat T(\mathbf u),\hat T(\mathbf v)] = 0 \Leftrightarrow \mathbf u^T \Omega \mathbf v \in \mathbb Z$.

Quantum unitaries generated by quadratic Hamiltonians are represented by $2m \times 2m$ real symplectic matrices $M \in \mathrm{Sp}(2m,\mathbb R)$ respecting $M^T\Omega M = \Omega$. 
We define the unitary-valued function $\hat Q$ that takes as input a symplectic matrix $M$ and outputs the corresponding quantum unitary. Its effect on the quadrature coordinates is given by $\hat Q^\dag (M)\hvx\hat Q(M) = M\hvx$, where $\hat Q$ acts by conjugation on each $\hat x_j$ individually and $M$ acts by matrix-vector multiplication on the vector $\mathbf x$. Commuting $\hat Q(M)$ through a translation operator yields
\begin{equation}\label{eq:commRelQofM}
\hat Q(M) \hT(\mathbf v) = \hT(M\mathbf v) \hat Q(M),
\end{equation}
and the operator-valued function $\hat Q$ is a homomorphism such that $\hat Q(M_1) \hat Q(M_2) = \hat Q(M_1 M_2)$. Given a $2m\times 2m$ real symmetric matrix $J$ such that $M=\exp\{\Omega J\}$, $\hat Q$ can be computed using
\begin{equation}
\hat Q(M) = \exp\left\{\frac{-i}{2} \hvx^T J \hvx\right\}.
\end{equation} 

An important subgroup of symplectic matrices are those that are also orthogonal, $O\in \mathrm{O}(2m,\mathbb R)\cap\mathrm{Sp}(2m,\mathbb R)$ respecting $O^T O = \mathbb I$. The quantum unitary associated with these matrices, $\hat Q(O)$, preserves the total excitation number in all modes, $[\hat Q(O), \hat n]= 0$ with $\hat n = \sum_j \had_j \ha_j = (\hvx \cdot \hvx - m)/2$. Geometrically, the fact that $\hat Q(O)$ preserves the photon number is equivalent to the fact that orthogonal matrices $O$ preserve the Euclidian distance in $\mathbb R^{2m}$. 

We define a rotation of the $j$th mode 
\begin{equation}
\hat R_j(\theta) = \expo{-i \theta \hat n_j},
\end{equation}
with $\hat n_j = \had_j \ha_j$, and its associated symplectic representation
\begin{equation}
R(\theta) = \begin{pmatrix}
\cos\theta  & -\sin\theta \\
\sin\theta & \cos\theta\\
\end{pmatrix},
\end{equation}
with $\hat R(\theta) = \hat Q[R(\theta)]$. Moreover, we also define a beamsplitter operation between two modes $j,k$, $\hat B_{j \rightarrow k} = \exp\{-i\pi(\hx_j \hp_k - \hp_j \hx_k)/4\}$ which has a symplectic representation
\begin{equation}
B_{j \rightarrow k} = \frac{1}{\sqrt 2} \begin{pmatrix}
1 & 0 &-1 & 0\\
0 & 1 & 0 & -1 \\
1 & 0 & 1 & 0 \\
0 & 1 & 0 & 1 \\
\end{pmatrix}.
\end{equation}
The arrow in the graphical representation of the beamsplitter operation, for example in \cref{fig:TesseractGates}c) below, matches the direction of the $j\rightarrow k$ arrow.

Finally, we denote logical operations acting on encoded qubits with an overhead bar, $\bar G$, and denote the corresponding multimode unitary as $\hat U(\bar G)$. As is usual for error-correcting codes, logical gates can have multiple equivalent representatives, and as a result the mapping $\hat U(\bar G)$ is not unique. The representative referred to with the mapping $\hU$ will be clear from the context. 

An example of the different mathematical objects mentioned above and the space they act on is summarized in \cref{table:exampleMapping}.

\begin{table}
\begin{tabular}{ |C{0.1\textwidth} | C{0.25\textwidth} | C{0.1\textwidth}| }
\hline
Symplectic & \multirow{2}*{Oscillator space}  & Logical\\
space &  & space\\ \hline
Acts on $\mathbb R^2$ & Acts on $L^2(\mathbb R)$ & Acts on $\mathbb C^2$\\ \hline
$R(\pi/2)$ &  $\hat Q[R(\pi/2)] = \hat R(\pi/2) = \hU(\bar H)$ & $\bar H$\\ \hline
\end{tabular}
\caption{Example of different representations for a logical Hadamard gate in the single-mode square GKP code, implemented by a rotation by $\pi/2$ in phase space. In general, symplectic matrices act on the real vector space $\mathbb R^{2m}$, quantum unitaries act on the set of square integrable functions $L^2(\mathbb R^{m})$ and logical operations act on a qubit Hilbert space $\mathbb C^2$.}\label{table:exampleMapping}
\end{table}

\section{Multi-dimensional Grid States}\label{sect:multimodeGKP}
In this section, we review and extend the theory on multimode GKP codes introduced in Ref.~\cite{Gottesman01a}. We refer the reader to \cref{sectApp:latticeTheory} and Ref.~\cite{Conway13c} for more details about lattice theory.

\subsection{General Theory}\label{subsect:generalMultimodeTheory}
The aim here is to encode a logical qudit in translation-invariant grid states. Taking $m$ oscillator modes, a QEC code is associated with a (classical) lattice $\Lambda$ in $2m$ dimensions, each mode contributing two quadrature coordinates, $\hx$ and $\hp$. The lattice $\Lambda$ is generated by a set of $2m$ linearly independent translations $\{\mathbf s_j\}$, which we arrange in a $2m\times 2m$ matrix $S$ where each row corresponds to one basis vector $\mathbf s_j$,  
\begin{equation}\label{eq:generatorMatrixS}
S = \begin{pmatrix}
\mathbf s_1 \\
\mathbf s_2 \\
... \\
\mathbf s_{2m}
\end{pmatrix}.
\end{equation}
The lattice points are then given by
\begin{equation}\label{eq:LambdaDef}
\Lambda = \left\{S^T \mathbf a \mid\, \mathbf a \in \mathbb Z^{2m}\right\}.
\end{equation}
The QEC grid code is defined by associating each generator of the lattice with a generator of the stabilizer group, $\mathbf s_j \rightarrow \hat T(\mathbf s_j)$. The (infinite) stabilizer group is then given by
\begin{equation}
\label{eq:stabGroup}
\mathcal S = \left\{\Pi_{j=1}^{2m} \hT(\mathbf s_j)^{a_j} \mid\, \mathbf a \in \mathbb Z^{2m}\right\},
\end{equation}
with each stabilizer associated with a point on the lattice $\Lambda$. Code words are defined to be in the simultaneous +1 eigenspace of all stabilizers.

The generators of the quantum translation, $\hat T$, associated with the generators of the stabilizer group are given by
\begin{equation}\label{eq:translationGenerators}
\mathbf{\hg} = -l S\Omega \hvx,
\end{equation}
such that the $j$th generator of the stabilizer group is given by $\hat T(\mathbf s_j) = \exp\{i \hg_j\}$. Measuring the stabilizer $\hat T(\mathbf s_j)$ is equivalent to measuring the modular quadrature coordinate $\hg_j \mod 2\pi$~\cite{Aharonov69d,Popescu10j}, and in particular eigenstates of $\hat T(\mathbf s_j)$ are also eigenstates of $\hg_j$. \\ \indent
In contrast to qubit QEC stabilizer codes, the stabilizers of the GKP code have a continuous spectrum. Restricting the code space to the $+1$ eigenspace of the stabilizers thus imposes an infinite amount of constraints, restricting the eigenvalues of $\hg_j$ to be within the countable set $g_j = 0 \mod 2\pi$ within the uncountable eigenstates with eigenvalues $g_j \in \mathbb R$.
The intersection of the $+1$ eigenspace of each stabilizer is a finite dimensional space, with a dimension $d$ specified below.
Quadrature coordinate eigenstates are equivalent to infinitely squeezed states which contain an infinite amount of energy, making the code words of ideal GKP codes unphysical. We consider the realistic, finite-energy version of the code words and stabilizer group $\mathcal S$ below in \cref{subsect:finiteEnergyGKP}.

Another difference between GKP codes and qubit stabilizer codes is the number of stabilizers needed to specify the code space. Indeed, $n - k$ stabilizers are needed to specify how to encode $k$ logical qubits into $n$ physical qubits. In contrast, $2m$ translation operators are needed to describe the GKP code space, irrespective of the code dimension.

The symplectic Gram matrix of the lattice $\Lambda$, setting the pairwise commutation relations of the stabilizer generators, is given by
\begin{equation}\label{eq:GramMatrix}
A = S\Omega S^T,
\end{equation}
such that $\hat T(\mathbf s_j) \hat T(\mathbf s_k) = \hat T(\mathbf s_k) \hat T(\mathbf s_j) \expo{2\pi i A_{jk}}$.
In order for the stabilizers to commute with each other, we should therefore have $\Lambda$ symplectically integral, meaning that $A$ should only contain integers. 

Associated with the lattice $\Lambda$, we define the symplectic dual lattice $\Lambda^*$ as the ensemble of points that have an integer symplectic form with the lattice points $\Lambda$,
\begin{equation}\label{eq:dualLatticeDef}
\Lambda^* = \{\mathbf v \mid\, S\Omega\mathbf v \in \mathbb Z^{2m} \}.
\end{equation}
In the rest of this work, we simply refer to $\Lambda^*$ as the dual lattice instead of the more precise term symplectic dual lattice.
Since the lattice is symplectically integral, the definition of $\Lambda^*$ implies that $\Lambda \subseteq \Lambda^*$.
One choice of generator matrix for the dual lattice $\Lambda^*$ is obtained from~\footnote{We only consider $2m$-dimensional lattices in $m$ modes, such that $S$ is full rank and $A$ is always invertible.}
\begin{equation}\label{eq:dualLatticeGenerator}
S^* = A^{-1} S.
\end{equation}

Associating a translation to each point in the dual lattice, \cref{eq:dualLatticeDef} implies that the set of translation operators $\{\hat T(\vlambda^*)| \vlambda^* \in \Lambda^* \}$ forms (modulo phases) the centralizer of $\mathcal S$ in the group of translations, i.e.\ corresponds to all translations that commute with all elements of the stabilizer group. 
In a QEC code, logical operators correspond to operators that leave the stabilizer group invariant. Here, we associate all translations by a dual lattice vector to a logical Pauli operator. Since translations that differ by a lattice vector are equivalent in the logical subspace, the logical information is encoded in the dual quotient group $\Lambda^* / \Lambda$ and the number of distinct logical operators is given by the number of dual lattice points inside the fundamental parallelotope of $\Lambda$. Equivalently, the number of distinct logical operators is given by the ratio of volumes between the fundamental parallelotope of the base and dual lattices.
Defining the determinant of a lattice to be $\text{det}(\Lambda) = \text{det}(A)$, the condition to encode a $d-$level qudit with $d^2$ logical Pauli operators is that
\begin{equation}\label{eq:codeDimension}
\text{det}(\Lambda) = d^2.
\end{equation}
The links between classical lattices and quantum grid codes are summarized in \cref{table:ClassicalQuantumMapping}.

To investigate multimode codes, we first search for the different codes that can be obtained by (only) scaling the lattice size by a constant $c \in \mathbb R$, starting with a symplectically integral lattice $\Lambda$.
Importantly, the condition that the translations $\{\hat T(\mathbf s_j)\}$ commute imposes constraints on the attainable code dimension $d$. 
The determinant of the lattice, which sets the code dimension, scales as 
\begin{equation}\label{eq:volumeScalingLattice}
\mathrm{det}(c\Lambda) = c^{4m} \mathrm{det}(\Lambda).
\end{equation}
On the other hand, the symplectic Gram matrix, which sets the commutation relations of the stabilizers, scales as 
\begin{equation}\label{eq:areaScalingLattice}
A(c\Lambda) = c^2 A(\Lambda).
\end{equation}
In other words, the code dimension scales as a volume, while the commutation relations between the stabilizers scale as an area. Note that the single-mode case is special since area and volume coincide, and scaling a single-mode lattice by any constant $c = \sqrt a$ with $a\in \mathbb Z$ an integer results in a valid code.
However, multimode ($m \geq 2$) GKP codes are more constrained than their single-mode counterpart since the conditions given by \cref{eq:volumeScalingLattice,eq:areaScalingLattice} do not coincide.

We denote $\Lambda_0$ the smallest integral lattice that can be built by scaling $\Lambda$, and we denote the associated symplectic Gram matrix $A_0$~\footnote{$\Lambda_0$ can be found by a deflation $\Lambda_0 = \Lambda/c'$, where $c' = \sqrt{\text{gcd}(A)}$ with $\text{gcd}(A)$ the greatest common divisor among all the matrix elements of $A$.}. 
Under the scaling by $c$, the elements of $A$ should remain integers, which imposes the constraint that $c^2 = a \in \mathbb Z$ be an integer. Combining with \cref{eq:volumeScalingLattice}, we get that a lattice $\Lambda_0$ allows for codes of dimension 
\begin{equation}
\label{eq:dimensionCondition}
d = a^{m} \det(S_0).
\end{equation}
In particular, if $\det(S_0) = 1$ such that the lattice encodes a single logical state, scaling the size of the lattice allows to encode ensembles of $m$ qudits of dimension $a$. For a single-mode (two-dimensional lattice), any lattice can be rescaled such that $\mathrm{det}(\Lambda_0) = 1$, and since $m=1$ all code dimensions are possible.

While scaling a lattice allows one to build codes out of known symplectically integral lattices, it is not the only allowed operation. Consider the basis $S' = SO$ where $O \in \mathrm{O}(\mathbb R,2m) - \mathrm{Sp}(\mathbb R,2m)$ is orthogonal but \emph{not} symplectic. For example, a two-mode rotation in the quadrature coordinates $\hq_1,\hq_2$ that leaves their conjugate coordinates $\hp_1,\hp_2$ invariant is orthogonal but not symplectic. With respect to the Euclidean norm, $S$ and $S'$ are bases for two equivalent lattices, $S S^T = S' S'^T$, but the symplectic Gram matrix associated with these bases are not the same, $S\Omega S^T \neq S'\Omega S'^T$.
This means that for a general basis $S$ where $A = S\Omega S^T$ is not integral, there can be an orthogonal transformation such that $A' = S'\Omega S'^T$ is integral. In other words, viewing the symplectic form as a sum of areas, there can be special ``rotations'' of the lattice where these areas add up to integers for each pair of stabilizers.
To search for codes that do not respect \cref{eq:dimensionCondition}, our strategy is to scale a lattice to the desired volume, and then search for an orthogonal transformation $O$ such that $A'$ is integral, see \cref{sect:IntegralLatticeSearch}. Note that once a solution is found using this procedure, say for dimension $d'$, then all solutions of dimension $d' a^m$ are also valid per \cref{eq:dimensionCondition}.

\begin{table}
\begin{tabular}{ |c |c | c| }
\hline
 & Classical Lattice & Grid Code\\ \hline
\multirow{2}*{Stabilizers} & $\Lambda =$ & $\mathcal S =$ \\
 & $\left\{S^T \mathbf a \mid \mathbf a \in \mathbb Z^{2m}\right\}$ & $\left\{\Pi_{j=1}^{2m} \hT(\mathbf s_j)^{a_j} \mid \mathbf a \in \mathbb Z^{2m}\right\}$ \\ \hline
Condition on  & \multirow{3}*{$A$ integral} & \multirow{3}*{$[\hat T(\mathbf s_j), \hat T(\mathbf s_k)] = 0$} \\
commutation &  &  \\ 
of stabilizers &  &  \\ \hline
Code dimension & $\det A = d^2$ & $d$ logical states \\ \hline
Logical  & \multirow{2}*{$\mathbf p \in \Lambda^*$} & \multirow{2}*{$\hU(\bar P) = \hat T(\mathbf p)$} \\
Pauli operators & & \\
  \hline
\end{tabular}
\caption{Expression of different QEC concepts for lattices in $\mathbb R^{2m}$ and grid codes in $m$ modes.}\label{table:ClassicalQuantumMapping}
\end{table}

In this work we mainly focus on encoding a qubit $d=2$ in multiple modes. Each vector of the dual lattice $\Lambda^*$ is associated with a logical Pauli operator $\bar P$, $\bar P \in \{\bar X,\bar Y,\bar Z\}$. For each logical Pauli operator $\bar P$, we choose a base representative $\mathbf p_0$ such that any translation by $\mathbf p$ with $\hat T(\mathbf p) = \hU(\bar P)$ is expressed as $\mathbf p = \mathbf p_0 + \vlambda$ for some $\vlambda \in \Lambda$. We define the set of points $P = \mathbf p_0 + \Lambda$.
We define a $3 \times 2m$ matrix where each row sets one of these base representatives
\begin{equation}\label{eq:L0definition}
L_0 = \begin{pmatrix}
\mathbf x_0 \\
\mathbf y_0 \\
\mathbf z_0 
\end{pmatrix},
\end{equation}
with $\mathbf x_0,\mathbf y_0$ and $ \mathbf z_0$ associated with the logical Pauli operators $\bar X,\bar Y$ and $\bar Z$, respectively. Without loss of generality, we choose the representatives of minimum length with respect to the Euclidean norm for each Pauli operator, such that $\hat T(\mathbf p) = \hU(\bar P) \Rightarrow |\mathbf p| \geq |\mathbf p_0|$. The logical identity operator is omitted from \cref{eq:L0definition} since it corresponds to the stabilizers with the trivial base representative $\mathbf 0$. 

We will also make use of GKP codes encoding a single state, $d=1$. We refer to such GKP codes as ``qunaught'' states since they carry no quantum information, and we associated them with a subscript $\varnothing$~\cite{Walshe20q,Noh22b}. The single-mode square qunaught state is sometimes referred to as a sensor state since it can be used to precisely measure translations in two conjugate quadrature coordinates simultaneously~\cite{Duivenvoorden17a}.

Finally, when it is clear from the context, we sometimes abuse the notation and refer to the vectors $\mathbf s_j,\mathbf p \in \mathbb R^{2m}$ as the quantum translation operators they are associated with. 

\subsection{Finite-Energy Multimode GKP}\label{subsect:finiteEnergyGKP}
In the previous section, we considered grid states that extend infinitely in phase space. In other words, the eigenstates of translation operators are superpositions of infinitely squeezed states containing an infinite amount of energy. Here, we consider the
finite-energy GKP states that are obtained by taking a $m$-mode envelope of the form
\begin{equation}\label{eq:envelopeOperator}
\hat E_\beta = \exp\left(-\sum_{j=1}^m \beta_j \hat n_j\right),
\end{equation}
where $\beta_j$ parametrizes the size of the GKP in the $j$th mode. One can interpret this envelope as the multiplication of the grid state by the density matrix of a $m$-mode thermal state, motivating the notation choice ``$\beta$''. Alternatively, \cref{eq:envelopeOperator} can be interpreted as a gaussian envelope in phase space since $\hat n_j = (\hq_j^2 + \hp_j^2 + 1)/2$. The finite-energy logical code words $\bar \psi=0,1$ are then defined as 
\begin{equation}
\ket{\bar \psi_\beta} = \mathcal N_{\psi,\beta} \hat E_\beta \ket{\bar \psi_0},
\end{equation}
with $\mathcal N_{\psi,\beta}$ a normalization constant and $\ket{\bar \psi_0}$ the ideal, infinite-energy code words.

Generalizing the approach from Ref.~\cite{Royer20a}, we define finite-energy stabilizers from the similarity transformation induced by the envelope,
\begin{equation}\label{eq:finiteEnergyStabilizers}
\begin{aligned}
\hat T_{j,\beta} &= \hat E_\beta \hat T(\mathbf s_j) \hat E_\beta^{-1},\\
&=  \exp\{i \hat E_\beta \hg_j \hat E_\beta^{-1}\}.
\end{aligned}
\end{equation}
Finite-energy code words are exact +1 eigenstates of these operators, $\hat T_{j,\beta} \ket{\bar \psi_\beta} = \ket{\bar \psi_\beta}$.
From \cref{eq:translationGenerators}, the generators of translations of the stabilizer group transform to 
\begin{equation}
\hat E_\beta \mathbf{\hg} \hat E_\beta^{-1} = -l S\Omega \left\{\cosh\left[\mathrm{Diag}(\pmb \beta)\right]+ i\, \sinh\left[\mathrm{Diag}(\pmb \beta)\right] \Omega\right\}\hvx,
\end{equation}
where we have defined the $2m$-dimensional vector $\pmb \beta = (\beta_1,\, \beta_2,\,...\beta_m ) \otimes (1,1)$ and Diag corresponds to the operation of building a diagonal matrix from a vector. In the limit $\beta \rightarrow 0$, we recover the translation operators, $\hat T_{j,0} = \hat T(\mathbf s_j)$. 

We consider a homogeneous envelope size $\beta_j = \beta$ for all $j$ in the rest of the paper, such that $\hat E_\beta = \exp(-\beta \hat n)$, with $\hat n = \sum_j \hat n_j$ the total excitation number. For this choice, operations that commute with $\hat n$ also commute with the envelope, and we refer to such operations as \emph{envelope-preserving}. Gaussian operations in that category, $\hat Q(O)$ with $O \in \mathrm{Sp}(2m,\mathbb R)\cap \mathrm{O}(2m,\mathbb R)$, can be implemented by a combination of beam-splitters and phase shifters, i.e. passive linear optics. Envelope-preserving gates also include non-linear gates such as the unitaries generated by Kerr and cross-Kerr interactions. Importantly, envelope-preserving operations are exact for finite-energy GKP codes.

Taking the logarithm of the stabilizers $\hat T_{j,\beta}$, and in analogy with continuous-variable cluster states~\cite{Menicucci07e,Gu09f}, we define the finite-energy nullifiers of the code
\begin{equation}\label{eq:finiteEnergyNullifier}
\hd_j = \left(\frac{\left[\mathbf s_j^T\Omega\hvx\;\mathrm{ mod }\;l/\cosh\beta\right]}{\sqrt{2|s_j|\tanh(\beta)}} - i \sqrt{\frac{\tanh(\beta)}{2|s_j|}} \mathbf s_j^T\cdot \hvx \right).
\end{equation}
The finite-energy code words are then also defined by $\hd_j \ket{\bar \psi_\beta} \approx 0\; \forall\; j$~\footnote{This equation is only approximately true. In particular, the logarithm of the finite-energy stabilizers is not well-defined since the operators $\hat T_{j,\beta}$ have eigenvalues over the branch cut of the complex logarithm.}. Without the modular part of the first term, $\hat d_j$ corresponds to the nullifier of a finitely squeezed state.

\subsection{Gauge Choices}\label{sect:GaugeChoice}
As demonstrated by \cref{eq:LambdaDef,eq:stabGroup}, the points of the lattice $\Lambda$ are in one-to-one correspondence with the elements of the stabilizer group $\mathcal S$. However, we remark that $\vlambda \in \Lambda$ does not imply that $\hT(\vlambda) \in \mathcal S$, and the correspondance between elements of $\Lambda$ and $\mathcal S$ requires an additional phase. For example, take two stabilizers $\hT(\mathbf s_1)$ and $\hT(\mathbf s_2)$ ``only'' commuting by $2\pi$, i.e. such that $\mathbf s_1^T \Omega \mathbf s_2 = 1$. Following \cref{eq:commRelDisplacements}, we have $\hT(\mathbf s_1 + \mathbf s_2)\ket{\psi} = - \hT(\mathbf s_1)\hT(\mathbf s_2)\ket{\psi} = -\ket{\psi}$, for all logical code words $\ket{\psi}$. Within the lattice points $\Lambda$, we thus distinguish between two subsets $\Lambda_\pm \subseteq \Lambda$ such that, for states $\ket{\psi}$ in the code space,
\begin{equation}
\Lambda_\nu = \left\{\vlambda \in \Lambda\, \mid\, \hT(\vlambda)\ket{\psi} = \nu \ket{\psi}\right\},
\end{equation}
with $\nu = \pm$. For example, for the single-mode square qunaught state
 with generator matrix $S_\varnothing = \mathbb I_2$, we have the generator vectors $\mathbf s_1, \mathbf s_2 \in \Lambda_+ $, while their sum $\mathbf s_1 + \mathbf s_2 =(1\,; 1)\in \Lambda_-$. For a qubit encoded in a single-mode, all lattice vectors associated with the stabilizer group necessarily belong to $\Lambda_+$ since there are only two generators and the dimension condition imposes that $|\mathbf s_1^T \Omega \mathbf s_2| = 2$.

As will become clear below, it is useful to allow different gauge choices $\pmb \mu \in \mathbb Z_2^{2m}$ such that for the generators of the stabilizer group $\hT(\mathbf s_j)\ket{\psi} = (-1)^{\mu_j}\ket{\psi}$ for states $\ket{\psi}$ in the code space. The stabilizer group is correspondingly updated to
\begin{equation}
\label{eq:stabGroupMu}
\mathcal S_{\vmu} = \left\{\Pi_{j=1}^{2m} [(-1)^{\mu_j}\hT(\mathbf s_j)]^{a_j} \mid\, \mathbf a \in \mathbb Z^{2m}\right\},
\end{equation}
 Different choices for $S$ or $\pmb \mu$ can lead to different subsets $\Lambda_\pm$ for the same base lattice $\Lambda$~\footnote{Note that the eigenvalues of the translation operators $\hT(\vlambda)$ lie on the complex unit circle, and in all generality we could have defined a continuous gauge $\mu_j \in [0,2)$. Here we restrict ourselves to the choice $\mu_j = 0,1$ for simplicity, as most operations of interest preserve this structure.}.

Defining $A_\text{\lltriangle}$ as the lower triangular part of the symplectic Gram matrix $A$ given by \cref{eq:GramMatrix}, we classify the lattice vectors, $\vlambda \in \Lambda$, according to the gauge $\vmu$
\begin{equation}\label{eq:nuFunction}
\begin{aligned}
\nu_{\vmu}(\vlambda) ={}& \exp\left\{i\pi\vlambda^TS^{-1}\left[A_\text{\lltriangle}(S^{-1})^T\vlambda + \vmu\right]\right\}.
\end{aligned}
\end{equation}
Since $\Lambda$ is symplectically integral, we have that $\nu_{\vmu}(\vlambda) \in \{\pm1\}$. Moreover, if $\vlambda \in \Lambda_\nu$, then its inverse is also in the same set, $-\vlambda \in \Lambda_\nu$.

In a similar fashion to the stabilizer gauge $\vmu$, we define a gauge for the logical Pauli operators $\pmb \upsilon \in \mathbb Z_2^3$, which is equivalent to a so-called Pauli frame~\cite{Knill05a}. Accordingly, we define the eigenstate of the logical Pauli operator $\bar P$ such that $(-1)^{\upsilon_p} \hT(\mathbf p_0)\ket{\psi_{+P}} = \ket{\psi_{+P}}$. We note that one of the three elements of $\vupsilon$ is redundant, as the gauge is fully set by $\vupsilon_x$ and $\vupsilon_z$.
Here we choose to keep all three elements for convenience.
In a similar manner to the lattice vector subsets $\Lambda_\pm$, we define the subsets $P_\pm$ for each Pauli operator $P\in\{X,Y,Z\}$ as
\begin{equation}
P_{\nu} = \left\{\mathbf p \in P\, |\, \hT(\mathbf p)\ket{\psi_{+P}} = \nu \ket{\psi_{+P}}\right\},
\end{equation}
which depends on the stabilizer gauge $\vmu$, the Pauli frame gauge $\pmb \upsilon$ and the base representatives $\{\mathbf p_0\}$.
The sign associated with a particular vector $\mathbf p \in P$ is computed using
\begin{equation}\label{eq:nuFunctionPauli}
\begin{aligned}
\nu^P_{\pmb \mu,\pmb \upsilon}(\mathbf p) ={}& \expo{i\pi\left[\mathbf p_0^T \Omega \mathbf p + \upsilon_p\right]}\nu_{\vmu}(\mathbf p - \mathbf p_0),
\end{aligned}
\end{equation}
and logical Pauli operators are given by
\begin{equation}
\hU(\bar P) = \nu^P_{\mathbf \vmu,\pmb \upsilon}(\mathbf p) \hat T(\mathbf p),
\end{equation}
for all $\mathbf p \in P = \mathbf p_0 + \Lambda$. 

In order for the Pauli eigenstates $\ket{\psi_{\pm P}}$ to have eigenvalue $\pm 1$, we note that we should have $\nu_{\vmu}(2\mathbf p_0) = 1$, where $2\mathbf p_0 \in \Lambda$ by construction. Indeed, the eigenvalues of translations by $\hat T(\mathbf p_0)$ are constrained to be $\sqrt{\nu_{\vmu}(2\mathbf p_0)}$, such that $\nu_{\vmu}(2\mathbf p_0) = -1$ implies that $\hat T(\mathbf p_0)\ket{\psi} = \pm i \ket{\psi}$. 
To constrain the eigenvalues of the Pauli operators to be real, we add the condition that
\begin{equation}\label{eq:gaugeValidityCondition}
2 A^{-1}\vmu\; \mathrm{ mod }\; 2 = \mathbf 0.
\end{equation}
This condition is obviously always respected in the trivial gauge $\vmu = \mathbf 0$, and in the special case of a single-mode GKP qubit, this is the only gauge respecting this condition. However, for multimode lattices, there are multiple gauges allowed. 

\emph{Gauge Updates}\textemdash
It is useful to consider the effect of different operations on the gauge $\vmu$. For example, consider a displacement by half a lattice vector $\hT(\pmb \tau/2)$ with $\pmb \tau \in \Lambda$, an operation useful for QEC purposes; see \cref{sect:ErrorCorrection}. While code words are not (generally) eigenstates of $\hT(\pmb \tau/2)$, the displaced state $\ket{\psi'} = \hT(\pmb \tau/2) \ket{\psi}$ is still an eigenstate of all stabilizers.
As a result, $\ket{\psi'}$ can still be considered inside the code space after updating the gauges to
\begin{subequations}\label{eq:gaugeUpdateDisplacement}
\begin{align}
\vmu' &= (\vmu + S\Omega \pmb \tau) \,\,\mathrm{mod}\,\, 2,\\
\pmb \upsilon' &= (\pmb \upsilon + L_0\Omega \pmb \tau) \,\,\mathrm{mod}\,\, 2.
\end{align}
\end{subequations}
The above equation also means that given a state in a gauge $\vmu$, we can set the gauge to the desired value $\vmu_\mathrm{target}$ by applying a translation $\hat T(\pmb \tau/2)$ to the state, with
\begin{equation}\label{eq:gaugeSettingDisplacement}
\pmb \tau = -\Omega S^{-1}[(\vmu_\mathrm{target} + \vmu)\,\,\mathrm{mod}\,\, 2] .
\end{equation}

It was already pointed out that single-mode GKP code words are similar to the Landau levels of an electron in a two-dimensional plane with a uniform magnetic field. Correspondingly, translations in phase space are similar to magnetic translation operators~\cite{Gottesman01a,Rymarz21g}. Carrying this comparison further, the gauge $\vmu$ we defined in this section plays a role similar to the electromagnetic gauge fields, which should be taken into account when considering physical symmetries of the electronic system~\cite{Tassie64b}.

\subsection{Logical Pauli and Clifford Gates}
In GKP codes, logical Pauli operations can be realized in a fault-tolerant manner by applying translations $\hat T(\mathbf p)$. However, one potential issue with performing gates this way is that the energy contained in the GKP state increases. Equivalently, translation operators are not envelope-preserving, $[\hat T(\mathbf p),\hat E_\beta] \neq 0$. This issue can be partly mitigated through various strategies such as choosing random representatives of a given Pauli, symmetrizing the direction of the displacements~\cite{Tzitrin20a}. One can also use the error-correction circuits introduced below in \cref{sect:ErrorCorrection} to correct the ``envelope errors'' caused by these translations.
A more attractive option is to perform logical Pauli operations in software by updating $\vupsilon$. For single-mode codes, this approach can be generalized to the whole single-qubit Clifford group by performing computations in the so-called Clifford frame~\cite{Aaronson04y,Chamberland18v,Grimsmo21t}, adapting logical two-qubit gates depending on the frame.\\ \indent
Next, consider a symplectic operation $M$ mapping lattice vectors to other lattice vectors, $M\Lambda = \Lambda$, by which we mean that $M\vlambda\in \Lambda$ for all $\vlambda \in \Lambda$. In other words, $M$ is a lattice symmetry that preserves the commutation relations of the modes.
Since $M$ leaves the lattice invariant, it maps dual lattice vectors to other dual lattice vectors. Equivalently, its associated quantum representation, $\hat Q(M)$, corresponds to a logical Clifford operation in the code space, mapping logical Pauli operators to other logical Pauli operators. Importantly, this operation does not necessarily preserve the lattice subsets $\Lambda_\pm$ and, in general, we have $M\Lambda_\nu \neq \Lambda_\nu$. However, this can be taken into account by properly updating the gauges, see \cref{sectApp:gaugeUpdates} for details.\\ \indent
We remark that, for a given lattice symmetry $M$, there can be special gauges where $M \Lambda_\nu = \Lambda_\nu$. Such gauges affect the symmetries of the code words, and are especially useful for non-Clifford operations. For example, one could perform a controlled-M gate by coupling the modes to an ancilla qubit. Without the condition that $M \Lambda_\nu = \Lambda_\nu$, applying a superposition of identity and $\hat Q(M)$ leaves the state in a superposition of $\pm 1$ eigenstates for some stabilizers, which is outside of the code space.

\subsection{Envelope-Preserving Non-Clifford Gates}\label{subsect:EnvelopePreservingNonCliffordGates}
In this section, we investigate how the symmetries of a lattice can allow easier implementation of non-Clifford gates. More precisely, we show how the symmetries of the lattice $\Lambda$ are reflected in the code words, and how these properties can be used to design non-Clifford gates.
We focus in particular on lattice symmetries which are isometries, i.e.\ operations $M$ where $M \Lambda = \Lambda$ and $M\in \mathrm{O}(2m,\mathbb R)\cap\mathrm{Sp}(2m,\mathbb R)$. We remark that for $M$ to correspond to a non-trivial logical operation, dual lattice vectors correponding to distinct logical operators must have the same length. For example, an envelope-preserving logical Hadamard gate can only be implemented in a code where $|\mathbf x_0| = |\mathbf z_0|$.

Below, we assume that the gauge is set such that no gauge update is required after applying $\hat Q(M)$, i.e.\ we set $\vmu$ such that $M\Lambda_\nu = \Lambda_\nu$. This condition ensures that applying superpositions of different powers of $M$ leaves the state inside the code space.

Since distances are preserved by $M$ and the number of lattice points of a given length is finite, $M$ has finite order and there is an integer $p$ such that $M^p = I$. Moreover, there is an operator $\hat F$ with integer spectrum such that $\hat Q(M) = \exp\{i (2\pi /p) \hat F\}$. Indeed, the identity operator can be expressed as $\hat I = \hat Q(I) = \hat Q(M^p) = \hat Q(M)^p = \exp\{i 2\pi  \hat F\}$, which forces the eigenvalues of $\hat F$ to be integers. We label the eigenstates of $\hat F$ by their eigenvalue $j \in \mathbb Z$ and a degeneracy index $k$, $\hat F\ket{j,k}_F = j\ket{j,k}_F$. For single-mode codes, the only possible isometries are rotations by an angle $2\pi/p$, with generator operator $\hat F = \hat n$. In this case, the $F-$basis reduces to the standard Fock basis and due to the crystallographic theorem $p$ is restricted to $p \in \{1,2,3,4,6\}$.

Knowing that $M$ acts as a Clifford gate $\bar G$ in the logical subspace, the operation $\hat Q(M) = \hU(\bar G)$ has finite order $g$ within the code space. Since $\hat Q(M^p) = \hU(\bar G)^p = \hat I$, we have that $g$ divides $p$, i.e.\ $p = b g$ for some integer b. 
In the (possibly non-Pauli) basis where $\bar G$ is diagonal, $\{\ket{\bar G},\ket{\expo{i\frac{2\pi}{g}}\bar G}\}$ , the code words have a special structure with respect to the $F-$basis,
\begin{subequations}\label{eq:GdiagonalStates}
\begin{align}
\ket{\bar G} &= \sum_j \sum_k c_{0jk} \ket{pj,k}_F,\\
\ket{\expo{i\frac{2\pi}{g}}\bar G} &= \sum_j \sum_k c_{1jk} \ket{pj + b,k}_F,
\end{align}
\end{subequations}
and we compute that
\begin{equation}
\begin{aligned}
\hat Q(M)\ket{\expo{i\frac{2\pi a}{g}}\bar G} &= \expo{i \frac{2\pi}{p} \hat F}\ket{\expo{i\frac{2\pi a}{g}}\bar G},\\
&= \expo{i\frac{2\pi a}{g}}\ket{\expo{i\frac{2\pi a}{g}}\bar G},\\
&= \hat U(\bar G)\ket{\expo{i\frac{2\pi a}{g}}\bar G},
\end{aligned}
\end{equation}
with $a \in \{0,1\}$. The value of the coefficients $c_{ajk}$ depend on the envelope size $\beta$, and do not affect the properties derived in this section.

Consider the example of a single-mode square GKP code encoding a qubit. That code is based on the square lattice which is invariant under a quarter-turn rotation $M = R(\pi/2)$. This isometry has order $p=4$, and within the code space $\hat Q(M) = \hat R(\pi/2)$ acts as a logical Hadamard gate with order $g=2$. Rotations in phase space are generated by the photon number operator $\hat R(\pi/2) = \exp\{i(2\pi/4)\hat n\}$ and, following \cref{eq:GdiagonalStates}, the eigenstates of the Hadamard gate can be expressed in the Fock basis as
\begin{subequations}
\begin{align}
\ket{\bar H_{\square}} &= \sum_j c_{0j} \ket{4j},\\
\ket{-\bar H_{\square}} &= \sum_j c_{1j} \ket{4j + 2}.
\end{align}
\end{subequations}

From the properties above, we now show how to perform a non-Clifford gate $\sqrt{\bar G}$ when $\bar G$ has even order $g \in 2\mathbb Z$. This non-Clifford gate is based on a quartic interaction $\hat F^2$.
Indeed, leveraging the special support of the states \cref{eq:GdiagonalStates} in the $F-$basis, we have
\begin{equation}\label{eq:singleQubitNonClifford}
\hU\left(\sqrt{\bar G}\right) = \expo{i \frac{g\pi}{p^2}\hat F^2}.
\end{equation}
Applying this gate in the diagonal basis of $\bar G$, we obtain 
\begin{equation}
\begin{aligned}
\hU\left(\sqrt{\bar G}\right)\ket{\expo{i\frac{2\pi a}{g}}\bar G} &= \expo{i \frac{g\pi}{p^2}\hat F^2}\ket{\expo{i\frac{2\pi a}{g}}\bar G},\\
&= \sum_j \sum_k c_{ajk} \expo{i \frac{g\pi}{p^2} p^2 (j + \frac{a}{g})^2}\ket{pj + ba,k},\\
&= \expo{i \frac{\pi a}{g}}\ket{\expo{i\frac{2\pi a}{g}}\bar G},
\end{aligned}
\end{equation}
as desired and with $a\in \{0,1\}$. 

Returning to the example of the Hadamard gate in the single-mode square code where $\hat F = \hat n$, $p=4$ and $g=2$, we can implement a square root of Hadamard gate through a Kerr unitary 
\begin{equation}
\hU_{\square}\left(\sqrt{\bar H}\right) = \expo{i \frac{\pi}{8}\hat n^2}.
\end{equation}
Although the special Fock state support of the Hadamard eigenstates was originally mentioned in Ref.~\cite{Gottesman01a}, the fact that a Kerr gate can implement a $\sqrt{\bar H}$ gate has, to the best of our knowledge, not been pointed out before.

Using the same properties as above, we also derive entangling non-Clifford gates between two qubits A and B,
\begin{equation}\label{eq:twoQubitNonClifford}
\hU(C^{(\bar G)}\bar G) = \expo{i \frac{2\pi g}{p^2}\hat F_A\hat F_B},
\end{equation}
which can be understood as effecting $\bar G$ on qubit B conditioned on qubit A being in the state $\ket{\expo{i\frac{2\pi}{g}}\bar G}$. For two single-mode square code, we obtain a Hadamard-controlled Hadamard gate based on a cross-Kerr interaction~\cite{Grimsmo20a}, 
\begin{equation}
\hU_{\square}(C^{(\bar H)}\bar H) = \expo{i \frac{\pi}{8}\hat n_A\hat n_B}.
\end{equation}

A crucial property of the gates derived above is that they are \emph{exact} for finite-energy GKP codes. Indeed, since by construction the unitaries $\exp\{i\theta \hat F\}$ preserve the excitation number $\hat n$ for any angle $\theta$, the generator of that unitary commutes with the excitation number, $[\hat F,\hat n] = 0$, which then implies that $[\exp\{i\theta \hat F^2\},\hat E_\beta] = 0$. As a result, the gates defined by \cref{eq:singleQubitNonClifford,eq:twoQubitNonClifford} preserve the code space irrespective of the size of the envelope. This property is in stark contrast to the cubic phase gate~\cite{Gottesman01a}, whose fidelity is intrinsically limited in finite-energy GKP states~\cite{Hastrup21a}. 

In order for the code words to exhibit the $F-$basis structure shown in \cref{eq:GdiagonalStates}, the gauge should be set such that $M\Lambda_\nu = \Lambda_\nu$. This enforces that applying a superposition of $M$ and identity leaves the state inside the code space. However, there are generally two inequivalent gauges where this is respected, where the code words are $\pm 1$ eigenstates of the modified stabilizers. Importantly, the logical Clifford operation implemented by the same unitary can differ in the two gauges, impacting the order of the gate $g$. Assuming that the gates \cref{eq:singleQubitNonClifford,eq:twoQubitNonClifford} are implemented by evolving the system under an Hamiltonian $\hH \propto \hat F^2$, it is desirable to choose the gauge where $g$ is minimal in order to reduce the interaction time. 

For example, in the single-mode square code, the rotation $\hat R(\pi/2)$ leaves the gauges $\vmu = \mathbf 0$ and $\vmu = (1,1)$ intact. In the former gauge, this rotation corresponds to a logical Hadamard gate with order $g=2$, while in the latter gauge it corresponds to a logical $\sqrt{\bar Y}$ gate with order $g=4$.

The properties above are reminiscent of rotation-symmetric codes~\cite{Grimsmo20a}, a class of bosonic codes designed such that code words have support on specific Fock states. However, as defined in Ref.~\cite{Grimsmo20a}, a rotation $\hat R$ in these codes implements a logical \emph{Pauli} operation, and a Kerr (cross-Kerr) interaction corresponds to a Clifford $\bar S$ ($C\bar Z$) gate. In contrast, for the square GKP code, a rotation $\hat R(\pi/2)$ corresponds to a Clifford gate, and a Kerr (cross-Kerr) interaction corresponds to a non-Clifford gate. These properties are summarized in \cref{table:GateComparison}, where we see that for a given Hamiltonian order, GKP codes allow gates deeper in the Clifford hierarchy compared to rotation-symmetric codes~\cite{Gottesman99e}. In this table we focus on the comparison between the square GKP code and the four-legged cat code~\cite{Leghtas13a,Grimsmo20a} since they have the same order of rotation symmetry $\hat R(\pi/2)$. In principle, a non-Clifford gate can be realized in rotation-symmetric codes by going to higher-order interactions, for example by implementing an octic Hamiltonian $\hH \propto \hat n^4$, and gates based on higher powers of $\hat F$ can also be derived for GKP codes. However, any non-Clifford gate is sufficient to perform universal quantum computation. Moreover, the propagation of errors worsens with higher-order gates, and they are also more difficult to engineer, so that going to higher order is not likely to be the optimal route to universality.

\begin{table}
\begin{tabular}{ |c|c |c | c | c| }
\hline
\multirow{2}*{Hamiltonian} & \multirow{2}*{Unitary} & \multicolumn{2}{c|}{Square GKP code} & \multirow{2}*{Cat Code}\\ 
&& $\vmu = (0;0)$ & $\vmu = (1;1)$ & \\ \hline
Quadratic&$\hat R(\pi/2)$ & $\bar H$ & $\sqrt{\bar Y}$ & $\bar Z$ \\ \hline
\multirow{4}*{Quartic}&$\expo{i\frac{\pi}{4} \hat n^2}$ & $\bar H$ & $\sqrt[4]{\bar Y}$ & $\bar Z$ \\ 
&$\expo{i\frac{\pi}{8} \hat n^2}$ & $\sqrt{\bar H}$ & - & $\bar S = \sqrt{\bar Z}$ \\ 
&$\expo{i\frac{\pi}{2} \hat n_A\hat n_B}$ & $\bar I$ & $C^{(\bar Y)}\sqrt{\bar Y}$ & $\bar I$ \\ 
&$\expo{i\frac{\pi}{4} \hat n_A\hat n_B}$ & $C^{(\bar H)}\bar H$ & - & $C\bar Z$ \\ \hline
Octic&$\expo{i\frac{\pi}{64} \hat n^4}$ & $\sqrt[4]{\bar H}$ & - & $\bar T = \sqrt[4]{\bar Z}$\\ \hline
\end{tabular}
\caption{Comparison of gates in the GKP square code and the four-legged cat state. The gate $C^{(\bar Y)}\sqrt{\bar Y} = \exp\{i\frac{\pi}{8}(3(\bar Y_A + \bar Y_B) + 5(\bar I + \bar Y_A \bar Y_B))\}$ corresponds to the equivalent of a controlled-phase gate in the $\bar Y$ basis, and the gate $C^{(\bar H)}\bar H = \exp\{i\frac{\pi}{4}(3 + \bar H_A + \bar H_B + 3\bar H_A \bar H_B)\}$ corresponds to the equivalent of a controlled-Z gate in the Hadamard basis.}\label{table:GateComparison}
\end{table}

Since the gates introduced above are not Gaussian, they generally propagate errors in an unfavourable way. For example, an excitation loss before the Kerr gate \cref{eq:Tgate} leads to a large rotation error, $\expo{i\pi \hat n^2/8}\ha = \ha \expo{i\pi (\hat n-1)^2/8} \propto \ha \expo{i\pi \hat n/4}\hat U(\sqrt{\bar H})$.
These non-linear gates rely on precise ``refocusing'' of the lattice peaks at the end of the gate, or equivalently refocusing of the F-states phases, a process which can be impeded by errors. As a result, general errors during the non-linear gates leave the states far from the code space. This propagation of error could be mitigated by, for example, using these gates for magic state preparation and post-selecting only states where no errors are detected. Also, the fact that the non-linear gates commute with the total excitation number is beneficial to limit the propagation of errors. For instance, it implies that they commute with the ``no-jump'' part of the amplitude damping channel evolution, see $\hat K_0$ in \cref{eq:KrausOpsPhotonLoss} below.
The propagation of amplitude damping errors during cross-Kerr gates is investigated more thoroughly in Ref.~\cite{Grimsmo20a} in the context of rotation-symmetric codes.

\subsection{Codes and Lattice Packings}\label{sect:latticePackings}
A standard problem in lattice theory is to search for the densest possible lattice packings, and designing QEC grid codes based on these lattices generally leads to good error correction properties. Placing a sphere at each lattice point and increasing their radius until the spheres touch, the lattice packing ratio $\Delta$ is defined as the fraction of space covered by the spheres. This ratio is maximized when the sphere radius is maximized.\\ \indent
Consider a Gaussian translation error model, where a translation error $\hat T(\mathbf e)$ is applied with a probability density $P(\mathbf e) = \exp\{-|\mathbf e|^2/(2\sigma^2)\}/(2\pi\sigma^2)^{m}$. In some contexts, this error model can be a good approximation to more physically relevant error channels such as amplitude damping~\cite{Gottesman01x,Albert18a,Noh19b,Wang19k,Noh20b,Conrad21t}. Each dual lattice vector is associated with a logical Pauli operator, such that maximizing the distance between dual lattice points minimizes the probability that a translation error causes a logical error. Maximizing the lattice packing ratio of the dual lattice therefore means that errors with larger norm can be corrected, and we can (in principle) correct all translation errors where
\begin{equation}
|\mathbf e| \leq 2 \left(\frac{\Delta}{dV_{2m}}\right)^{\frac{1}{2m}},
\end{equation}
where we have defined $V_{2m}$ as the volume of a unit radius sphere in $\mathbb R^{2m}$. An error larger than the bound above can enter the sphere of a point associated to a different logical operator, leading to a logical error. We can get a simple estimate for the probability that an error occurs by computing the probability that the norm of the error is larger than the radius of the spheres in the sphere packing problem, 
\begin{equation}
P(\mathrm{error}) \approx \mathrm{erfc}\left[\frac{1}{\sigma \sqrt{m}}\left(\frac{\Delta}{dV_{2m}}\right)^{\frac{1}{2m}}\right].
\end{equation}
Considering lattices with denser lattice packing (larger $\Delta$) leads to smaller error probabilities. Increasing the dimension of the code $d$ leads to larger error probabilities since the dual lattice points are closer together.
\\ \indent
Although designing codes based on dense lattice packings intuitively leads to good error-correcting codes, we note that the simple model above ignores several important design considerations, amongst which we mention three. First, some translation errors outside of the considered spheres can be corrected, and a more precise estimate for the error probability is obtained by considering the Voronoï cell of the dual lattice. Secondly, the model above assumes an ideal error correction procedure. In practice, the stabilizer measurements required for error correction are themselves noisy, which can affect different lattices in different ways. Thirdly, the geometry of the lattice is important in the sense that some neigbouring dual lattice points can be logically equivalent. This second consideration is only relevant when $m \geq 2$, since for two-dimensional lattices neighboring dual lattice points are always associated to distinct logical operators.

There is extensive literature on integral lattices with good lattice packing ratios~\cite{Conway13c}, defined such that the (standard) Gram matrix $G = S S^T$ contains only integers. Although one might hope to use them as a starting point to build interesting QEC codes, integral lattices are not generally \emph{symplectically} integral, and vice versa. 
The mapping from a classically integral lattice to a QEC code is further complicated by the fact that it is not unique. Indeed, as mentioned above in \cref{subsect:generalMultimodeTheory}, one can rotate the lattice around different axes which can change the symplectic form of generator vectors. Applying these rotations on the quadrature coordinates, the commutation relations are not preserved, and these rotations do not correspond to physical operations.
More precisely, $G = SS^T$ stays invariant under orthogonal transformations $S' = SO$ with $O\in \mathrm{O}(\mathbb R,2m)$, while $A =S\Omega S^T$ does not when $O\notin \mathrm{Sp}(\mathbb R,2m)$. 

One family of interesting codes are given by symplectic lattices, i.e. lattices that admit a generator matrix that is also a symplectic matrix ($S\Omega S^T =\Omega$). These lattices encode a single state, $d=1$, and by extension can also encode ensembles of $m$ qudits, \cref{eq:volumeScalingLattice}~\cite{Harrington01a,Harrington04t}. 

\subsection{Lattice Examples}\label{sect:latticeExamples}

We review below a few common lattices with good sphere-packing properties~\cite{Conway13c}, and the QEC codes they allow. In \cref{table:qubitLattices}, we summarize the main lattices studied in this work to encode qubits. 
In \cref{sect:qubitCodeConstruction}, we provide an explicit construction of symplectically integral lattices starting from qubit stabilizer codes, and show how the lattices of the current section relate to this construction. 

\emph{Single-mode Lattices}
\textemdash
First, we review some of the single-mode codes based on two-dimensional lattices.
We refer to the single-mode square lattice encoding a single state, $d=1$, as the square qunaught state with generator matrix $S_\varnothing = \mathbb I_{2}$. We further define the rectangular qunaught state with generator matrix
\begin{equation}
S_{\varnothing_\eta} = \begin{pmatrix} \eta & 0 \\
0 & 1/\eta 
\end{pmatrix},
\end{equation}
with $\eta \in \mathbb R^+$. The associated (single) code word is given in the position basis as
\begin{equation}
\label{eq:qunaughtStatePositionBasis}
\ket{\varnothing_\eta} \propto \sum_{j \in \mathbb Z} \ket{{j \eta l}}_x.
\end{equation}
The square GKP code encoding a qubit has a basis generator matrix $S_{\square} = \sqrt 2 \mathbb I_{2}$. We also make use of a rotated square qubit code, with generator matrix
\begin{equation}
S_{\Diamond} = \begin{pmatrix} 1 & 1 \\
1 & -1 
\end{pmatrix}.
\end{equation}
Although formally equivalent to the square code, we refer to it as the diamond qubit code.

It is interesting to note that the individual code words of the square qubit code are given by rectangular qunaught states $\ket{+Z_\square} = \ket{\varnothing_{\sqrt 2}}$ and $\ket{+X_\square} = \ket{\varnothing_{1/\sqrt 2}}$. The $-1$ code words can be obtained by setting the gauge $\vmu = (0,1)$ and $\vmu = (1,0)$, respectively.
The $\pm \bar Y$ Pauli eigenstates are, up to a $\pi/4$ rotation of phase space, given by the square qunaught state with $\eta=1$, but choosing the mixed gauge $\vmu = (0/1,1/0)$. Equivalently, the $\pm \bar Y$ Pauli eigenstates of the diamond qubit code are related to the square qunaught state $\eta=1$ through a gauge choice.

Finally, we also refer to the hexagonal qubit code with generator matrix
\begin{equation}
S_{\hexagon} = \frac{2}{\sqrt[4]{3}}\begin{pmatrix} 1 & 0 \\
-1/2 & \sqrt3/2 
\end{pmatrix}.
\end{equation}

\emph{Hypercubic Lattice}
\textemdash The simplest multimode lattice is the hypercubic lattice $\Lambda = \mathbb Z^{2m}$, with generator matrix $S = \mathbb I_{2m}$ for which $\mathrm{det}(\Lambda_{0}) = 1$. According to \cref{eq:dimensionCondition}, valid encodings can be obtained by scaling the hypercubic lattice are of dimension $a^m$, which is equivalent to encoding $m$ qudits of dimension $a$. In this situation, the generators become $S = \sqrt{a} \,\mathbb I_{2m}$, with the dual lattice given by $S^* = -\Omega/\sqrt{a}$, see \cref{eq:dualLatticeGenerator}. 

One approach to quantum computing based on GKP states is to concatenate the hypercubic lattice code encoding $m$ qubits with another qubit code~\cite{Kitaev03a,Menicucci14a,Vuillot19a,Noh20b,Terhal20a}. In this approach, the information is discretized at the single-mode level, and the upper level code is mostly treated as a standard qubit code, potentially incorporating the continuous nature of the single-mode error syndromes to improve the decoding procedure~\cite{Fukui17a,Fukui18q,Fukui18r,Vuillot19a,Noh20b,Noh22b,Bourassa21g}. We explore further the link between concatenated GKP codes and ``genuine'' multimode lattices in \cref{sect:qubitCodeConstruction}. 

\begin{figure}[t]
    \centering
      \includegraphics[scale=1]{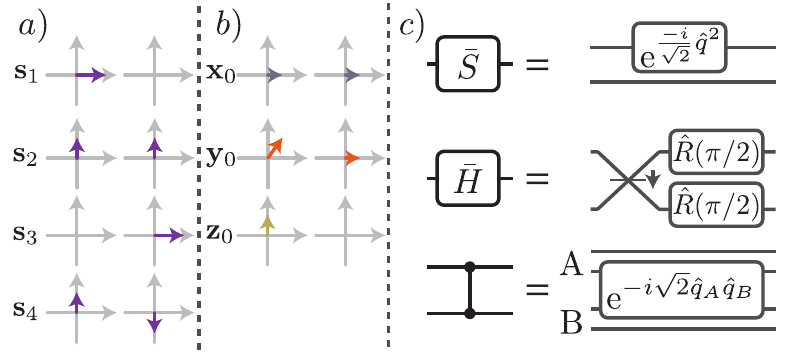}
    \caption{Tesseract code. a) Phase space representation of the stabilizer generators. The left and right graphs represent the projection onto the (q,p) phase space of the first and second mode, respectively. b) Similar representation for the logical operators. c) Logical Clifford operations.
     The $\bar S$ gate can be effected by a shearing gate in either mode (with only one choice represented). The $\bar H$ gate is implemented by a beamsplitter followed by a phase shift of $\pi/2$ in both modes. The logical $C\bar Z$ gate is implemented by a rescaled SUM gate. We represented this gate as an operation between the second mode of A and the first mode of B, but choosing any pair of modes from the codes A and B is equivalent.}
    \label{fig:TesseractGates}
\end{figure}

However, we note that in this concatenated construction, the logical qubit is not defined by a hypercubic lattice. In two modes, it is possible to encode a single hypercubic qubit using the basis
\begin{equation}\label{eq:Stess}
S_{\mathrm{tess}} = \sqrt[4]2\begin{pmatrix} 1 & 0 & 0 & 0 \\
0 & \frac{1}{\sqrt 2} & 0 & \frac{1}{\sqrt 2}\\
0 & 0 & 1 & 0 \\
0 & \frac{1}{\sqrt 2} & 0& -\frac{1}{\sqrt 2} 
\end{pmatrix},
\end{equation}
which is related to $\mathbb I_4$ by a scaling factor $\sqrt[4]2$ and a rotation by $\pi/4$ in the $p_1,p_2$ plane, a non-symplectic transformation. We refer to that qubit code as the tesseract code, which has logical operators of length $\sqrt[4]2$ times larger than than the single-mode square code. From the intuition of the Gaussian translation error model, this code is therefore more robust than the square code against errors. 
We explore the tesseract code further in \cref{sect:Cubiclattice}, and its stabilizer generators are illustrated in \cref{fig:TesseractGates}a).

Beyond the encoding of a qubit, $d=2$, we find that all code dimensions that can be expressed as the sum of three squares
\begin{equation}\label{eq:SumOfThreeSquares}
d = a^2 + b^2 + c^2,
\end{equation}
for $a,b,c \in \mathbb Z$, can be implemented as a two-mode hypercubic code, see \cref{sect:IntegralLatticeSearch}. Legendre's three-square theorem specifies which numbers cannot be written as a sum of three squares. Applied to \cref{eq:SumOfThreeSquares}, we have that there exists codes of all dimensions that \emph{cannot} be written as $d = 4^f(8g + 7)$ for $f,g$ non-negative integers. All hypercubic codes of size $d \leq 20$ are therefore possible except for $d=7,15$.

\emph{D-Type Lattices}\textemdash 
The $D_{2m}$ root lattices are obtained by starting with the hypercubic lattice $\mathbb Z^{2m}$, coloring points in two colors in a checkerboard pattern, and then removing all points of one color. Equivalently, the $D_{2m}$ lattice is generated by taking all points whose sum of coordinates in $\mathbb Z^{2m}$ is even. The determinant of $D_{2m}$ lattices in the standard integral basis is given by $\mathrm{det}(\Lambda_0) = 4$ for all $m \geq 2$~\cite{Conway13c}, which means that one can encode a single qubit in $m$ modes using this lattice.
A standard choice for the generator matrix is given by
\begin{equation}\label{eq:DlatticeGenerators}
S_{D_{2m}} = \begin{pmatrix}
1 & -1 & 0 & ... & 0 & 0\\
0 & 1 & -1 & ... & 0 & 0 \\
0 & 0 & 1 & ... & 0 & 0 \\
... & ... & ... & ... & ... & ... \\
0 & 0 & 0 & ... & 1 & -1\\
0 & 0 & 0 & ... & 1 & 1 \\
\end{pmatrix},
\end{equation}
which reduces to the diamond code in two dimensions (single-mode).
As we show in \cref{sect:qubitCodeConstruction}, the $D_{2m}$ lattices are closely related to the concatenation of diamond GKP codes with a repetition code, and we explore the two-mode lattice $D_4$ in more detail in \cref{sect:D4lattice}. In particular, $D_4$ gives the densest lattice packing in 4 dimensions, and the dual of the $D_4$ lattice is also a $D_4$ lattice.

For the $D_4$ code, there exists an orthogonal transformation that allows for a $d=1$ code~\cite{Harrington04t},
\begin{equation}\label{eq:SD4qunaughtState}
S_{D_4,\varnothing} = 2^{1/4}\begin{pmatrix} 1 & 0 & 0 & 0 \\
\frac{-1}{2} & \frac{-1}{\sqrt 2} & \frac{1}{2} & 0 \\
0 & \frac{1}{\sqrt 2} & 0 & \frac{1}{\sqrt 2}\\
0 & \frac{1}{\sqrt 2} & 0& \frac{-1}{\sqrt 2} 
\end{pmatrix}.
\end{equation}
More generally, we found codes based on the $D_4$ lattice for all code dimensions $d \leq 20$ except $d=14$, see \cref{sect:IntegralLatticeSearch}.

\emph{A-Type Lattices}
\textemdash Another common lattice are the A-type root lattices, with determinants that depend on the dimension in the standard integral basis, $\mathrm{det}(\Lambda_0) = 2m+1$. The $A_2$ lattice corresponds to the hexagonal lattice, which allows for the densest sphere packing in two dimensions. 

\emph{E-Type Lattices}
\textemdash
In 6 and 8 dimensions ($m=3,4$), the densest lattice packings are obtained by using E-type lattices~\cite{Viazovska17j}. However, for $E_6$, we have $\mathrm{det}(\Lambda_0) = 3$ which is not a perfect square. As a consequence, this lattice cannot be used to construct quantum codes. On the other hand, the 8-dimensional lattice $E_8$ has $\mathrm{det}(\Lambda_0) = 1$, and can be used to encode 4 qudits of dimension $a$. One choice of lattice generators is
\begin{equation}\label{eq:E8latticeGenerators}
S_{E_8} = \begin{pmatrix}
2 & 0 & 0 & 0 & 0 & 0 & 0 & 0\\
-1 & 1 & 0 & 0 & 0 & 0 & 0 & 0\\
0 & -1 & 1 & 0 & 0 & 0 & 0 & 0\\
0 & 0 & -1 & 1 & 0 & 0 & 0 & 0\\
0 & 0 & 0 & -1 & 1 & 0 & 0 & 0\\
0 & 0 & 0 & 0 & -1 & 1 & 0 & 0\\
0 & 0 & 0 & 0 & 0 & -1 & 1 & 0\\
1/2 & 1/2 & 1/2 & 1/2 & 1/2 & 1/2 & 1/2 & 1/2\\
\end{pmatrix}.
\end{equation}
For an ensemble of qubits, $a = 2$ leading to $d = 2^4$, the logical operators have minimum length $\min(|\mathbf p|) = 1$, which improves over the hypercubic lattice encoding 4 qubits where $\min(|\mathbf p|) = 1/\sqrt 2$.

\begin{table}
\begin{tabular}{ |c |c | c| c | }
\hline
Lattice Name & \# of modes & min($|\mathbf s_j|$) & min($|\mathbf p|$)\\ \hline
Square & 1 & $\sqrt 2 \approx 1.41$ & $1/\sqrt 2 \approx 0.71$ \\
Hexagonal & 1 & $2/\sqrt[4]{3} \approx 1.52$ & $1/\sqrt[4]{3} \approx 0.76$ \\
Tesseract & 2 & $\sqrt[4]{2} \approx 1.19$ & $1/\sqrt[4]{2} \approx 0.84$ \\
$D_{4}$ & 2 & $\sqrt 2 \approx 1.41$ & 1 \\
  \hline
\end{tabular}
\caption{Examples of qubit codes based on common lattices.}\label{table:qubitLattices}
\end{table}

\emph{Leech Lattice}
\textemdash
The Leech lattice has been shown to be the optimal lattice packing in 24 dimensions (12 modes)~\cite{Cohn17i}, with a determinant $\mathrm{det}(\Lambda_0) = 1$. This lattice can be used to encode an ensemble of 12 qubits, with logical operators of minimum length $\min(|\mathbf p|) = \sqrt 2$, which is twice as large as the square code.

\subsection{Code Switching}\label{sect:codeSwitching}
Consider a situation where we want to switch back and forth between a lattice $\Lambda_C$ defined on $m_C$ modes and two separable lattices $\Lambda_A$ and $\Lambda_B$ with disjoint support on $m_A$ and $m_B$ modes, respectively, with $m_A + m_B = m_C$.
In particular, we are interested in the situation where $\Lambda_C^* \subseteq \Lambda_A^* \oplus \Lambda_B^*$, such that any logical state of the code defined on $\Lambda_C$ corresponds to a logical state of the split code defined on $\Lambda_A \oplus \Lambda_B$. In other words, we are interested in situations where the logical space of the C code is a subspace of the logical space of the $AB$ code. For example, one class of lattices that can be combined together are the $D_{2m}$ family of lattices, where $D^*_{2m_C} \subset D^*_{2m_A} \oplus D^*_{2m_B}$. In particular, two diamond GKP codes ($D_2$) can be combined together to form the $D_4$ code.
Since for integral lattices the stabilizer lattice forms a subset of the dual lattice, $\Lambda \subseteq \Lambda^*$, we have the hierarchy
\begin{equation}\label{eq:latticeSurgeryHierarchy}
\Lambda_A \oplus \Lambda_B \subseteq \Lambda_C \subseteq \Lambda_C^* \subseteq \Lambda_A^* \oplus \Lambda_B^*.
\end{equation}
We refer to the operation $\Lambda_C \rightarrow \Lambda_A\oplus \Lambda_B$ as splitting, and refer to the converse operation $\Lambda_A\oplus \Lambda_B \rightarrow \Lambda_C$ as merging. We outline below how to perform these operations, and leave the details of the gauge updates to \cref{sect:appGaugeCodeSwitching}.

\emph{Lattice Splitting}\textemdash
While we give the operation of splitting a specific name, no physical operation is actually performed, and in practice splitting two lattices consists only in software updates. The only caveat to the previous statement is that if one requires that the eigenvalues of the logical Pauli operators remain real after the splitting (the condition in \cref{eq:gaugeValidityCondition}) a translation set by \cref{eq:gaugeSettingDisplacement} might be required.

\emph{Lattice Merging}\textemdash
The next operation we consider is lattice merging, where we combine two lattices into one, $\Lambda_A\oplus \Lambda_B \rightarrow \Lambda_C $. We focus in particular on the case where all lattices $A,B,C$ encode a qubit, such that the merging operation projects a two-qubit subspace into the single qubit subspace of the $C$ lattice. 
To perform the merging operation, we choose a vector $\vlambda_m \in \Lambda_C$ in the $C$ lattice stabilizers that is not in the $AB$ lattice, $\vlambda_m \notin \Lambda_A \oplus \Lambda_B$. Due to the hierarchy \cref{eq:latticeSurgeryHierarchy}, we have $\vlambda_m \in \Lambda_A^* \oplus \Lambda_B^*$, i.e.\ the vector $\vlambda_m$ corresponds to a (non-identity) logical Pauli operator of the $AB$ code. If the condition \cref{eq:gaugeValidityCondition} is respected, then in the pre-merging state $\hat T(2\vlambda_m)\ket{\psi} = \ket{\psi}$, and the eigenvalues of $\hat T(\vlambda_m)$ are constrained to be $\pm 1$. 
To merge the two lattices, we measure the eigenvalue of $\hat T(\vlambda_m)$ non-deterministically, setting it to a definite value $\nu_m =\pm 1$ and projecting the two-qubit subspace ($d=4$) of the $AB$ code onto the single qubit subspace ($d=2$) of the $C$ code.
This measurement can be done with circuits similar to those required for error correction, and we delay the discussion on its implementation to \cref{sect:ErrorCorrection}. 

\subsection{Mappings Between Lattices}
Given a state encoded in a lattice $S$, there always exists a Gaussian transformation that allows to encode the state in a different lattice $S'$ provided that both codes are of the same prime dimension, $\det(S) = \det(S')$, and $d = \det(S)$ is prime as we show below. 
When these conditions are respected, the statement above translates to the fact that we can relate two lattice generator matrices through
\begin{equation}
S' = R S M,
\end{equation}
where $M\in \mathrm{Sp}(2m, \mathbb R)$ is a symplectic matrix and $R\in \mathrm{GL}(2m,\mathbb Z)$ is an unimodular matrix, i.e.\ an invertible integer matrix with det$(R) = \pm 1$. Essentially, $M$ represents lattice deformations and rotations, while $R$ represents changes of basis for identical lattices. Under such a transformation, the symplectic Gram matrix transforms to
\begin{equation}\label{eq:topologicalAequivalence}
\begin{aligned}
A' &= R A R^T.
\end{aligned}
\end{equation}
Given two generator matrices $S$ and $S'$ such that there exists $R$ where the equation above is respected, then $S$ and $S'$ are related through the symplectic transformation $M = S^{-1} R^{-1} S'$. As shown in Ref.~\cite{Gottesman01a}, there always exists a unimodular transformation such that
\begin{equation}\label{eq:AdiagonalForm}
A' = RAR^T = \begin{pmatrix}0_{m\times m} & D\\ -D & 0_{m\times m}\end{pmatrix},
\end{equation}
where $D$ is a $m\times m$ diagonal matrix. The code dimension is set by $d = \det(D) = \Pi_j D_{jj}$, where each element is a non-zero integer, and we can always choose the elements of $D$ to be positive and in a non-increasing order. As a result, for codes of prime dimension $d$, the form \cref{eq:AdiagonalForm} is unique, which then implies that all symplectic Gram matrices are related by some unimodular matrix $R$. In other words, for codes of prime dimension $d$, there always exist a symplectic operation which can be implemented by a Gaussian circuit such that $S = S' M$.
In particular, this implies that we can always encode a multimode qubit through a gaussian circuit and starting from the separable lattice with a qubit encoded in the first mode and qunaught states in the other modes, $S_{\square}\oplus S_\varnothing^{\oplus m-1}$.

For code sizes which are not prime, the form of $D$ in \cref{eq:AdiagonalForm} is not unique, and therefore not all symplectic Gram matrices are equivalent, which then implies that not all lattices are related through a gaussian transformation. For example, the four-dimensional code given by $S = (2S_\varnothing)\oplus S_\varnothing$ cannot be mapped to the two-qubit code $S' = (\sqrt 2 S_\varnothing)\oplus (\sqrt 2 S_\varnothing)$, although both codes are of the same dimension, $\det S = \det S'= 4$.

\section{Error correction}\label{sect:ErrorCorrection}
There are two strategies to correct errors in single-mode GKP codes, and in this section we generalize both of these approaches to multimode codes. In continuous variable systems, a non-Gaussian resource is required in order to perform error-correction, and the two strategies differ in the resource used.
The first one leverages the intrisic non-linearity of an ancilla qubit coupled to the oscillator modes. The second approaches rather relies on ancilla GKP codes and homodyne measurements, leveraging the non-Gaussianity of a fresh ancilla GKP state to correct errors. Note that the latter strategy requires a method to reliably prepare ancilla GKP states and when discussing this second strategy, we assume that we have access to a deterministic source of fresh ancilla single-mode GKP states. We refer the reader to Refs.~\cite{Travaglione02a,Pirandola04a,Vasconcelos10a,Terhal16a,Motes17a,Shi19a,Su19a,Weigand20a,Noh22b,Hastrup21j} for proposals on how to prepare them.

\subsection{Error correction with ancilla qubits}
\begin{figure}[t]
    \centering
    \includegraphics[scale=1]{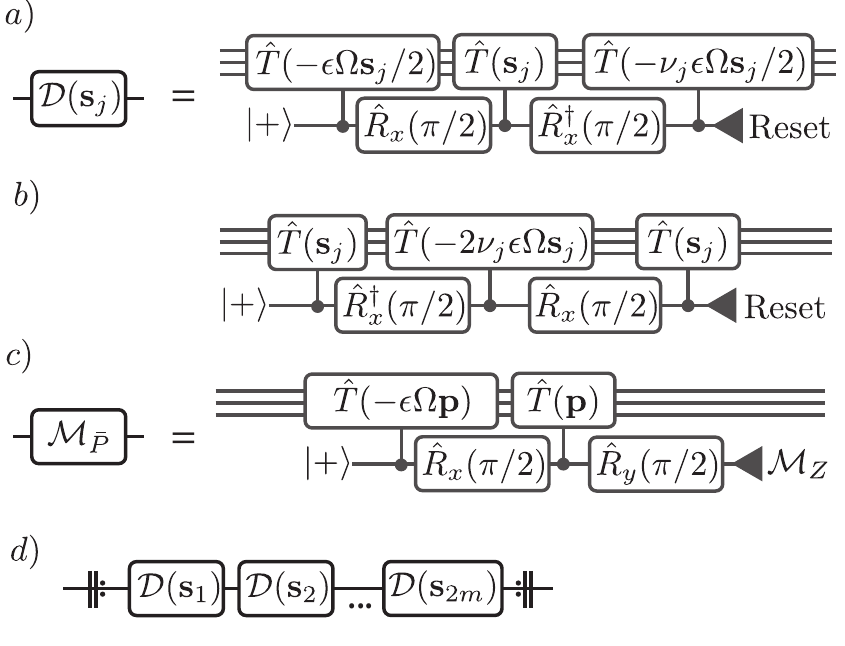}
    \caption{Autonomous QEC circuits based on a qubit ancilla. The first three wires represent the ensemble of $m$ modes, and the last wire corresponds to the two-level ancilla. a) Generalized small-Big-small protocol, which we represent at the logical level as applying an operation $\mathcal D(\mathbf s_j)$. b) Generalized Big-small-Big protocol. In both protocols, the sign of the ``correction'' controlled translations depends on the gauge,  $\nu_j = (-1)^{\mu_j}$. c) Finite-energy Pauli operator measurement. d) The full code space can be stabilized by cycling through the dissipation rounds for all generators $\{\mathbf s_j\}$.}
    \label{fig:StabilizationCircuits}
\end{figure}

First, we describe the error correction strategy based on ancilla qubits coupled to the harmonic oscillators containing the logical information. The protocols we propose generalize the single-mode protocols from Refs. \cite{Campagne-Ibarcq20a,Royer20a,de-Neeve20a} to the multimode case. The generalized small-Big-small (sBs) and Big-small-Big protocols (BsB)~\cite{Royer20a,de-Neeve20a} are shown in \cref{fig:StabilizationCircuits} (a) and (b), respectively, and the generalized logical measurement circuit~\cite{Royer20a,Hastrup21c} is shown in \cref{fig:StabilizationCircuits}c). 

Inspired by the fact that the code space is a zero eigenspace of the code space nullifiers, $\{\hat d_j\}$ from \cref{eq:finiteEnergyNullifier}, we aim to engineer an ensemble of $2m$ dissipators,
\begin{equation}\label{eq:contDissipationStab}
\dot \rho = \sum_{j=1}^{2m} \mathcal D[\hat d_j]\rho,
\end{equation}
where $\mathcal D[\hat o]\bullet = \hat o \bullet \hat o^\dag - \{\hat o^\dag \hat o,\bullet\}/2$ is the standard dissipation superoperator and $\rho$ is the multimode state encoding the logical quantum information. This strategy relies on the fact that the steady state of the master equation above is given by $\hat d_j\ket{\psi} = 0 $ for all $j$, which precisely corresponds to the code space.

Instead of implementing directly the continuous dissipators of \cref{eq:contDissipationStab}, we discretize the oscillators-bath interaction and replace the baths by a single qubit which is frequently reset~\cite{Brun02a,Bouten09a,Ciccarello17a,Gross18a,Royer20a}. This can be achieved through repeated oscillators-bath interactions of the form
\begin{equation}\label{eq:UnitaryStabilization}
\hU_{\mathrm{stab}}^{(j)} = \expo{-i \sqrt{\Gamma_j}(\hat d_j \sigma_- + \hat d_j^\dag \sigma_+)},
\end{equation}
where $\Gamma_j$ is an effective (dimensionless) cooling rate.
Resetting the ancilla qubit to its ground state and repeating this interaction, the entropy from the oscillators is removed in such a way as to cool the system towards the +1 eigenspace of the exact $\hat T_{j,\beta}$ stabilizer of the finite-energy GKP code. We refer to one interaction and reset cycle as a dissipation round, and
the full code space is stabilized by alternating dissipation rounds for each of the $2m$ stabilizer generator $\hat T_{j,\beta}$. 

In practice, implementing the unitary \cref{eq:UnitaryStabilization} is challenging due to the fact that the nullifiers $\hat d_j$ contain a modular multimode quadrature. To simplify this unitary, we trotterize the interaction and leverage the intrinsic modularity of the ancilla qubit, leading to the circuits illustrated in \cref{fig:StabilizationCircuits}. In these circuits, the size of the GKP code envelope is set by 
\begin{equation}
\epsilon = \frac{\sinh \beta}{2},
\end{equation}
and the effective dissipation rate is given by $\Gamma_j \approx |\mathbf s_j|\pi \epsilon /\sqrt 2$. Since this rate is proportional to $\epsilon$, errors are corrected at a slower rate for larger GKP codes (smaller $\beta$). However, larger GKP codes offer better protection against larger errors, such that there is an optimal GKP size that balances between faster correction and larger error protection for a given error channel.

Since an $m$-mode code has $2m$ stabilizer generators, one must apply a combination of at least $2m$ dissipation circuits. However, we note that one could choose to apply more than $2m$ dissipation circuits, especially for lattices which have more than $2m$ lattice vectors of minimum length (counting once a vector and its inverse). For example, in the single-mode hexagonal code, one can choose to cycle among the three stabilizers of the same length $\mathbf s_1 = r_{\hexagon}(1;0)$, $\mathbf s_2 = r_{\hexagon}[-1/2;\sqrt 3/2]$ and $\mathbf s_1 +\mathbf s_2 = r_{\hexagon}[1/2;\sqrt 3/2]$, with $r_{\hexagon} = 2/\sqrt[4]{3}$~\cite{Campagne-Ibarcq20a}. 

We note that the lattice generators $\{\mathbf s_j\}$ do not necessarily have support on all modes. As a result, it is possible perform the error correction circuits using multiple qubit ancillas, each being coupled to a subset of oscillators. Dissipation circuits for generators with disjoint mode support could also be performed in parallel. We illustrate in \cref{fig:StabilizationCircuits}d) the dissipation circuit sequence used in this work, where we choose to simply cycle through dissipation rounds of all stabilizers. We leave the full optimization of the general ``dissipation schedule'' for future work, anticipating that it will depend on the error model, the particular code under study, and the available couplings between ancilla and oscillator modes given the physical architecture.

The main building block of the circuits in \cref{fig:StabilizationCircuits} is a symmetrized controlled multimode translation, 
\begin{equation}
C\hat T(\mathbf v) = \expo{i \frac{\hat \sigma_z}{2}l \hvx^T\Omega \mathbf v},
\end{equation}
which effects a translation by $\pm \mathbf v/2$ if the ancilla is in $\ket{g}$ or $\ket{e}$, respectively. This type of operation was demonstrated in a single microwave cavity in Ref.~\cite{Campagne-Ibarcq20a} and its multimode generalization can be realized by using a qubit coupled dispersively to multiple modes. We consider a qubit-oscillators Hamiltonian
\begin{equation}\label{eq:dispersiveHam}
\hH = \sum_{j=1}^{m}\left[\frac{\chi_j}{2}\had_j \ha_j \hat{ \sigma}_z + \mathcal E_{j}(t)\had_j + \mathcal E_{j}^*(t)\ha_j\right],
\end{equation}
with $\chi_j$ the dispersive interaction between the $j$th mode and the ancilla, and $\mathcal E_j(t)$ is the classical drive applied to the $j$th mode. In short, the classical drives $\{\mathcal E_j\}$ displace the state of the different oscillators, which then rotate in different directions depending on the state of the ancilla. With suitable echo pulses, this strategy can be used to generate any controlled translation, see \cref{sect:app:CT} for more details. 
Interestingly, the controlled displacement rate in each mode depends on the drive amplitude and the dispersive shift, $\chi_j \mathcal E_j$, such that this type of interaction can be implemented in systems with small dispersive coupling by considering strong drives. Moreover, the drive amplitude in each mode can be independently adjusted, and the dispersive shifts of the ancilla qubit to each mode need not be matched.
Finally, there is no specific restriction on the oscillator mode frequencies.

Multimode controlled translations can also be implemented in other platforms such as in the motional modes of trapped ions. In this architecture, controlled translations are generated by a laser which activates a state-dependent force~\cite{Haljan05a,Fluhmann19a}. Multiple state-dependent forces in different modes can then be activated using multiple lasers, a type of interaction which has already been realized in the context of (ion-ion) multi-qubit gates~\cite{Choi14l}.

A cartoon of possible physical implementations for two-mode grid codes are represented in \cref{fig:PracticalImplementation}, where (a) a single ion is coupled to two motional modes and (b,c) a single transmon qubit in the Y-mon geometry is coupled to (b) two microwave single-post cavities~\cite{Gao19d} or (c) phononic-crystal-defect resonators~\cite{Arrangoiz-Arriola19i}. 

In principle, a multimode controlled translation could be obtained through sequential pairwise interactions, performing single-mode controlled translations in series. While the final operation is equivalent, we note that the effect of an ancilla decay event during the controlled translations will be different. We now study the effect of ancilla decay errors when controlled translations are generated from a dispersive interaction, \cref{eq:dispersiveHam}. As we show in \cref{sect:app:CT}, an ancilla qubit decay error during a controlled translation results in an effective rotation and translation
\begin{equation}\label{eq:CTerror}
C\hat T_\mathrm{err} = \hat \sigma_{i\mathrm{err}} \hat T[\mathbf e] \Pi_j \hat R_j(\varphi_{j}),
\end{equation}
where the rotation in each mode is upper bounded by $|\varphi_{j}| \leq |\chi_j T|$ with $T$ the interaction time and $i\mathrm{err} = \pm$. The translation error $\hat T(\mathbf e)$ occurs on a line parametrized by the time of the error, $t_{err} \in [0,T]$. In the limit where the dispersive coupling is small, $\chi_j \rightarrow 0$, and where the controlled translation is generated in a straight line, the rotation error disappears and the translation error is colinear with the desired translation, $\mathbf e = \eta \mathbf s_j$ with a ratio $\eta \in [0,1/2]$. The rotation error can also be reduced by considering more than one echo pulse during the controlled translation. To simplify the analysis below, we assume that ancilla errors are given by bit flips rather than qubit decay, propagating as errors of the form $\mathbf e = \eta \mathbf s_j$ with $\eta \in [0,1]$. 

For the small-big-small protocol, the middle large controlled translation leaves the state effectively displaced by $\pm\mathbf s_j/2$. For a single-mode GKP, this effectively applies a logical Pauli operation, but in general it leaves the state outside of the code space. One solution to recover the original subspace is to apply the dissipation circuit twice, effectively displacing the state by 0 or $\pm \mathbf s_j$. More efficiently, this can also be taken care of in software by updating the gauge, using the translation update rule \cref{eq:gaugeUpdateDisplacement}. The sign in front of the effective translation does not impact the gauge change, such that a superposition of translations,  $\hat T(+\mathbf s_j/2) \pm \hat T(- \mathbf s_j/2)$, has the same effect as $\hat T(\mathbf s_j/2)$ with respect to the gauge. To account for the gauge change, we modify the error-correction circuits such that the state is steered towards the $-1$ eigenspace of the corresponding stabilizer.
As shown in \cref{fig:StabilizationCircuits}, this is achieved by changing the sign of the last small controlled translation. Alternatively, the sign of the ancilla $\pi/2$ rotation can be inverted. 

We have described the multimode generalization of the higher-order sBs and BsB protocols, but a similar analysis could be applied to the sharpen-trim protocol of Ref.~\cite{Campagne-Ibarcq20a}.

\emph{Classical Dissipation Model}\textemdash
In order to better understand the effect of the qubit-based error correction procedure described above, we build a simple classical model which is much faster to simulate than the full multimode quantum model. More precisely, we consider the effect of the dissipation rounds on points in a classical phase space $\mathbf x \in \mathbb R^{2m}$. 
We define the modular vector $\mathbf q_m = lS\Omega \mathbf x \,\,\mathrm{mod}\,\, 2\pi$, which projects the point $\mathbf x$ on each generator of translation, $\mathbf g_j$, associated with a stabilizer $\mathbf s_j$.
From the potential
\begin{equation}
\Phi = \frac{1}{2}\mathbf q_m \cdot \mathbf q_m,
\end{equation}
we approximate the quantum dissipative map generated by the gadgets in \cref{fig:StabilizationCircuits} by setting the evolution of a classical point $\mathbf x$ as
\begin{equation}\label{eq:dissipationEqClassical}
\dot{\mathbf x} = - \nabla \Phi.
\end{equation}
This equation is motivated by the fact that each dissipation circuit can be understood as approximately measuring a generator of translation $\hat g_j$ and applying a correction pulse towards $\hat g_j \,\,\mathrm{mod}\,\, 2\pi = 0$. We choose \cref{eq:dissipationEqClassical} to be linear for simplicity, a choice that does not affect the position of the fixed points which are the main objects of interest. Moreover, for simplicity, \cref{eq:dissipationEqClassical} does not model the random shot noise of the measurements.

To obtain an estimate of the effect of a single translation error $\mathbf e$, we apply the evolution map after setting $\mathbf x(t = 0) = \mathbf e$. From the position at long times $\mathbf x(t\rightarrow \infty) = M(\mathbf e)$, we infer the logical operation applied by checking to which set of points in the dual lattice $M(\mathbf e) \in \Lambda^*$ belongs, with a successful correction when the logical identity is applied, $M(\mathbf e) \in \Lambda$. Since finite-energy states have a finite width, we approximate the probability that the translation error $\mathbf e$ is converted into a logical error as
\begin{equation}\label{eq:classicalErrorProbability}
\mathbb P(\mathrm{error}|\mathbf e) = 1 - \int d\mathbf r I_{M(\mathbf r) \in \Lambda} G(\mathbf e,\sigma)
\end{equation}
where $I_{M(\mathbf e) \in \Lambda}$ is an indicator function that equals 1 if $M(\mathbf e) \in \Lambda$ and 0 otherwise, and $G(\mathbf e,\sigma)$ is a Gaussian distribution with mean $\mathbf e$  and standard deviation $\sigma$ approximately given by
\begin{equation}
\sigma = \sqrt{\tanh\left(\frac{\mathrm{arcsinh}(2\epsilon)}{2}\right)} \approx \sqrt{\epsilon}.
\end{equation}
In two dimensions the integral in \cref{eq:classicalErrorProbability} can be performed analytically, otherwise we use Monte Carlo integration methods.

\begin{figure}[t]
    \centering
    \includegraphics[scale=1]{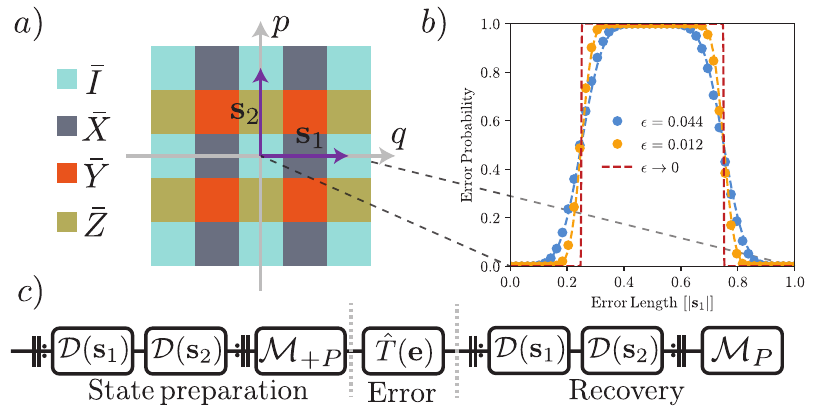}
    \caption{Effect of translation errors in the square code. a) Final pauli error (colors) for different initial translation error $\mathbf e = (q,p)$ according to the classical model. b) Logical error probability for translation errors of the form $\mathbf e = \eta \mathbf s_1$. We compare the classical model \cref{eq:classicalErrorProbability} (dashed lines) and the full quantum model (dots), computed using the circuit in panel c). We also compare different GKP size (colors), and in the infinite size limit $\epsilon \rightarrow 0$ we expect a sharp transition from correctable to uncorrectable error at $\mathbf e = \mathbf s_1/4$ and $\mathbf e = 3\mathbf s_1/4$ (red dashed line).}
    \label{fig:ErrorPropagation2d}
\end{figure}

In \cref{fig:ErrorPropagation2d}a), we show the final logical operation applied in the single-mode code space of the square code as a function of an initial translation error $\mathbf e = (q,p)$. As we show, translation errors around the stabilizers $\mathbf 0,\mathbf s_1,\mathbf s_2$ are fully corrected (light blue regions). 
On the other hand, translation errors around the logical Pauli operators, $\mathbf x = \mathbf s_1/2$, $\mathbf y = (\mathbf s_1 + \mathbf s_2)/2$ and $\mathbf z = \mathbf s_2/2$, are uncorrectable by design.

When applying the dissipation circuits of \cref{fig:StabilizationCircuits} and in the limit that $\chi \rightarrow 0$, decay errors on the qubit propagate as translation errors $\mathbf e = \eta \mathbf s_j$ with $\eta \in [0,1]$. In panel \cref{fig:ErrorPropagation2d}b), we show the logical error probability as a function of $\eta$ for the first stabilizer $\mathbf s_1$.
We find good agreement between the full quantum model (full dots) and the classical model (dashed lines) without any fit parameters. We also compare different GKP sizes, and show that the transitions are sharper at the edges for larger GKP sizes due to their smaller quadrature fluctuations. In the limit of infinite GKP size, $\epsilon \rightarrow 0$, we expect a sharp transition between correctable and uncorrectable errors.

When a qubit decay event occurs, we take the probability that it propagates as a logical error as
\begin{equation}\label{eq:errorProbPropagation}
\begin{aligned}
\mathbb P(\mathrm{error}|\mathrm{qubit\, decay}) &= \int_0^1 d\eta\, \mathbb P(\mathrm{error}|\eta\mathbf s_j)\mathbb P(\eta),
\end{aligned}
\end{equation}
and we assume that the probability of a qubit decay is small such that $\mathbb P(\eta)$ is approximately given by a uniform distribution. In the case of the square code, the probability of a logical error is constant $\mathbb P(\mathrm{error}|\mathrm{qubit\, decay}) \approx 1/2$ and does not depend on the GKP size, an estimate that agrees well with numerical simulations of the full quantum model~\cite{Royer20a}. For single-mode GKP codes, the lifetime of the logical information is therefore limited by the lifetime of the ancilla, a situation that is not improved by increasing the GKP size. We show below that multimode codes, on the other hand, have increased robustness against the propagation of ancilla errors.

Note that in the classical model presented above, the effective dissipation rate depends (in general) on the position of the initial point $\mathbf x$, not just its length $|\mathbf x|$. Indeed, inside a unit cell \cref{eq:dissipationEqClassical} can be rewritten as $\dot{\mathbf x} = - l^2 \Omega S^T S \Omega \mathbf x$, which leads to $\mathbf x(t) = \exp\{- l^2 \Omega S^T S \Omega t\}\mathbf x(0)$. As a result, the effective dissipation rates are proportional to the eigenvalues $\{\lambda_j\}$ of the Hessian matrix $H(\Phi) = -l^2 \Omega S^T S \Omega$. Intuitively, the break-even point is related to the ratio between the smallest dissipation rate and the oscillator physical error rate, $\mathrm{min}(\{\lambda_j\})/\kappa \delta t$, which explains why the square code can perform better than the hexagonal code in qubit-based error correction~\cite{Royer20a,de-Neeve20a}. Moreover, it also explains why the tesseract code performs better than the $D_4$ code, see \cref{fig:photonLifetime} and the sections below.

\subsection{Homodyne Error correction}
The second approach to QEC for GKP codes is to use fresh ancilla modes prepared in generalized GKP qunaught states. Measuring the stabilizers of the ideal GKP code is equivalent to measuring the modular quadratures $\{\hg_j\}$, and to simplify the analysis we choose an envelope-preserving unitary $\hU_B$ such that 
\begin{equation}
\hx_{2m} = \hU^\dag_B \frac{\hg_j}{|\mathbf g_j|} \hU_B,
\end{equation}
where $\mathbf g_j = l \Omega \mathbf s_j$. The task of measuring the jth stabilizer then reduces to the task of measuring $\hx_{2m} \mod 2\pi/|\mathbf g_j|$, for which single-mode GKP circuit have already been developed. In particular, we make use of the SUM gate 
\begin{equation}\label{eq:SUMgate}
\mathrm{SUM}_{j\rightarrow k} = \expo{-i\hx_j \hp_k},
\end{equation}
which maps the quadrature coordinates $\hx_j \rightarrow  \hx_j$, $\hp_j \rightarrow  \hp_j - \hp_k$, $\hx_k \rightarrow \hx_k + \hx_j$ and $\hp_k \rightarrow \hp_k$. This gate also acts as a logical CNOT gate between two single-mode square GKP codes.

Applying the SUM gate between a mode in the state $\ket{\psi}$ and an ancilla in the ideal qunaught state $\ket{\varnothing_\eta}$, we obtain 
\begin{equation}
\expo{-i\hx_1 \hp_2} \int dq\, \psi(q)\ket{q}\otimes \ket{\varnothing_\eta} = \sum_j\int dq\, \psi(q)\ket{q}\otimes \ket{j\eta l + q},
\end{equation}
where we have written the state in the position basis and used the expression for the qunaught state \cref{eq:qunaughtStatePositionBasis}. Measuring the ancilla in the position basis, $\hx$, then yields a measurement result $q + j\eta l$ for some random $j \in \mathbb Z$, which is equivalent to measuring $\hq \, \mathrm{mod}\, \eta l$. In order to measure $\hq \, \mathrm{mod}\, 2\pi / |\mathbf g_j|$, one should therefore set $\eta = 1/|\mathbf s_j|$. In the case of error correction for the square GKP code, the procedure described above is equivalent to the approach originally developed in~\cite{Gottesman01a}. Indeed, for the square code we have $|\mathbf s_j| = \sqrt 2$, and we have shown in \cref{sect:latticeExamples} that $\ket{\varnothing_{1/\sqrt 2}} = \ket{+ X_{\square}}$. 

The SUM gate \cref{eq:SUMgate} combined with the $\hat U_B$ allows to measure the quadratures $\{\hg_j\}$ by propagating translation errors from the data modes to the ancilla. In a similar fashion, translation errors in the ancilla mode propagate to the data modes.
More precisely, noise in the ancilla $\hp$ quadrature propagates as errors $\hat T(\sigma \mathbf s_j)$ in the data mode, with $\sigma$ a random variable taken from a normal distribution with standard deviation set by the level of squeezing of the ancilla. We remark that in both homodyne and qubit-based error correction, noise in the ancilla propagates as translation errors colinear with $\mathbf s_j$, such that minimizing the effect of these errors is desirable for both error correction methods.

The protocol laid out above is not unique, as it is possible to choose different ancilla lattice constants $\eta \rightarrow c \eta$ if the measurement results and the SUM gate are properly rescaled, $\expo{-i \hx_j \hp_k} \rightarrow \expo{-i \hx_j \hp_k/c}$. Moreover, while we choose here to map the multimode quadrature $\hat q_j$ to a single-mode, one could instead perform the combined unitary $\hU^\dag_B \expo{-i \hx_{2m} \hp_a}\hU_B$ as a series of (rescaled) SUM gates.

\begin{figure}[t]
    \centering
    \includegraphics[scale=1]{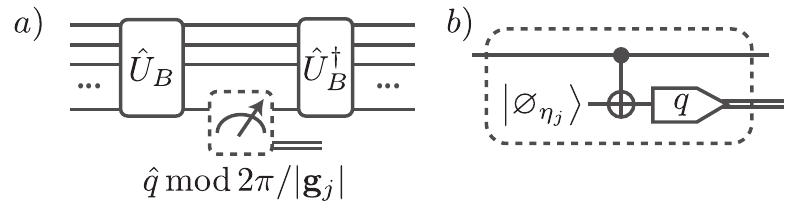}
    \caption{Steane-type error correction circuits based on ancilla GKP states. a) The desired quadrature is mapped to a single-mode on which a modular quadrature measurement is performed. b) A modular quadrature measurement can be performed by preparing a single-mode qunaught state with $\eta_j = 1/|\mathbf s_j|$ and applying a SUM gate between the mode to be measured and the ancilla. Alternatively, a rescaled SUM gate could be applied between the ancilla and each mode over which $\mathbf s_j$ has support.}
    \label{fig:ErrorCorrectionCV}
\end{figure}

\begin{figure}[t]
    \centering
    \includegraphics[scale=0.7]{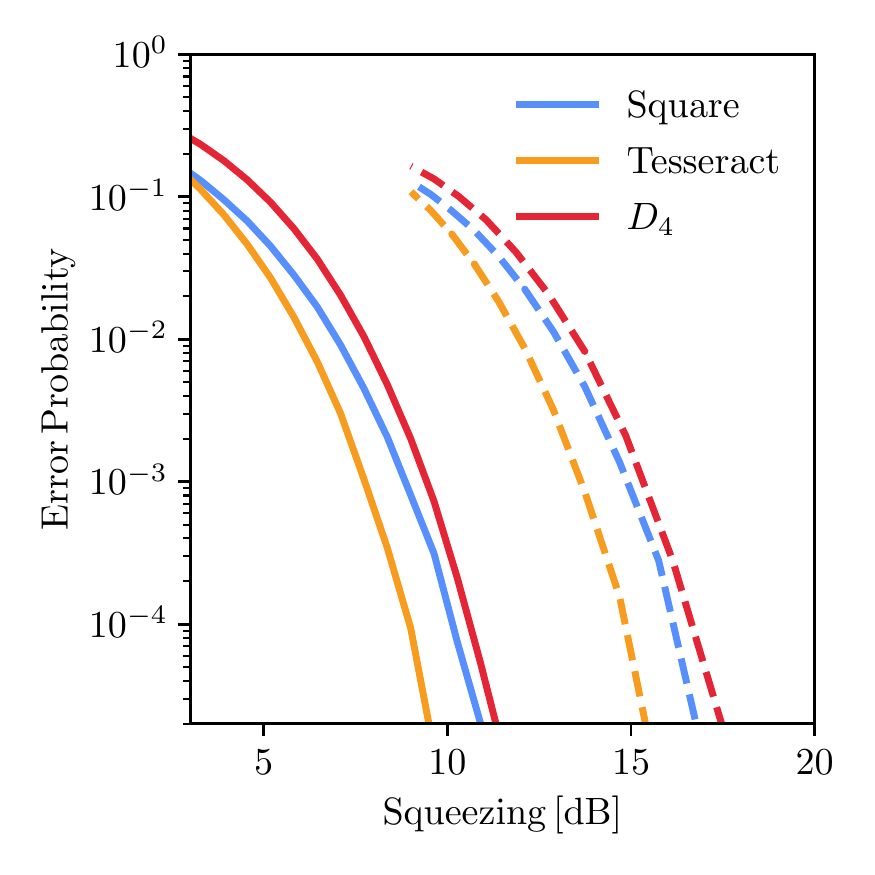}
    \caption{Logical error probability for different GKP codes against the Gaussian translation error channel, with a translation noise standard deviation expressed in decibels. Full line correspond to measurement performed with ideal ancillas GKP states, and dashed lines correspond to noisy ancillas.}
    \label{fig:ErrorProbHomodyne}
\end{figure}

Consider a translation error $\hat T(\mathbf e)$ affecting a logical state. After measuring each stabilizer generator, for example with the circuits of \cref{fig:ErrorCorrectionCV}, we get a syndrome $\pmb \xi$ related to the error through
\begin{equation}
\pmb \xi = -l^2 S \Omega \mathbf e \;\mathrm{mod} \; 2\pi,
\end{equation}
with the modulo part applied element-wise. 
To correct the error, one simple decoding choice is to take a correction translation $\pmb \delta = \Omega S^{-1} \pmb \xi/l^2$. After the correction, the initial state has been translated by $\hat T(\pmb \delta)\hat T(\mathbf e) = \expo{i\theta} \hat T(\mathbf e_m)$, with an irrelevant global phase $\theta$ and a remaining (uncorrectable) translation by $\mathbf e_m = l \Omega S^{-1}\mathbf a$ for some $\mathbf a \in \mathbb Z^{2m}$. Expressing the remaining displacement in the basis of the stabilizer lattice, we obtain $\mathbf b = A^{-1} \mathbf a$, a logical error occurs if $\mathbf b \notin \mathbb Z^{2m}$.
For a single-mode square code, this decoder is equivalent to the standard memoryless decoder, which corrects all translation errors smaller than $|\mathbf e| < 1/\sqrt{8}$ in the ideal code limit. \Cref{fig:ErrorProbHomodyne} shows the error probability after a full round of error correction with GKP ancilla states and the decoding strategy laid out above. We consider a Gaussian translation error channel where modes are translated by a random amount taken from a multivariate normal distribution with standard deviation $\sigma$, $\mathbf e \sim \mathcal N(\sigma)$, which we express in decibels
\begin{equation}
\sigma^{(\mathrm{dB})} = 10 \log_{10}\left(\frac{1/2}{\sigma^2}\right).
\end{equation}
This measure of noise does not distinguish between a mixture of random translations and the quantum fluctuations due to a finitely-squeezed state. Moreover, we remark that the variance of the total translation error scales linearly with the number of modes, $E[|\mathbf e|^2] = m \sigma^2$.

Full lines correspond to error correction with ideal ancillas, i.e.\ only data modes are affected by the error channel. Dashed lines, on the other hand, correspond to the more realistic situation of finite-energy ancillas, and we choose the ancilla noise to be identical to the data noise. We compare the performance of different GKP codes (colors), and discuss the features of the tesseract and $D_4$ code below in \cref{sect:Cubiclattice,sect:D4lattice}, respectively.

In this work, we mostly focus on qubit-based error correction methods, and we leave the optimization of decoders for GKP-ancilla based QEC for future work. For example, Ref.~\cite{Vuillot19a} discusses a general decoder that could be applied to multimode lattices. Moreover, the method presented above does not preserve the envelope since the ideal stabilizers are measured, not the finite-energy stabilizers of \cref{eq:finiteEnergyStabilizers}. As a result, each QEC round adds energy to the modes of the system. It would be interesting to develop general teleportation-based error-correction circuits similar to those that were developed for single-mode GKP codes~\cite{Glancy06a}.

\section{Tesseract code} \label{sect:Cubiclattice}
In this section, we study in detail the two-mode tesseract code with generator vectors given by \cref{eq:Stess} and illustrated in \cref{fig:TesseractGates}a). 

\subsection{Code Words}
The tesseract code is based on a four-dimensional hypercubic lattice, such that the stabilizer generators are all of equal length, $|\mathbf s_j| = \sqrt[4]{2}$ for all $j$, and are all orthogonal to each other, $\mathbf s_j \cdot \mathbf s_k = 0$ for $j \neq k$.
As illustrated in \cref{fig:TesseractGates}b), we choose the base representative of the logical Pauli operators for the tesseract code
\begin{equation}\label{eq:L0Tesseract}
L_{0,\mathrm{tess}} = \begin{pmatrix}
\mathbf x_0 \\
\mathbf y_0 \\
\mathbf z_0 \\
\end{pmatrix} = \sqrt[4]{2}\begin{pmatrix}
1/2 & 0 & 1/2 & 0 \\
1/2 & 1/\sqrt 2 & 1/2 & 0 \\
0 & 1/\sqrt 2 & 0 & 0\\
\end{pmatrix}.
\end{equation}
Similar to the square code, the $\bar Y$ operator of the tesseract code has a length $\sqrt 2$ larger than the $\bar X,\bar Z$ operators. However, compared to the square code, the tesseract code has longer logical operators by a factor $\sqrt[4]{2} \approx 1.19$, while the stabilizers are smaller by a factor $\sqrt[4]{2}$. Another important distinction between the tesseract code and all single-mode GKP codes is that logical operators of the tesseract code are not colinear with a stabilizer vector. More precisely, we have $\mathbf x_0 = (\mathbf s_1 + \mathbf s_3)/2$, $\mathbf z_0 = (\mathbf s_2 + \mathbf s_4)/2$ and $\mathbf y_0 = (\mathbf s_1 + \mathbf s_2 + \mathbf s_3 + \mathbf s_4)/2$, as illustrated in \cref{fig:TesseractGates,fig:ErrorPropagationTesseract}a). We contrast this with single-mode codes where we can always choose $\mathbf x_0 = \mathbf s_1/2$, $\mathbf z_0 = \mathbf s_2/2$ and $\mathbf y_0 = (\mathbf s_1 + \mathbf s_2)/2$.

With the choice of Pauli operators given by \cref{eq:L0Tesseract}, the logical $+\bar Z$ eigenstate is given by the separable state
\begin{equation}
\ket{+\bar Z} = \ket{\varnothing_{\sqrt[4]{2}}}^{\otimes 2}.
\end{equation}
The logical $-\bar Z$ eigenstate is given by the same tensor product of qunaught states, but choosing a mixed gauge $\vmu = (0,1)$ for both modes. While we chose in \cref{eq:L0Tesseract} the base $\bar Z$ representative to have support only on the first mode, an equivalent representative has support only in the second mode, $\mathbf z' = \sqrt[4]{2}(0,0,0,1/\sqrt 2)$. In particular, a measurement in the $\bar Z$ basis can be performed by accessing a single one of either mode comprising the tesseract code.
Alternatively, we can measure both representatives and implement a two-bit repetition code to mitigate the effect of different faults such as oscillator errors, measurement errors or ancilla decay during the measurements. Since $\mathbf z_0$ and $\mathbf z'$ have support on different modes, faults in one measurement do not affect the other. The circuit shown in \cref{fig:StabilizationCircuits}c) is also quantum non-demolition (QND), such that multiple measurements could be realized to mitigate the effect of measurement errors.

\subsection{Gates}
In the following, we outline how to perform any operation in the Clifford group generated by $\langle \bar S, \bar H, C\bar Z\rangle$, with results summarized in \cref{fig:TesseractGates}c). Since the tesseract code is the four-dimensional generalization of the square code, we show that the logical gates in both codes are qualitatively similar.

\emph{Hadamard Gate}\textemdash The $\bar X$ and $\bar Z$ operators have the same length, $|\mathbf x_0| = |\mathbf z_0|$, and there is an envelope-preserving operation that exchanges them, corresponding to the effect of a Hadamard gate in the code space. The Hadamard logical gate is implemented from a beamsplitter followed by a $\pi/2$ phase shift of both modes,
\begin{equation}\label{eq:HadamardTesseract}
\hat U_{\mathrm{tess}}(\bar H) = \hat R(\pi/2)^{\otimes 2}\hat B.
\end{equation}
In the square code, a Hadamard gate is obtain by a phase shift $\hU_{\square}(\bar H) = \hat R(\pi/2)$.

\emph{Phase Gate}\textemdash The phase gate can equivalently be understood as a diagonal gate in the $\bar Z$ basis, $\bar S = \mathrm{Diag}[1;i]$, or as mapping the Pauli operators $\bar X \rightarrow \bar Y$, $\bar Y \rightarrow -\bar X$ and $\bar Z \rightarrow \bar Z$. Taking the later view shows that $\bar S$ cannot be implemented by an envelope-preserving operation since $|\mathbf x_0| \neq |\mathbf y_0|$, see \cref{eq:L0Tesseract}.
However, taking the former view shows that the $\bar S$ gate can be realized through a single-mode unitary since logical $\bar Z$ operators have single-mode support. More precisely, the logical $\bar S$ gate can be implemented by a shearing gate in either mode,
\begin{equation}
\hat U_{\mathrm{tess}}(\bar S) = \expo{\frac{-i}{\sqrt 2}\hx^2 }\otimes \hat I\,\,\mathrm{or}\,\,\hat I\otimes\expo{\frac{-i}{\sqrt 2}\hx^2}.
\end{equation}
Note that any gate which is diagonal in $\bar Z$ basis can, in principle, be obtained through a single-mode unitary in the tesseract code.
In the square code, the phase gate is also implemented via a shearing gate, $\hat U_{\square}(\bar S) = \exp\{-i \hx^2/2\}$, which does not preserve the envelope.

\emph{Controlled-$\bar Z$ Gate}\textemdash We now present a logical two-qubit $C\bar Z$ gate realized between two tesseract codes A and B, each comprised of two modes (four modes in total).
This $C\bar Z$ gate is realized through a two mode rescaled SUM gate
\begin{equation}
\hat U_{\mathrm{tess}}(C\bar Z) = \expo{-i\sqrt 2\hx_A\hx_B },
\end{equation}
where $\hx_{A/B}$ is the $\hx$ quadrature of either mode comprising the $A/B$ tesseract code. This gate does not preserve the envelope.

\emph{Non-Clifford Gates}\textemdash
One potential approach to universality originally developed in Ref.~\cite{Gottesman01a} for single-mode codes is to perform a cubic gate of the form $\hat V = \exp\{i\gamma \hq^3\}$ for some specially tuned $\gamma$, which allows to perform a logical $\bar T = \mathrm{Diag}[1;\expo{i\pi/4}]$ when combined with gaussian operations. Since the $\bar T$ gate is diagonal in the $\bar Z$ basis, this approach can also be deployed in the tesseract code by applying $\hat V$ to a single-mode and choosing $\gamma = 1/(2l\sqrt[4]{2})$. However, we note that the cubic gate is designed for ideal GKP codes and does not commute with the envelope operator, $[\hat V,\hat E_\beta]\neq 0$. As a result, applying the cubic gate to realistic GKP states does not map code words to other code words, and it was shown that the fidelity of this gate is intrinsically limited for finite-energy square GKP codes~\cite{Hastrup21a}. We expect that these limitations also apply to the tesseract code.

To solve this issue, we propose an exact square-root-Hadamard gate following the approach laid out in \cref{subsect:EnvelopePreservingNonCliffordGates}. Starting with the logical Hadamard gate (order $g=2$), we compute that the gate \cref{eq:HadamardTesseract} has order $p=8$. As a result, we can implement an envelope-preserving non-Clifford gate in the tesseract code,
\begin{equation}\label{eq:sqrtHadamardTess}
\hU_{\mathrm{tess}}\left(\sqrt{\bar H}\right) = \expo{i\frac{\pi}{32}\hat F^2},
\end{equation}
with $\hat F = \hat q_1^2 + \hat p_1^2 + \hat q_2^2 + \hat p_2^2 - 2  + \hq_1 \hp_2 - \hp_1 \hq_2$. Although we do not have a concrete proposal to realize this gate, we point out that tunable quartic Hamiltonians can be realized in microwave cavities~\cite{Rosenblum18a,Ye21v,Zhang21y}.

An alternative option to universal quantum computing that does not require any non-Clifford gate is through magic state injection~\cite{Bravyi05s}, which reduces the challenge of realizing non-Clifford gates to preparing magic states. In particular, Ref.~\cite{Baragiola19b} introduced a method to probabilistically prepare magic states through the homodyne error-correction procedure presented in \cref{sect:ErrorCorrection}. This approach could be deployed for the tesseract code by first preparing single-mode magic states, then 
using code switching techniques to prepare the two-mode magic state. This approach is better suited to optical implementations of GKP codes, where precise homodyne measurements are readily available.

In the microwave domain, an option for magic state preparation is to first prepare a magic state in the ancilla used for control of the oscillator mode and then teleport the ancilla state to the bosonic code, as was realized in Ref.~\cite{Campagne-Ibarcq20a}. Another option to realize a direct non-Clifford gate would be to realize a controlled version of the Hadamard gate, with the control qubit being the ancilla. A similar type of controlled-beamsplitter has already been demonstrated between two microwave cavities~\cite{Gao19d}. One downside of these two approaches is that they are limited by the lifetime of the qubit ancilla. When using a transmon-type qubit as the ancilla, this issue could be partly mitigated by making use of the third level of the ancilla through path-independent gates~\cite{Ma20y}. 

\subsection{Error Correction}
Having presented the tesseract code and how to perform logical operations, we now study the robustness of this code against different types of errors.

\begin{figure}[t]
    \centering
      \includegraphics[scale=0.5]{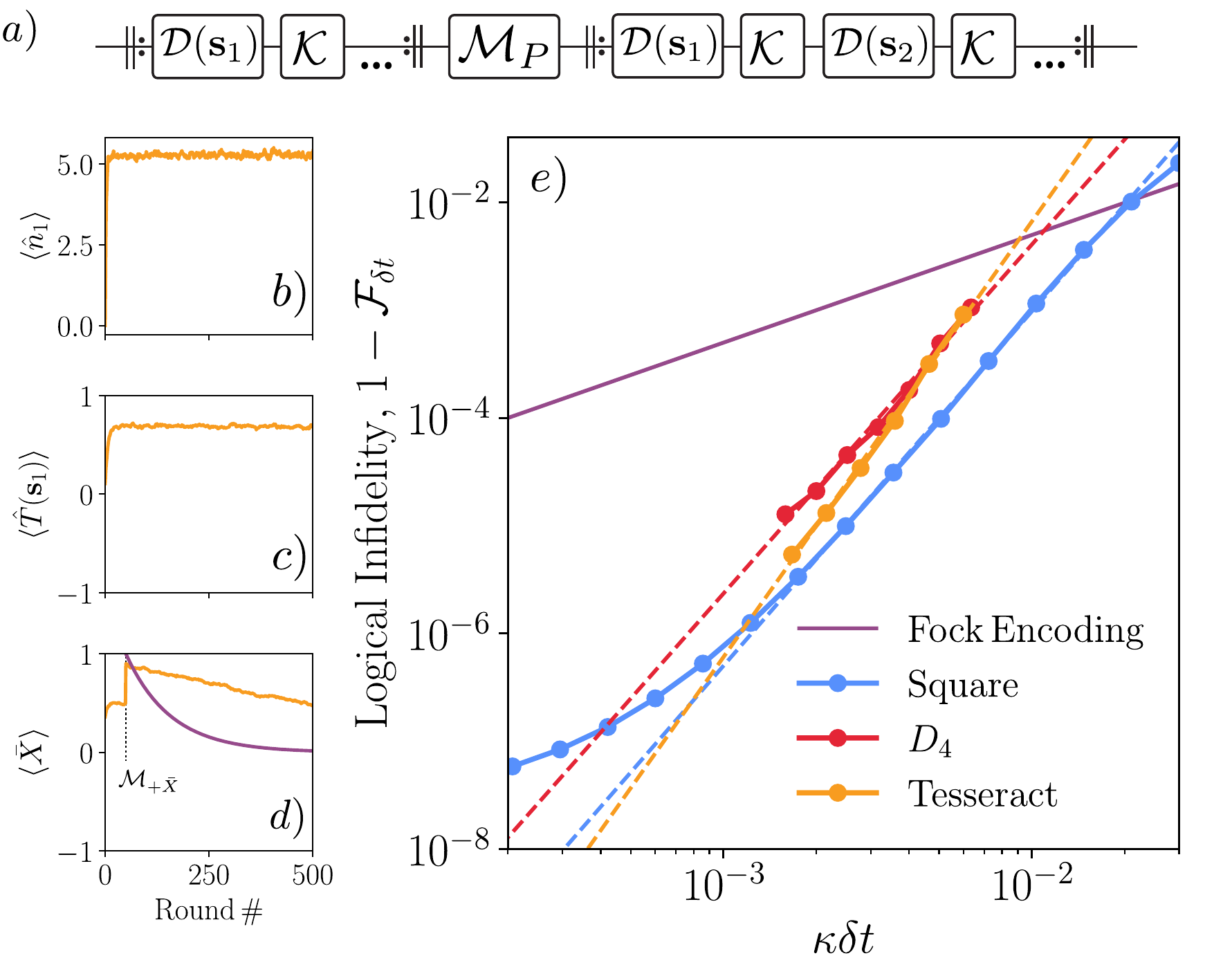}
    \caption{Logical channel infidelity in the presence of amplitude damping. a) Logical circuit used to extract lifetimes. The dissipation circuits are applied in series to project the modes into the code space, a logical measurement is performed to prepare a specific Pauli eigenstate, and the dissipation circuits are applied in series to stabilize the code space. (b) Average excitation number in the first mode as a function of time, reported as the number of dissipation rounds. (c) Expectation value of the stabilizer $\hat T(\mathbf s_1)$  as a function of time. (c) Expectation value of the logical Pauli operator $\bar X$ as a function of time. For reference, we show the decay of the $\bar X$ logical operator in the Fock encoding (purple), which decays at a rate $\kappa/2$. Panels (b,c,d) are computed using an amplitude damping rate of $\kappa \delta t = 4.6 \times 10^{-3}$. (e) Infidelity as a function of the amplitude damping rate $\kappa \delta t$ for different codes. Dots are computed from simulations and dashed lines correspond to an exponential fit.}
    \label{fig:photonLifetime}
\end{figure}

\emph{Amplitude Damping}\textemdash First, we investigate its robustness against resonator amplitude damping when correcting errors using the qubit-based method. For now, we assume that no errors occur on the ancilla qubit. As shown in \cref{fig:photonLifetime}a), we extract the lifetime of the code words by preparing the state inside the code space, projecting the state onto a Pauli eigenstate using the circuit in \cref{fig:StabilizationCircuits}c) and analyzing the subsequent decay of the Pauli expectation value. 
Between each dissipation circuit $\mathcal D(\mathbf s_j)$, we apply an amplitude damping channel $\{\hat K_k\}$ with decay rate $\kappa$ for a time $\delta t$ through its Kraus representation, 
\begin{equation}\label{eq:KrausOpsPhotonLoss}
\hat K_k = \left(\frac{\gamma}{1 - \gamma}\right)^{k/2}\frac{\ha^k}{\sqrt{k!}}(1 - \gamma)^{\hat n/2},
\end{equation}
with $ k\geq 0$ and $\gamma = 1 - \expo{-\kappa \delta t}$. 

\Cref{fig:photonLifetime} shows an example of the time evolution of the expectation value for (b) the excitation number in the first mode, (c) the stabilizer $\langle \hat T(\mathbf s_1) \rangle$ and (d) logical operator $\langle \bar X \rangle$ as a function of the round number for $\kappa \delta t = 4.6\times 10^{-3}$. With respect to panel c), the other stabilizers exhibit very similar behavior, and the fact that the average value is lower than one is partly due to the finite-energy nature of the state, and partly due to the finite ratio between the rate of error correction and the rate at which errors occur. The evolution of the expectation values for the other logical Pauli operators (when the logical qubit is projected onto the corresponding eigenstate) follows a behavior very similar to panel d).

\Cref{fig:photonLifetime}e) compares the resulting channel infidelity for different GKP codes as a function of the unitless amplitude damping rate, $\kappa \delta t$. The channel fidelity is computed from 
\begin{equation}\label{eq:channelFidelity}
\mathcal F_t = \frac{1 + \sum_{\alpha \in \{x,y,z\}} \expo{-\gamma_\alpha t}}{4},
\end{equation}
where the rates $\gamma_\alpha$, $\alpha \in \{x,y,z\}$, are the decay rates for each Pauli eigenstate. Theses rates are obtained by fitting an exponential decay to the time evolution of the associated logical Pauli operator after the projection measurement.
Each logical damping rate is obtained by averaging over 200 trajectories with different realizations of the noise and ancilla states at the end of the dissipation circuits. \Cref{fig:photonLifetime}e) shows the fidelity of the logical channel for a time $\delta t$ that corresponds to the time between each dissipation circuit.
Dots correspond to full numerical simulations extracted using the circuit in panel a) with full lines being guides for the eye, and dashed lines correspond to an exponential fit $i\mathcal F = 1 - \mathcal F = (\kappa \delta t/\alpha)^\delta$, with fitting parameters $\alpha$ and $\delta$. 

We interpret $\delta$ as a measure of the distance of the code, in analogy to qubit QEC codes where the logical error probability scales as $i\mathcal F \sim (p/p_*)^{(d+1)/2}$ for some physical error probability $p$, threshold value $p_*$ and distance $d$. For qubit codes, the distance is given by the smallest support of the logical operators, which directly sets the support of the smallest uncorrectable error. In the present context, the distance is linked to the length of the smallest translation error that cannot be corrected.

We define the \emph{break-even} point of error correction as the crossing point between the infidelity of the Fock encoding (purple) and the GKP code. Intuitively, the break-even point sets the physical error rate below which the error-correction code is useful, i.e.\ when the correction properties overcome the additional errors due to the increase in excitation number.

We remark that both parameters $\alpha$ and $\delta$ depend on the code size. Increasing the envelope size (smaller $\beta$) causes $\delta$ to increase and $\alpha$ to decrease. This latter fact implies that $\alpha$ is not a threshold, and errors cannot be arbitrarily reduced by considering larger envelopes since $\alpha \rightarrow 0$ as $\beta \rightarrow 0$. Indeed, increasing the envelope implies that there are more excitations in the GKP code, and thus more excitation loss errors for a fixed $\kappa$. Moreover, the correction rate decreases for larger GKP sizes. In contrast, for qubit codes, the error increase caused by adding qubits is counterbalanced by the additional measurements that are performed in parallel, such that $p_*$ does not depend on the number of qubits.

Here, we choose the size of the GKP codes, set by $\epsilon$ in the dissipation circuits, such that the average photon number per mode is $\langle \hat n_j\rangle \approx 5.5$. For reference, we show in purple the infidelity of the Fock (single rail) encoding, which is the best code in the absence of error correction for this error model. 

\Cref{fig:photonLifetime}e) shows that the tesseract code (orange) has a larger distance than the square code (blue), $\delta_{\mathrm{tess}} = 4.0> \delta_{\square} = 3.3$, which we attribute to the fact that the Pauli operators of the tesseract code are longer by a factor $\sqrt[4]2 \approx 1.2 \approx \delta_{\mathrm{tess}}/\delta_{\square}$.

However, the break-even point of the tesseract code is lower than the square code. We mainly attribute this decrease to the fact that we analyzed the ``minimal working example'' of the two-mode code, i.e.\ a single ancilla qubit coupled to both modes. 
In this implementation, only a single bit of information is extracted at each step, and the dissipation circuits are performed in series. As a result, after a full round of dissipation circuits, the noise acted for four steps (4$\delta t$) in the tesseract code, against two steps (2$\delta t$) for the square code.
 By coupling more ancilla qubits to the modes, one could perform the dissipation circuits in parallel, which would correspondingly increase the break-even point and the fidelity. We also attribute part of the decrease in the break-even point to the added excitation number of the two-mode tesseract code. Indeed, by fixing the average excitation number per mode, we obtain a total excitation number $\langle \hat n\rangle_{\mathrm{tess}} \approx 2 \langle \hat n\rangle_{\square}$, which leads to a larger probability of error at each time step.

Due to the large tail in Fock basis of the GKP states, a large Hilbert space for each mode is required to perform the simulations. As a result, we were not able to accurately compute the robustness of codes with a number of excitations per mode much larger than $\langle \hat n_j\rangle \approx 5.5$. Here, we choose to perform the simulations in the Fock basis, truncating the Hilbert space at $65$ excitations for both modes.

\begin{figure}[t]
    \centering
    \includegraphics[scale=1]{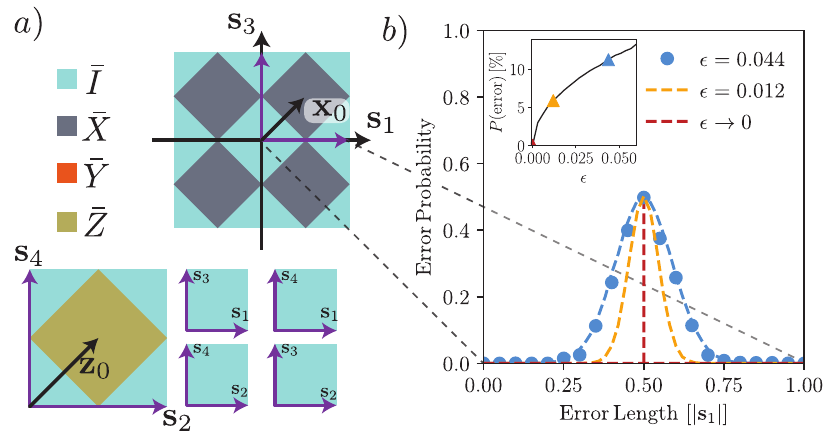}
    \caption{Effect of translation errors in the tesseract code. a) Logical operation applied (colors) as a function of the initial translation error in the planes given by span($\mathbf s_j,\mathbf s_k$). b) Logical error probability as a function of the length of a translation error along $\mathbf s_1$. We compare the full quantum model (blue dots) and the classical model (dashed line). Inset: Integrated logical error probability given that an error $\eta \mathbf s_1$ occured, with $\eta$ sampled uniformly in the interval $[0,1]$.}
    \label{fig:ErrorPropagationTesseract}
\end{figure}

\emph{Ancilla errors}\textemdash Next, we investigate how ancilla decay errors propagate to the logical subspace, when correcting errors using an ancilla qubit. Before discussing full numerical simulations where we extract the logical lifetime, we first use the classical model presented in \cref{sect:ErrorCorrection} to gain more intuition into the effect of these ancilla decay events.
As shown in \cref{sect:ErrorCorrection}, different translation errors propagate as different logical errors after error correction. In \cref{fig:ErrorPropagationTesseract}, we illustrate the effect of these translation errors in the tesseract code. Since four-dimensional spaces are difficult to visualize directly, we illustrate the effect of translation errors in different two-dimensional cuts of the full four-dimensional space. More precisely, we illustrate the effect of translation errors in the two-dimensional spaces spanned by each pair of lattice generators $\{\mathbf s_j\}$.
Crucially, we show in panel a) that the identity regions (light blue) around each stabilizer lattice point are directly connected through an ``isthmus''. As a result, for an ideal GKP state, one can correct any translation error of the form $\mathbf e = \eta \mathbf s_j$ for $\eta \in [0,1]$, which are the errors caused by an ancilla decay event.
The only exception is a translation error of exactly $\mathbf e = \mathbf s_j/2$, an error set of measure 0 which (taking $\mathbf s_1$ for example) would correct randomly to $\bar I$ or $\bar X$.  
This is in stark contrast to single-mode GKP codes, for which a translation error in the range $\eta \in [1/4,3/4]$ is uncorrectable as illustrated in \cref{fig:ErrorPropagation2d}.

In practice, each peak of the GKP state has a finite width, such that a translation error near the isthmus leaves finite-energy states with a partial overlap inside the uncorrectable region. In \cref{fig:ErrorPropagationTesseract}b), we compute the probability of a logical error given that a translation error $\mathbf e = \eta \mathbf s_j$ occured. We compare the results of the classical model (dashed lines) and of the full quantum model (dots) computed using the circuit in panel a), and find excellent agreement between the two without any fit parameter. We show in this figure that errors around $\mathbf e \approx \mathbf s_j/2$ lead to a finite probability of logical error, with a maximum of a $50\%$ probability of error at $\eta = 1/2$.
Crucially, the probability that an ancilla decay error propagates as a logical error in the tesseract code is reduced compared to the square GKP code and, for $\epsilon = 0.044$, we estimate that $\mathbb P(\mathrm{error}|\mathrm{qubit\, decay}) \approx 11\%$. 

Moreover, the state overlap inside the uncorrectable region can be reduced by decreasing the quadrature fluctuations of each individual peak of the GKP state. In other words, logical errors due to ancilla decay can be reduced by increasing the GKP size, which is in stark contrast to single-mode codes that feature a constant $\mathbb P(\mathrm{error}|\mathrm{qubit\, decay}) \approx 50\%$, see \cref{fig:ErrorPropagation2d}.
In the inset of \cref{fig:ErrorPropagationTesseract}b), computed using the classical model, we show the probability that an ancilla error propagates as a logical error as a function of the GKP size, set by the parameter $\epsilon$ in the dissipation circuits in \cref{fig:StabilizationCircuits}. The width of the quadrature coordinates fluctuations for each peak of the GKP state scale as $\sigma \approx \sqrt{\epsilon} \appropto 1/\sqrt{\bar n}$, with $\bar n$ the average excitation number in the GKP state. As a result, we obtain $\mathbb P(\mathrm{error}|\mathrm{qubit\, decay}) \appropto 1/\sqrt{\bar n}$. Through a fit we find $\mathbb P(\mathrm{error}|\mathrm{qubit\, decay}) \approx 0.53 \sqrt{\epsilon}$, which improves over the square code constant $\mathbb P(\mathrm{error}|\mathrm{qubit\, decay}) \approx 0.5$. \\ \indent
This last feature is particularly interesting for many experimental platforms, such as microwave cavities, where the physical lifetime of the modes is typically much larger than the physical lifetimes of the ancilllas. Indeed, longer oscillator lifetimes mean that the modes can host larger GKP states, which in turn are less sensitive to the propagation of ancilla errors. In contrast, for single-mode codes, the lifetime of the logical GKP qubits is directly limited by the ancilla lifetime.

We remark that the analysis above only holds to first order. In general, a second ancilla decay error cannot be corrected when it occurs before the state has recovered from the first error. Moreover, we showed earlier that the effective correction rate scales inversly with the GKP size, so that more dissipation circuits are required to recover from ancilla errors as the GKP size increases. As a result, second-order effects are not completely negligible for realistic ancilla decay rates. 

\begin{figure}[t]
    \centering
      \includegraphics[scale=0.5]{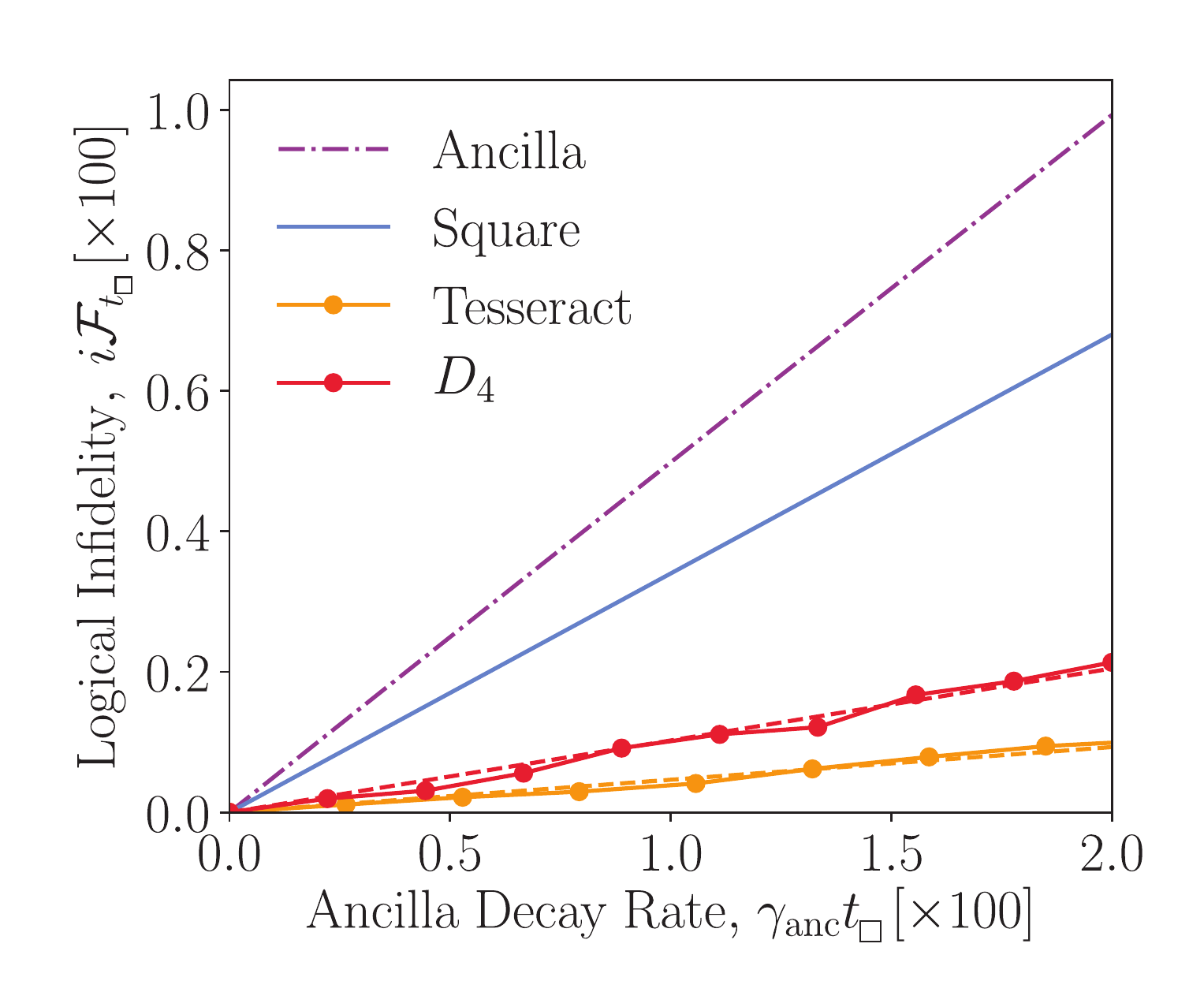}
    \caption{Logical process infidelity, $i\mathcal F$, in the presence of qubit ancilla decay, $\gamma_{\mathrm{anc}}$. The infidelity is computed for a time $t_\square$, i.e.\ , the time required to perform a controlled translation $C\hat T(\mathbf s)$ in the single-mode square code. The dotted-dashed purple line corresponds to the infidelity of the bare ancilla. Dots are the results of simulations, and dashed lines are linear fits. Although the infidelity of all GKP codes is proportional to the decay rate of the ancilla, multimode codes are much more robust to ancilla errors.}
    \label{fig:qubitLifetime}
\end{figure}

We now turn to full numerical simulations where we extract the lifetime of the logical information in the presence of finite ancilla lifetime.
\Cref{fig:qubitLifetime} shows the logical infidelity, computed using \cref{eq:channelFidelity}, as a function of the ancilla lifetime for the different GKP codes, in the absence of oscillator errors. For concreteness, we consider a model where the qubit ancilla is dispersively coupled to both oscillator modes.
In order to reduce the Hilbert space required to perform the simulations, we do not model the qubit directly. Rather, at each step, we apply the Kraus operators corresponding to a perfect dissipation circuit with probability $\expo{-\gamma_\mathrm{anc} t_{\mathrm{tot}}/2}$, and apply an error circuit with probability $1 - \expo{-\gamma_\mathrm{anc} t_{\mathrm{tot}}/2}$, with $\gamma_\mathrm{anc}$ the decay rate of the ancilla and $t_{\mathrm{tot}}$ the total time of the sBs dissipation circuit, see \cref{app:qubitDecay}. To compute the error circuit, we replace one of the three controlled translations in the sBs circuit with the unitary \cref{eq:CTerror}, taking an error time $t_{\mathrm{err}}$ randomly from a uniform distribution. 
We have defined above $t_{\mathrm{tot}} = 2t_{\epsilon} + t_{s}$, where $t_\epsilon$ is the time taken to perform the small controlled translations and $t_{s}$ is the time taken to perform $C\hat T(\mathbf s_j)$, with $t_{\epsilon}/t_{s} = \epsilon$. 
We pick which controlled translations to replace with probability $\{t_\epsilon/t_{\mathrm{tot}},t_s/t_{\mathrm{tot}},t_\epsilon/t_{\mathrm{tot}}\}$, respectively. We assume perfect qubit rotation pulses, as well as perfect qubit state preparation. These type of errors can only cause correctable errors, such that their effect on the logical lifetime is expected to be negligible compared with decay events during the large controlled translations~\cite{Royer20a}.
Note that we do include the effects of finite dispersive shifts between the ancilla and the oscillator modes, which can cause rotation errors. More concretely, we take parameters similar to Ref.~\cite{Campagne-Ibarcq20a} and choose $\chi t_s = 0.08$, which for $\chi/2\pi = 28 $ kHz and displacements to 900 photons corresponds to performing large controlled translations in $t_s \approx 450$ ns. For simplicity we choose identical parameters for both modes, but stress that this is not a requirement.
Although decay events lead to imperfect echoing of the dispersive interaction and additional oscillator rotations, large logical lifetimes are still attainable in the regime of small dispersive shift $\chi t_s \ll 1$.

The logical channel infidelity, $i\mathcal F = 1 - \mathcal F$, obtained from the simulations described above, is shown in \cref{fig:qubitLifetime}. There, we compare results for the square code (blue) and the tesseract code (orange). For reference, we show the infidelity of the bare ancilla (purple dot-dashed line), i.e.\ the lifetime of the information without any oscillator mode. Extracting the relation between ancilla decay rate and infidelity through a linear fit (dashed lines), we obtain that for the tesseract code $i\mathcal F_{t_\square} \approx  0.054 \gamma_{\mathrm{anc}}t_\square$, a more than six-fold improvement over the square code for which $i\mathcal F_{t_\square} \approx  0.34\gamma_{\mathrm{anc}} t_\square$. This graph shows one of the main results of this paper: multimode lattices can allow a greater robustness against ancilla errors, which is typically the limiting factor for bosonic codes. 

We now make a few remarks to better undertand the results above. First, we discuss our choice of normalization to make the decay rates dimensionless, $\gamma_{\mathrm{anc}} t_{\square}$. Here, $t_{\square}$ corresponds to the time required to perform a controlled translation $C\hat T(\mathbf s)$ for $|\mathbf s| = \sqrt 2$, i.e.\ the length of the stabilizers for the single-mode square code. The length of the stabilizers in the tesseract code is given by $|\mathbf s| = \sqrt[4]{2}$, such that the probability of a decay event during a single controlled translation $C\hat T(\mathbf s_{\mathrm{tess}})$ is given by $1 - \exp\{-\gamma_{\mathrm{anc}} t_{\square}/\sqrt[4]{2}\}$. 

Our choice of normalization is consistent with performing the dissipation circuits in series, comparing different codes but keeping a fixed ancilla decay rate. 
This is the choice we made when considering the error model for oscillator amplitude damping (see \cref{fig:photonLifetime}). Performing the dissipation circuits in parallel would lead to an increase of logical error rate due to ancilla decay, but a decrease in error rate due to oscillator errors. Performing the circuits in series is more adapted to implementations where the physical lifetimes of oscillators are longer than the lifetimes of the qubit ancillas used for control, which is typically the case for microwave cavities coupled to transmon qubit ancillas. 

Moreover, our choice for $t_{\square}$ is somewhat arbitrary as the time taken to perform multimode controlled translations depends on their physical implementation. Indeed, taking the displacement rate to be the same as the single-mode rate in all modes, one could perform multi-mode controlled translations faster as the relevant quantity would be the maximal length of the single-mode projection for each stabilizers. Here we choose the global displacement rate to be the same irrespective of the number of modes involved, such that the only quantity setting the time $t_s$ is the length of the stabilizer, $|\mathbf s_j|$.
Another option would be to perform the controlled translations of each mode sequentially, in which case multimode controlled translation would take a longer time. However, performing translations in such a way would negate the effects of the isthmus feature and lead to poorer protection against ancilla decay.

Although our simple classical model predicts that the robustness to ancilla errors scales with the size of the GKP code words, we have not shown this in \cref{fig:qubitLifetime}. This is due to the large Hilbert space required to accurately compute the logical lifetime, which combined with the modest scaling of the error rate with photon number, $i\mathcal F \propto \sqrt{\bar n}$, makes it difficult to observe numerically.

Finally, another important remark is that the isthmus feature of the tesseract code does not make the protocol fault-tolerant to ancilla decay, in the sense that the logical lifetime is still proportional to the lifetime of the qubit ancilla, $i\mathcal F \propto \gamma_{\mathrm{anc}} t_\square$. However, the logical error rates shown in \cref{fig:qubitLifetime} correspond to the ``worst case scenario'', where the cavity spends all its time entangled with ancilla. In practice, the ancilla spends a large portion of its time unentanged with the oscillator modes due to, for example, finite ancilla reset times, delays in data processing or even intentional idle times. As a result, we expect that the lifetime of the logical information in the tesseract code can be more than twenty times larger than the physical lifetime of the ancilla, $i\mathcal F \geq  0.05 \gamma_{\mathrm{anc}}t_\square$.

Performing a simulation where we include both qubit decay errors and oscillator amplitude damping for realistic parameters, we compute that a $\bar Z$ logical lifetime of $2.1$ ms is possible. For this simulation we considered an oscillator lifetime of $1/\kappa = 500$ $\mu$s, a qubit $T_1$ of $1/\gamma_{\mathrm{anc}} = 50$ $\mu$s, a time $\delta t = 2$ $\mu$s between dissipation circuits and parameters for the controlled translations similar to those stated earlier. Reducing the time required for controlled translations to 200 ns, increasing the lifetime of the ancilla to $100$ $\mu s$ and increasing the oscillator lifetime to 1 ms, we compute a logical lifetime for the $Z$ eigenstates of $14$ ms, a more than two orders of magnitude increase over the ancilla lifetime. In principle, the logical lifetime could be further increased by considering larger GKP sizes, although we were not able to verify that fact. Regardless, the logical fidelity with respect to oscillator amplitude damping does depend on the GKP size~\cite{Royer20a}, such that this quantity should be optimized. We have not performed this optimization here.

The robustness to ancilla errors could be further increased by using biased-noised ancillas where bit flips are suppressed, for example Kerr cats~\cite{Puri19a,Grimm20a} or fluxonium qubits~\cite{Manucharyan09r}. Indeed, phase errors on the ancilla commute with the Hamiltonian generating the controlled translations, such that they propagate to the logical subspace as correctable errors, $\hat T(\epsilon)$ or as full lattice translations, $\hat T(\mathbf s_j)$, both of which are correctable~\cite{Royer20a}.

\emph{Homodyne Error Correction}\textemdash Finally, we briefly investigate the performances of the tesseract code when error correction is performed using GKP ancilla states and homodyne measurements. 
Full lines in \cref{fig:ErrorProbHomodyne} are computed taking noiseless (ideal) ancillas, and we compute that the tesseract code (orange) outperforms the square code (blue) for all squeezing levels except for very low squeezing where the overhead of using two modes instead of one dominates. We attribute this improvement to the longer logical operators of the tesseract code, which translates to lower error probabilities. Interestingly, this improvement is maintained in the more realistic situation of noisy ancillas (dashed lines), where we compute an order of magnitude reduction in error rate at $\sigma^{(\mathrm{dB})}\approx 15$ dB. As shown earlier, back-propagation of ancilla translation errors happens in the form $\hat T(\sigma\mathbf s_j)$, such that the isthmus feature of the tesseract code also plays a beneficial role when considering homodyne error correction.

\section{D4 code} \label{sect:D4lattice}
In this section, we study in detail a two-mode GKP code based on the $D_4$ lattice, which allows the densest lattice packing in four dimensions. As we show, this particular code has several interesting features, particularly with respect to logical operations.
Indeed, in contrast to single-mode GKP codes and the tesseract code, all single qubit Clifford gates can be performed with passive gaussian operations (envelope-preserving gates). The $D_4$ code also allows exact non-Clifford gates through envelope-preserving Kerr-type interactions.

To help with the visualization of the four-dimensional lattice, we mention that the $D_4$ is a laminated lattice. In a similar fashion to the 3-dimensional hexagonal close-packed lattice which can be built by stacking layers of two-dimensional honeycombs lattices (stacking oranges), we can build the $D_4$ lattice by ``stacking'' layers of 3-dimensional body-centered cubic lattices in the fourth dimension.
\subsection{Code Words}
Choosing a set of generators which all have support on both modes, we set
\begin{equation}\label{eq:SD4}
S_{D_4} = \begin{pmatrix}
1 & 0 & 1 & 0\\
1 & 0 & 0 & -1 \\
0 & 1 & -1 & 0 \\
1 & 0 & 0 & 1 \\
\end{pmatrix}.
\end{equation}
However, we remark that this choice is not unique since the $D_4$ lattice has 12 vectors of minimal length (not counting those that differ only by a sign). 
We choose the base representatives of the logical operators to be
\begin{equation}
L_{0,D_4} = \begin{pmatrix}
\mathbf x_0 \\
\mathbf y_0 \\
\mathbf z_0 \\
\end{pmatrix} = \begin{pmatrix}
1/2 & 1/2 & 1/2 & 1/2\\
-1/2 & 1/2 & 1/2 & 1/2 \\
1 & 0 & 0 & 0 \\
\end{pmatrix}.
\end{equation}
Similar to the stabilizers, this choice is not unique and for each logical Pauli operator there are 4 equivalent representatives of minimum length (not counting those that differ only by a sign). For example, the $\bar Z$ logical operator can be represented as $\mathbf z = (\pm1,0,0,0);(0,\pm1,0,0);(0,0,\pm1,0)$ or $(0,0,0,\pm1)$, such that $\bar Z$ eigenstates can be described as hypercubic qunaught states $S_Z = \mathbb I_4 = S_{\varnothing}^{\oplus 2}$. In the trivial gauge $\vmu = \mathbf 0$ and with the choice of basis in \cref{eq:SD4}, we have
\begin{equation}\label{eq:D4Zvarnothingcodewords}
\ket{\pm \bar Z} = \ket{\pm\varnothing}^{\otimes 2},
\end{equation}
with $\ket{\varnothing}$ the single-mode square qunaught state. We have also defined the ``negative'' qunaught state $\ket{-\varnothing}$ as the -1 eigenstate of the translation operators associated to the generators of $S_{\varnothing}$, i.e.\ the square qunaught state with gauge $\vmu = (1,1)$. As in the tesseract code, both code words are separable states of the two modes, and as a result can be prepared in independent modes.

Measurements in the $\bar Z$ basis of the $D_4$ code can be performed in both modes separately, yielding a simple two-bit repetition code that can allow the detection of one measurement error. In principle, we could also measure all four orthogonal representatives and take a majority vote to determine the most likely measurement result. However, we note that in this case all measurement are not independent as, for example, an ancilla decay error during the $\mathbf z = (1,0,0,0)$ measurement can propagate as a measurement error in the $\mathbf z = (0,1,0,0)$ measurement.

Due to the four-fold rotation symmetry of the qunaught states, they have a definite excitation number modulo four, and they can be expressed in the Fock number basis as
\begin{subequations}\label{eq:D4Zcodewords}
\begin{align}
\ket{+\varnothing} &= \sum_{j}c_{j+} \ket{4j},\\
\ket{-\varnothing} &= \sum_{j}c_{j-} \ket{4j+1}.
\end{align}
\end{subequations}
The fact that $n = 0 \,\mathrm{ mod } \,4$ for the positive qunaught state can be computed by directly applying the rotation operator to the state \cref{eq:qunaughtStatePositionBasis}. By expressing the negative qunaught state as a translated positive qunaught state, we obtain
\begin{equation}
\begin{aligned}
\expo{i\frac{\pi}{2} \hat n}\ket{-\varnothing} &= \expo{i\frac{\pi}{2} \hat n} \hat T[(1/2;1/2)]\ket{+\varnothing}\\
&= \hat T[(-1/2;1/2)]\ket{+\varnothing}\\
&= \expo{i\frac{\pi}{2}}\hat T[(1/2;1/2)]\hat T[(-1;0)]\ket{+\varnothing}\\
&= \expo{i\frac{\pi}{2}}\ket{-\varnothing}
\end{aligned}
\end{equation}
which implies that $\ket{-\varnothing}$ has support only on Fock states $n = 1 \,\mathrm{ mod } \,4$. Importantly, since the envelope does not change the excitation number, this is also true for finite-energy states.

We remark that in contrast to single-mode GKP codes, the two finite-energy code words of the $D_4$ code are exactly orthogonal as evidenced by \cref{eq:D4Zcodewords}. However, this orthogonality has limited usefulness in practice, as logical measurements of the code words are performed through (controlled) translations which do not allow perfect distinguishability. In principle, one could perform logical measurement through excitation number measurements in a similar fashion to logical measurement of cat codes~\cite{Ofek16a,Wang20a}. However, this type of measurement is less robust against oscillator errors such as photon loss. Moreover, the distinguishability limit for GKP codes is much smaller than typical errors induced by practical measurement circuits, such that we expect measurement of translation properties to remain optimal.

Interestingly, the $D_4$ code can be equivalently interpreted as the concatenation of two single-mode diamond GKP codes with a two-qubit repetition code along the $\bar Y$ axis. Taking two single-mode qubits encoded in diamond GKP codes and restricting the subspace to the +1 eigenspace of the $\bar Y_{\Diamond}\bar Y_{\Diamond}$ logical operator, we obtain the $D_4$ code. An inspection of the logical operators in both $D_4$ and diamond codes then reveals the correspondence 
\begin{subequations}\label{eq:D4toDiamond}
\begin{align}
\bar X_{D_4} &= \bar Z_{\Diamond}\otimes \bar Z_{\Diamond}\; \mathrm{ or } \; \bar X_{\Diamond}\otimes \bar X_{\Diamond},\\
\bar Y_{D_4} &= \bar Z_{\Diamond}\otimes \bar X_{\Diamond}\; \mathrm{ or } \; \bar X_{\Diamond}\otimes \bar Z_{\Diamond},\\
\bar Z_{D_4} &= \bar Y_{\Diamond}\otimes \bar I \; \mathrm{ or } \; \bar I\otimes \bar Y_{\Diamond}.
\end{align}
\end{subequations}\indent
In particular, this implies that two diamond codes can be merged into a single $D_4$ code, or conversely that a $D_4$ code can be split into two (potentially entangled) diamond codes using the techniques of \cref{sect:codeSwitching}. We recall that a diamond code is simply a rotated square code, such that the same correspondences apply to square GKP codes. These relations can be used to prepare entangled states of square GKP codes, for teleportation-based error correction of GKP codes, and to purify magic states of square or $D_4$ GKP codes, see \cref{app:codeSwitchingD4}.

\subsection{Gates}\label{sect:D4Gates}
\begin{figure}[t]
    \centering
	    \includegraphics[scale=1]{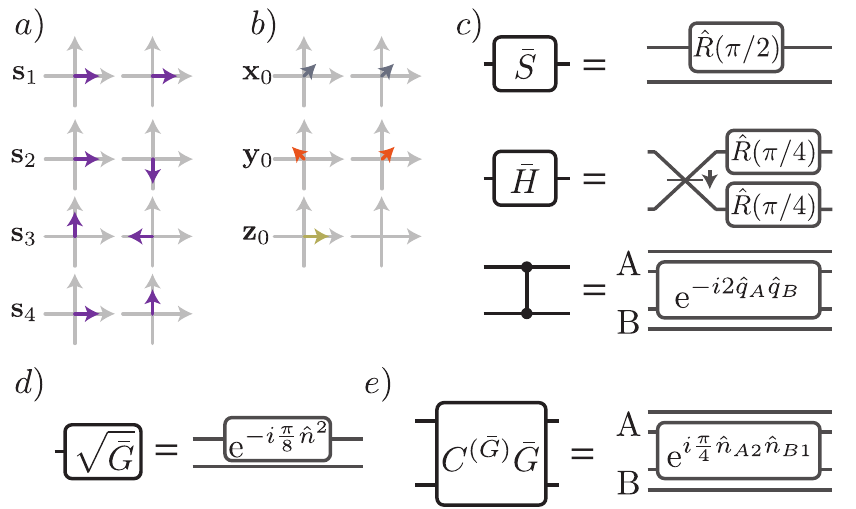}
    \caption{$D_4$ code. a) Phase space representation of the stabilizer generators. The left and right graphs represent the projection onto the (q,p) phase space of the first and second mode, respectively. b) Similar representation for the logical operators. c) Logical Clifford operations in the $D_4$ code.
         The $\bar S$ gate can be effected by a $\pi/2$ rotation in either mode, with a rotation in the first mode represented. The $\bar H$ gate is implemented by a beamsplitter followed by a rotation of $\pi/4$ in both modes. The logical $C\bar Z$ gate is implemented by a rescaled SUM gate. We represent this gate as an operation between the second mode of A and the first mode of B, but choosing any pair of modes from the codes A and B is equivalent. d) Non-Clifford $\sqrt {\bar G}$ logical gate, with $\bar G = \bar X \bar S = (\bar X + \bar Y)/\sqrt 2$. A Kerr gate in either mode realizes this logical operation. e) Non-Clifford $\bar G$-controlled $\bar G$ logical gate, which can be realized by a cross-Kerr gate involving any pair of mode from the two $D_4$ codes.}
    \label{fig:D4Gates}
\end{figure}

Interestingly, for the $D_4$ GKP code, all single-qubit Clifford gates can be physically realized using envelope-preserving Gaussian operations. In particular, logical Pauli gates can be implemented in an envelope-preserving manner in contrast to translations which do not commute with the envelope.
This is due to the high symmetry of the $D_4$ lattice, and in particular the fact that all Pauli operators have the same length, $|\mathbf x_0| = |\mathbf y_0| = |\mathbf z_0|$. We describe below the gate set $\langle \bar S,\bar H, C\bar Z\rangle$, with results summarized in \cref{fig:D4Gates}.

\emph{Phase Gate}\textemdash The $\bar S$ logical gate is given by a rotation of either mode by $\pi/2$,
\begin{equation}
\hU(\bar S) = \hat R(\pi/2)\otimes \hat I\; \mathrm{or}\; \hat I \otimes \hat R(\pi/2).
\end{equation}

\emph{Hadamard Gate}\textemdash As illustrated in \cref{fig:D4Gates}c), the logical Hadamard gate is obtained by combining a beam-splitter operation, $\hat B$, with a $\pi/4$ rotation of both modes, 
\begin{equation}\label{eq:HadamardGateD4}
\hat U_{D_4}(\bar H) = \hat R(\pi/4)^{\otimes 2}\hat B.
\end{equation}
Conveniently, the symplectic representation of $\bar H$ is proportional to a Hadamard matrix.
We remark that, in contrast to the $\bar S$ gate, the logical $\bar H$ gate modifies the trivial gauge, motivating the need for gauge updates.

\emph{Controlled-$\bar Z$ Gate}\textemdash We can realize a $C\bar{Z}$ gate between two $D_4$ code A and B by implementing a rescaled SUM gate of the form
\begin{equation}
\hat U_{D_4}(C\bar{Z}) = \expo{i2\hat o_{A}\hat o_{B}},
\end{equation}
where $\hat o_{\alpha}\in \{\hx_{\alpha,1},\hx_{\alpha,2},\hp_{\alpha,1},\hp_{\alpha,2}\}$ can be any quadrature coordinate of the two modes comprising the $D_4$ code $\alpha \in \{A,B\}$. This is due to the fact that a displacement in any of these four directions is equivalent to a logical $\bar Z$ gate. The choice $\hat U(C\bar{Z}) = \expo{i2\hx_{A,2}\hx_{B,1}}$ is illustrated in \cref{fig:D4Gates}c). While this $C\bar Z$ gate is not envelope-preserving and requires squeezing, it is qualitatively similar to the square and tesseract code $C \bar Z$ gates which also require (rescaled) SUM gates.

Note that it is impossible for any GKP-type qubit to have the full multi-qubit Clifford group implemented in a Gaussian and envelope-preserving manner. Take for example an entangling gate $\bar G$ such that $\bar P_2\otimes \bar P_3 = \bar G^\dag (\bar P_1\otimes \bar I ) \bar G$ with $\bar P_1,\bar P_2,\bar P_3$ all not identity. The symplectic representation of the previous equation reads $\mathbf p_2 \oplus \mathbf p_3 =  G(\mathbf p_1 \oplus \mathbf 0)$ for some $\mathbf p_1,\mathbf p_2,\mathbf p_3$ all not $\mathbf 0$.  Taking $\bar P_1$ such that its Pauli representative is of minimal length, we have $|\mathbf p_2 \oplus \mathbf p_3| > |\mathbf p_1 \oplus \mathbf 0|$ with respect to the Euclidian norm, implying that $G$ does not conserve distances. 

\emph{Non-Clifford Gates}\textemdash
In the following, we introduce envelope-preserving non-Clifford gates following the general approach laid out in \cref{subsect:EnvelopePreservingNonCliffordGates}. 
First, we introduce a logical $\bar T = \sqrt{\bar S} = \mathrm{Diag}[1;\expo{i\pi/4}]$ gate. Leveraging the fact that $\bar Z$ code words have a definite Fock number modulo four, see \cref{eq:D4Zcodewords}, we have
\begin{equation}\label{eq:Tgate}
\hU_{D_4}(\bar{T}) = \expo{i\frac{\pi}{4}\hat n_j^2},
\end{equation}
where $\hat n_j$ is the photon number operator of either mode comprising the $D_4$ code. 

We remark that the $\bar T= \sqrt{\bar S}$ is based on the $\bar S$ gate which is of order $g=4$. Changing the gauge to $\vmu = (0,1,1,1)$, a rotation $\hat R(\pi/2)$ rather implements a gate $\bar G = \bar X\bar S$ of order $g=2$. As a result, a Kerr unitary can also implement the gate
\begin{equation}\label{eq:Ggate}
\hU_{D_4}\left(\sqrt{\bar{G}}\right) = \expo{i\frac{\pi}{8}\hat n_j^2}.
\end{equation}
This gate can equivalently be expressed as $\sqrt{\bar{G}} = \bar T^\dag \bar H \bar S \bar H \bar T$.
Tunable Kerr interactions can be realized in a microwave cavity by driving a transmon qubit ancilla \cite{Zhang21y}, in which case the $\sqrt{\bar{G}}$ gate can be realized in half the time required for the $\bar T$ gate.

Leveraging the same Fock number properties of the $D_4$ code, a controlled-S gate can be realized using a cross-Kerr interaction
\begin{equation}
\hU_{D_4}(C\bar{S}) = \expo{i\frac{\pi}{2}\hat n_{A}\hat n_{B}},
\end{equation}
which is implemented in the $\vmu = \mathbf 0$ gauge. Updating the gauge of both modes such that the rotation $\hat R(\pi/2)$ implements a $\bar G$ gate, $\vmu = (0,1,1,1)$, we rather obtain
\begin{equation}
\hU_{D_4}(C^{(\bar G)}\bar{G}) = \expo{i\frac{\pi}{4}\hat n_{A}\hat n_{B}}.
\end{equation}
This gate can be equivalently expressed as $C^{(\bar G)}\bar{G} = (\bar T^\dag)^{\otimes 2}\times  \bar H\otimes \bar I\times\mathrm{CNOT}\times\bar H\otimes \bar I \times\bar T^{\otimes 2}$.
In microwave cavities, this gate could be realized with a tunable cross-Kerr coupling~\cite{Ye21v,Zhang21y}, assuming that higher order non-linearities are negligible.

We could also realize a $\sqrt{\bar H}$ in an envelope-preserving manner since the $\bar H$ is based on a lattice isometry. However, performing this gate is likely to prove more challenging than Kerr or cross-Kerr gates.

Finally, we note that there is a Gaussian operation that maps the $D_4$ code to the tesseract code (and the converse), in this case an independent squeezing and rotation of both modes. As a result, the non-Clifford gates presented in this section for the $D_4$ code could also be employed in the tesseract code by mapping one encoding to the other.

\subsection{Error Correction}
In this section, we follow the same structure as the tesseract error correction section and describe the robustness of the $D_4$ code against resonator amplitude damping, qubit ancilla errors and finally the Gaussian translation error channel using homodyne error correction.

\emph{Amplitude damping}\textemdash We consider a similar protocol as in the tesseract code, with the dissipation cicuits of the $D_4$ code applied in series and an excitation loss channel applied in-between each dissipation circuit. The results are shown by the red points in \cref{fig:photonLifetime}e). 

The obtained distance for the $D_4$ is similar to that of the square code, $\delta_{D_4} = 3.2 \approx \delta_{\Square} = 3.3$. Although the logical Pauli operators of the $D_4$ code have an increased length compared to the square code, $\min|\mathbf p_{0}| = 1$, the dissipation circuits are less efficient at correcting errors since the stabilizers are not orthogonal to each other. In particular, the smallest uncorrectable translation error in the $D_4$ code has a length $|\mathbf e| = 1/\sqrt 2$, which is equal to the length of the smallest uncorrectable error in the square code. As a result, the square and $D_4$ code have similar distances $\delta$.\\ \indent
Moreover, since the errors are corrected in series and the noise channel acts for twice longer than the square code before a full dissipation round is completed, the break-even point of the $D_4$ code is smaller than in the single-mode square code. The break-even point of the $D_4$ code is also smaller than the tesseract code, which can be understood through the classical model presented in \cref{sect:ErrorCorrection}. Computing the eigenvalues of the Hessian matrix $H(\Phi_{D_4})$, we get a smallest eigenvalue $\lambda = l^2 (2-\sqrt 3)\approx 1.7$, which is smaller than the tesseract code $\min_\lambda H(\Phi_{\mathrm{tess}}) = l^2 \sqrt[4]{2} \approx 8.9$. This is a result of the fact that the tesseract code basis treats the four-dimensional space homogeneously, while the $D_4$ code has preferred direction, such that there exists a direction in the $D_4$ code where errors are corrected at a slower rate.

\begin{figure}[t]
    \centering
    \includegraphics[scale=1.1]{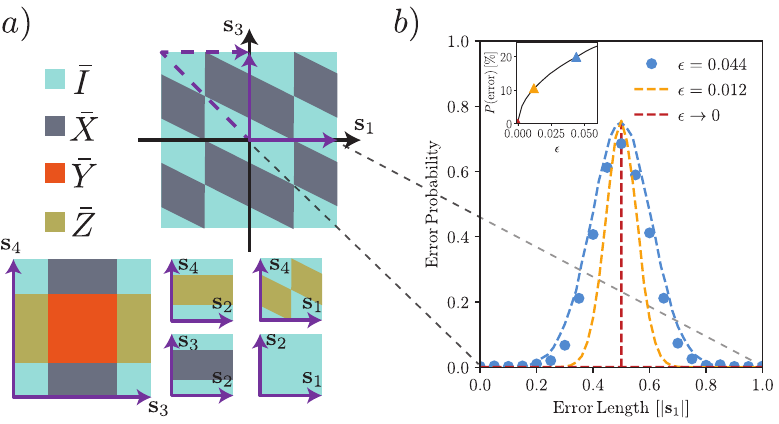}
    \caption{Effect of translation errors in the $D_4$ code. a) Logical operation applied (colors) as a function of the initial translation error in the planes given by span($\mathbf s_j,\mathbf s_k$). One could achieve the isthmus property by performing controlled translations following the purple dashed line instead of the more direct full line in the $\{\mathbf s_1,\mathbf s_3\}$ panel. b) Logical error probability as a function of the length of a translation error along $\mathbf s_1$. We compare the full quantum model (blue dots) and the classical model (dashed lines) Inset: Integrated logical error probability given that an error $\eta \mathbf s_1$ occured, with $\eta$ sampled uniformly.}
    \label{fig:ErrorPropagationD4}
\end{figure}

\emph{Ancilla Decay}\textemdash \Cref{fig:ErrorPropagationD4} illustrates how translation errors propagates as logical errors to the $D_4$ code. At first glance, the $D_4$ code does not possess the same ``isthmus'' properties that the tesseract code has. Indeed, while errors along $\mathbf s_2$ are always correctable and errors along $\mathbf s_1$ have the isthmus property, errors along $\mathbf s_3$ and $\mathbf s_4$ are not all correctable. As illustrated in the bottom left of \cref{fig:ErrorPropagationD4}a), translation errors in the plane spanned by $\{\mathbf s_3,\mathbf s_4\}$ propagate as logical errors in a similar fashion to the square GKP code, see \cref{fig:ErrorPropagation2d}. 
One solution to recover the isthmus property is to perform controlled translations in a zig-zag manner, for example decomposing $C\hat T(\mathbf s_3) = C\hat T(\mathbf s_3 - \mathbf s_1)C\hat T(\mathbf s_1)$ as shown by the dashed purple arrow in the $\{\mathbf s_1,\mathbf s_3\}$ plane. While logically equivalent to a direct controlled translation $C\hat T(\mathbf s_3)$, ancilla decay during these controlled displacements now propagate as translation errors along the dashed line, recovering the isthmus property. This decreases the probability that an ancilla decay error propagates as a logical error, at the cost of doubling the time required to perform controlled translations for 2 of the 4 stabilizers. \Cref{fig:ErrorPropagationD4}b) focuses on the effect of translation errors along the stabilizer $\mathbf s_1$. The points are computed using the full quantum model, while the dashed lines are computed using the simpler classical model. This stabilizer clearly exhibits the isthmus property, with an integrated error probability that decreases with increasing GKP size (smaller $\epsilon$). However, the $D_4$ isthmus is ``thinner'' compared to the tesseract code, with at worst an error probability of $75\%$ for $\mathbf e = \mathbf s_1/2$. The inset shows the integrated logical error probability given that a qubit decay occured during $\mathcal D(\mathbf s_1)$, computed using the classical model. The colored dots correspond to the dashed lines in the main panel. Through a fit of the black line, we find that $\mathbb P(\mathrm{error}|\mathrm{qubit\, decay}) \approx 0.94 \sqrt{\epsilon}$.

Using the full quantum model, the red points in \cref{fig:qubitLifetime} show the lifetime of the logical information as a function of the ancilla decay rate for the $D_4$ code (red points), using the zig-zag paths. We take the probability of an ancilla decay during the dissipation of $\mathbf s_3,\mathbf s_4$ double that of $\mathbf s_1,\mathbf s_2$ to account for the longer zigzag paths. These simulations show that an improvement over the single-mode square code is possible with the $D_4$ code, although the tesseract code performs better. Extracting the linear relation between ancilla decay rate and infidelity, we find $i\mathcal F \approx 0.10 \gamma_\mathrm{anc}t_\square$.

\emph{Homodyne Error-correction}\textemdash For the same reasons that the dissipation cicuits are less efficient in the $D_4$ code than in the tesseract code, homodyne error-correction for the $D_4$ code leads to poorer results than both square and tesseract code. This is due to the fact that, for this choice of basis, the homodyne measurements do not treat the two-mode phase space homogeneously.
However, we remark that our choice of basis and decoder is not optimal, and we expect that better performances are possible for this code.

The results above show that although in principle the $D_4$ code allows a better protection against errors, with logical operators being of longer length than the tesseract code, its performance is worse than the tesseract code for the error-correction schemes introduced.

\section{Lattices and Concatenated Codes}\label{sect:qubitCodeConstruction}
In this section, we explore further the general connections between multimode lattices and concatenated codes. Given the strong links between lattices and classical error-correcting codes~\cite{Conway13c}, it is reasonable to expect that good quantum codes are also related to interesting lattices. For example, early studies of multimode GKP codes~\cite{Harrington01a,Harrington04t} showed that symplectic multimode lattices can achieve encoding rates similar to single-mode GKP codes concatenated with CSS qubit codes for the Gaussian translation error channel. 
In this section, we lay out an explicit concatenated code construction for multimode lattices similar to construction A in Ref.~\cite{Conway13c}.

We start by taking $m$ oscillator modes and we define a separable lattice encoding $m$ qubits, $S = S_{\mathrm{base}}^{\oplus m}$ with $\det(S_\mathrm{base}) = 2$. A standard choice of base single-mode code is the square code, but we also provide examples below where other choices lead to interesting lattices. In general, one could choose a different two-dimensional lattice for each of the $m$ modes, and the results we present can easily be generalized to that case. The logical Pauli operators of the $m$-mode, $m$-qubit code can be expressed as 
\begin{equation}
L = (S_{\mathrm{base}}^*)^{\oplus m} = \frac{S_{\mathrm{base}}^{\oplus m}}{2}.
\end{equation}

We aim to concatenate these $m$ single-mode GKP qubits with a $[[m,k,d]]$ qubit stabilizer code $\mathcal C_q$. For each of the $m-k$ (qubit) stabilizers of $\mathcal C_q$, we replace a lattice generator of the base GKP code by the corresponding combination of base GKP Pauli operators. Out of the original $2m$ stabilizer generators each having support on a single-mode, only $m+k$ then remain. 

This replacement can be done by starting with the binary matrix representation of a Pauli stabilizer code~\cite{Calderbank97w}, which represents $\mathcal C_q$ as a $(m- k) \times 2m$ binary matrix. To make the rest of the construction easier, we swap the columns of the binary matrix such that the odd-numbered columns represent the presence of an $X$ operator, while the even-numbered columns represent the presence of a $Z$ operator.
 For example, the binary matrix of the five-qubit code, with stabilizer group generated by cyclic shifts of $XZZXI$, is mapped to
\begin{align}
\left(
\begin{array}{ccccc|ccccc}
1 & 0 & 0 & 1 & 0 &   0 & 1 & 1 & 0 & 0\\
0 & 1 & 0 & 0 & 1 &   0 & 0 & 1 & 1 & 0\\
1 & 0 & 1 & 0 & 0 &   0 & 0 & 0 & 1 & 1\\
0 & 1 & 0 & 1 & 0 &   1 & 0 & 0 & 0 & 1\\ 
\end{array}\right)\\
\rightarrow B = 
\begin{pmatrix}
1 & 0 & 0 & 1 & 0 & 1 & 1 & 0 & 0 & 0\\
0 & 0 & 1 & 0 & 0 & 1 & 0 & 1 & 1 & 0\\
1 & 0 & 0 & 0 & 1 & 0 & 0 & 1 & 0 & 1\\
0 & 1 & 1 & 0 & 0 & 0 & 1 & 0 & 0 & 1
\end{pmatrix}.
\end{align}
We then promote this binary matrix to a $2m \times 2m$ full rank integer matrix $T$ by adding rows with a single 2. This construction is motivated by the fact that, for single-mode codes, the Pauli operators $\mathbf x_0$ and $\mathbf z_0$ can always be chosen to be colinear with the stabilizers, $2S^* = S$. For a QEC code~\footnote{This construction is not necessarily full rank for QEC codes that do not correct all $\bar X$ and $\bar Z$ errors such as the repetition code. The position of the ``2s'' needs to be adjusted accordingly.}, $T$ can be constructed as
\begin{equation}\label{eq:concatenatedLatticeGenerators}
T = 
\begin{pmatrix}
\multicolumn{2}{c}{B}\\
0_{(m-k) \times (m+k)} & 2 \mathbb I_{m+k}
\end{pmatrix}.
\end{equation}
The generator matrix of the multimode lattice corresponding to the concatenated code can then be expressed as
\begin{equation}\label{eq:concatenatedDecomposition}
S_q = T L
\end{equation}
One can check that $\det(T) = 2^{2m - (m-k)}$, such that $\det(S_q) = 2^{k}$, encoding $k$ qubits as desired. In general, the decomposition above does not yield generators of minimal length, and it can be advantageous to redefine the generator matrix as $S_q = RTL$, for some unimodular matrix $R$ which can be found, for example, through the Lenstra-Lenstra-Lovász (LLL) algorithm~\cite{Lenstra82x}.

The logical Pauli operators of the final multimode code can either be obtained through a similar construction starting with the operators of the qubit code $\mathcal C_q$, or by following the approach outlined above in \cref{sect:multimodeGKP}, yielding a dual lattice generator matrix
\begin{equation}
S^*_q = - (L^{-1}T^{-1})^T \Omega.
\end{equation}
In this construction, the minimal length of the logical Pauli translations is given by
\begin{equation}\label{eq:distanceIncreaseMultimode}
|\mathbf p_{0,\mathrm{conc}}| = \min_{\bar P_{\mathcal C_q}} |\mathbf p| = \min_{\bar P_{\mathcal C_q}} \sqrt{\sum_{p\in \bar P_{\mathcal C_q}} |\mathbf p_{0,\mathrm{base}}|^2},
\end{equation}
where the minimization is over all representatives of the Pauli operator $\bar P$ in the code $\mathcal C_q$, and the sum is taken over all single-qubit Pauli operators comprising this representative of $\bar P$. Taking a hexagonal base code where $|\mathbf x_0| = |\mathbf y_0| = |\mathbf z_0|$, and defining $d_p$ the minimal support of the logical Pauli $\bar P$ in $\mathcal C_q$, the above equation simplifies to 
\begin{equation}
|\mathbf p_{0,\mathrm{conc}}| = \sqrt{d_p}|\mathbf x_{0,\hexagon}|.
\end{equation}
For asymmetric base codes such as the square code where $|\mathbf x_0| = |\mathbf y_0|/\sqrt 2 = |\mathbf z_0|$, the representative $\bar P$ which minimizes \cref{eq:distanceIncreaseMultimode} is not necessarily the representative of minimal support. We give below a few examples of multimode lattices that can be reframed in the concatenated construction.

\begin{figure}[t]
    \centering
      \includegraphics[scale=1]{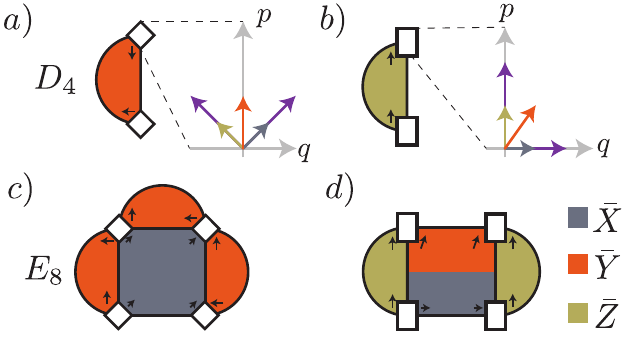}
    \caption{Correspondence between some multimode lattices and concatenated codes. Multimode stabilizers are represented by both a color coding, with reference on the bottom right of the figure, and by small black arrows. The diamonds and rectangles represent single-mode GKP code, with a phase space representation of these modes shown in panels a) and b). a) The $D_4$ lattice code is equivalent to a diamond code concatenated with a two-qubit repetition code. b) The tesseract code is equivalent to the concatenation of a rectangular code with a two-qubit repetition code along $\bar Z$. c) The $E_8$ lattice is equivalent to a stabilizer state with a base diamond code. d) Two tesseract code can be concatenated along the $\bar Y$ axis to form a four-mode code with equal length logical Pauli operators. }
    \label{fig:ConcatenationCorrespondance}
\end{figure}

\textit{Repetition Code and Diamond Code}\textemdash
Consider the base GKP code to be the diamond code and the qubit code $\mathcal C_q$ to be the $m$ qubit repetition code along the $\bar Y$ axis, $\ket{\bar 0/\bar 1} = \ket{\pm \bar Y_\Diamond}^{\otimes m}$. Following the concatenation construction, one obtains the $D_{2m}$ lattice family described in \cref{sect:latticeExamples}, with lattice generators given by, for example, \cref{eq:DlatticeGenerators}. This construction then yields the logical Pauli operators
\begin{equation}
L_{0,D_{2m}} = \begin{pmatrix}
1/2 & 1/2 & ... & 1/2 & 1/2\\
-1/2 & 1/2 & ... & 1/2 & 1/2 \\
1 & 0 & ... & 0 & 0 \\
\end{pmatrix},
\end{equation}
which reduce to the logical operators of the $D_4$ code in the two-mode case as illustrated in \cref{fig:ConcatenationCorrespondance}a).

As demonstrated by this example, the qubit code $\mathcal C_q$ need not be a full quantum error-correction code to yield an interesting lattice. The qubit code merely needs to increase the distance of some logical Pauli operator. In the case of the two-qubit repetition code $\mathcal C_q$, the distances for each logical Pauli operator are given by $(d_x,d_y,d_z) = (2,1,2)$. From \cref{eq:distanceIncreaseMultimode}, we obtain that the length of the logical Pauli operators is increased from $(|\mathbf x_0|,|\mathbf y_0|,|\mathbf z_0|) = (1/\sqrt 2,1,1/\sqrt 2)$ in the base diamond code to $(1,1,1)$ in the $D_4$ code. On top of the increased error-correction capabilities, this increased distance also symmetrizes the length of all logical Pauli operators, which in turn enables one to perform all single-qubit Clifford operations in an envelope-preserving manner. 

\textit{Repetition Code and Rectangular Code}\textemdash As illustrated in \cref{fig:ConcatenationCorrespondance}b), the tesseract code can be obtained from a base rectangular code,
\begin{equation}
S_{\vrectangle} = \sqrt[4]{2}\begin{pmatrix}
1 & 0 \\
0 & \sqrt 2\\
\end{pmatrix},
\end{equation}
concatenated with the two-qubit repetition code along the $\bar Z$ axis with $\mathbf z_0 = \mathbf s_2/2$.

The fact that the $D_4$ and tesseract code can be understood as two-qubit repetition codes explains why one of the logical Pauli operators can be measured in two separate modes, allowing the implementation of a two-bit measurement repetition code.  

In a similar fashion to the diamond GKP and qubit repetition code construction which symmetrizes the length of the logical Pauli operators, we can concatenate the tesseract code with a two-qubit repetition code along the $\bar Y_{\mathrm{tess}}$ axis to yield a four-mode lattice where all logical operators are of the same length.
The generator matrix of this four-mode is given by
\begin{equation}\label{eq:SfourModeCode}
S = \sqrt[4]{2}\begin{pmatrix}
1 & 0 & 0 & 0 & 0 & 0 & 0 & 0\\
0 & \frac{1}{\sqrt 2} & 0 & \frac{1}{\sqrt 2} & 0 & 0 & 0 & 0\\
0 & 0 & 1 & 0 & 0 & 0 & 0 & 0\\
0 & \frac{1}{\sqrt 2} & 0 & -\frac{1}{\sqrt 2} & 0 & 0 & 0 & 0\\
0 & 0 & 0 & 0 & 1 & 0 & 0 & 0\\
0 & 0 & 0 & 0 &  & \frac{1}{\sqrt 2} & 0 & \frac{1}{\sqrt 2}\\
0 & 0 & 0 & 0 & 0 & 0 & 1 & 0\\
1/2 & \frac{1}{\sqrt 2} & 1/2 & 0 & 1/2 & \frac{1}{\sqrt 2} & 1/2 & 0\\
\end{pmatrix},
\end{equation}
with logical operators
\begin{equation}\label{eq:LfourModeCode}
L_0 = \sqrt[4]{2}\begin{pmatrix}
1/2 & \frac{1}{\sqrt 2} & 1/2 & 0 & 0 & 0 & 0 & 0\\
1/2 & 0 & 1/2 & 0 & 0 & \frac{1}{\sqrt 2} & 0 & 0\\
0 & \frac{1}{\sqrt 2} & 0 & 0 & 0 & \frac{1}{\sqrt 2} & 0 & 0
\end{pmatrix}.
\end{equation}
\Cref{fig:ConcatenationCorrespondance}d) illustates this construction starting from the single-mode rectangular codes $S_{\vrectangle}$.
We compute that $|\mathbf x_0| = |\mathbf y_0| = |\mathbf z_0| = \sqrt[4]{2}$ in this code, such that
all single-qubit Clifford operations can be performed in an envelope-preserving manner, i.e.\ with passive Gaussian gates. As a result, we can prepare entangled Bell states of the tesseract code by first preparing two two-mode $\ket{\bar Y_{\mathrm{tess}}}$ states, then applying an envelope-preserving Clifford operation. 

We remark that in the four-mode code above, the logical Pauli operators have equal length, but do not have support on an equal number of modes, with $\mathbf y_0$ having support on three modes while $\mathbf x_0$ and $\mathbf z_0$ have support on two modes. Moreover, while the logical operators are of the same length, one stabilizer is of longer length, $\mathbf s_8 = \sqrt[4]{8}$ with $\mathbf s_{j \neq 8} = \sqrt[4]{2}$. This contrasts to the $D_4$ code which has both logical operators and stabilizers of equal length.

We remark that the $D_4$ and tesseract codes can be viewed as special cases of the dual-rail code construction of Ref.~\cite{Lau16m}, where arbitrary bosonic encodings are concatenated with a two-qubit repetition code, and the four-mode code is a special case of the quad-rail encoding introduced there. As a result, these GKP multimode codes also inherit the same properties as the dual- and quad-rails encoding, such as the ability to perform universal quantum computations using exponential-swap gates, provided the gauge is set correctly. However, in contrast to the general constructions of Ref.~\cite{Lau16m}, the special structure of the GKP codes allow one to perform Clifford gates using Gaussian operations, which are easier to implement than exponential-swap gates. Since the repetition code is the only non-trivial two-qubit code, it is not so surprising that concatenation-related constructions of two-mode bosonic codes are related. However, we emphasize that the choice of base lattice and the choice of repetition axis induce important differences as evidenced by, for example, the types of errors that can be corrected; see \cref{fig:ErrorPropagationTesseract,fig:ErrorPropagationD4}

\textit{$E_8$ Lattice}\textemdash
As a final example, we relate the $E_8$ lattice of \cref{sect:latticeExamples} to the construction above. In fact, the base $E_8$ lattice can be obtained by choosing the diamond code as the base GKP code and considering the stabilizer state with stabilizer group generated by $\langle \bar Y_1\bar Y_2,\bar Y_2\bar Y_3,\bar Y_3\bar Y_4,\bar Z_1\bar Z_2\bar Z_3\bar Z_4 \rangle$. As illustrated in \cref{fig:ConcatenationCorrespondance}c), this stabilizer state is one of the two code words comprising the $[[4,1,2]]$ error-detecting surface code. 

We finish this section by commenting on the differences between concatenated GKP codes and ``genuine'' multimode lattices. First, we remark that while all Pauli stabilizer codes correspond to a multimode lattice, the converse is not true. Indeed, starting from any concatenated decomposition \cref{eq:concatenatedDecomposition}, we can find new lattices by applying a symplectic transformation $M$,
\begin{equation}
S' = T L M.
\end{equation}
Except in special cases, the resulting lattice $S'$ cannot be decomposed as an integer matrix $T$ and a block-diagonal matrix $L$, and therefore does not correspond to a concatenated code. 

A second difference is that it is not necessarily obvious how qubit properties, such as the minimal support of logical operators, translate to the minimal length of Pauli operators in the multimode lattice. As a result, considering the properties of the multimode lattices can be helpful in the design of the code. For example, starting from single-mode rectangular codes instead of the more standard square code results in interesting codes such as the tesseract and four-mode code.

Finally, we argue that the main difference between concatenated codes and multimode lattice lies in the practical approach to the decoding procedure, not in the code words themselves. Since the code words themselves are identical, the maximum likelihood decoder is the same in both cases. However, maximum likelihood decoding is a hard problem that cannot be solved efficiently, such that efficient approximations must be made.
Traditionally, concatenated codes are decoded in a hierarchical fashion, correcting continuous translation errors at the single-mode level, then correcting discrete Pauli errors at the multimode level. In contrast, the decoding of ``genuine'' multimode lattices is done by identifying the most likely multimode translation corresponding to the syndrome and correcting it. In particular, concatenated decoders requires $3m-k$ measurements for a $k$-qubit code, while multimode decoders require only $2m$ measurements. While the former requires more measurements, the error correction also benefits from the added redundancy by passing information between levels of the hierarchy~\cite{Fukui17a,Vuillot19a,Noh20b,Hanggli20s,Noh22b}. We remark that the multimode decoder presented in \cref{sect:ErrorCorrection} is not optimal, as evidenced by the gap between the theoretical protection of the $D_4$ code and its practical performances shown in \cref{fig:ErrorProbHomodyne}. We leave the optimization of such decoders for future work. 

\section{Discussion and Conclusion}\label{sect:conclusion}
In this work, we have investigated bosonic quantum error correction codes based on multimode grid states. We first discussed how to design codes based on lattices in $\mathbb R^{2m}$, and what code dimensions are allowed by scaling and rotating the lattice. We showed for the four-dimensional hypercubic lattice that only code dimensions that can be written as a sum of three squares are possible. Looking ahead, we aim to generalize these relations and establish a link between arbitrary lattices and allowed code dimensions. For example, it would be interesting to investigate if it is possible to design a three-mode qubit code based on Bavard's $F_6$ lattice~\cite{Harrington04t}, or the $E_8$ lattice.

We investigated how lattice symmetries are reflected in the associated code words, and how to design non-Clifford gates based on these symmetries. Importantly, we showed that lattice symmetries which are isometries enable non-Clifford gates that are exact for finite-energy codes. 
For the single-mode square code, we showed that a Kerr Hamiltonian can generate a square-root-Hadamard logical gate.

We have introduced error-correction schemes for multimode grid codes based on qubit ancillas or single-mode GKP state ancillas. The latter approach follows closely the methods introduced originally in \cite{Gottesman01a}, and is qualitatively as hard as the error-correction circuits for single-mode codes. On the other hand, the error-correction circuits using two-level systems are based on building blocks that have already been demonstrated (separately) in different experimental platforms. As a result, we believe that the implementation of multimode grid codes is within reach in experimental platforms such as microwave cavities and the motion of trapped ions.

Although we focused in this work on qubit encodings, the error-correction circuits we introduced can also be applied to other code dimensions. For example, these circuits could be used to stabilize one dimensional grid states, which could allow more precise force measurements in multiple modes~\cite{Duivenvoorden17a,Zhuang20a}.

We have introduced in detail two qubit codes, namely the tesseract and the $D_4$ code which have support on two modes. We showed how to correct errors and how to perform a universal set of logical operations. The $D_4$ code has the interesting property that all Clifford operations can be performed using passive linear optics, i.e.\ beamsplitters and phase shifters. This property is enabled by the lattice geometry, in particular the fact that all Pauli operators are translations of equal length.

One the main result of our work is to show that both the $D_4$ and tesseract codes are more robust than single-mode codes to propagation of errors from the ancilla. This is linked to their ``isthmus'' property, where regions of correctable translation errors are directly linked in phase space. By only considering controlled translations that fall within correctable regions, we can recover from most ancilla decay errors. Moreover, the robustness of the grid codes to ancilla errors increases with code size. These properties are in stark contrast to single-mode codes, where the logical lifetime of the grid codes are directly limited by the physical lifetime of the ancilla.
Other techniques have been developed to increase the robustness of bosonic codes to errors in the ancilla such as path-independence~\cite{Ma20y,Ma22u} and Kerr cats~\cite{Puri19a}. These methods focus on improving the ancillas used for quantum control and are agnostic to the particular bosonic code used. As a result, these methods could be compatible with the multimode grid codes we introduced, which could lead to further improvements in the logical lifetimes of the GKP qubits.

Finally, we discussed several links between interesting lattices and single-mode GKP codes concatenated with qubit stabilizer codes. In particular, it is interesting to note that the lattices allowing the densest sphere packings in four ($D_4$) and eight ($E_8$) dimensions can be understood as the concatenation of a square qubit code with a two-qubit repetition code and a four-qubit stabilizer state, respectively. 

Although multimode codes promise the realization of long lifetimes for logical qubits, we do not expect that they will provide an efficient way to arbitrarily reduce the error rates of logical information. In order to build a fault-tolerant quantum computer, it would be interesting to study hybrid approaches where small multimode codes are concatenated with qubit codes. For example, it would be interesting to investigate the concatenation of the two-mode tesseract code with a surface code~\cite{Kitaev03a}. In particular, qubit codes are designed to exponentially suppress errors, such that an order of magnitude reduction in the error rate of the inner code (GKP codes) can lead to multiple orders of magnitude reduction of the logical error rate of the outer code (e.g. surface code).

\textit{Note added}\textemdash
While writing this manuscript, we became aware of similar work addressing multimode GKP codes~\cite{Conrad21h}. Amongst other results, this work discusses the gauge for multimode codes, the equivalence of symplectic lattices and the link between multimode lattices and concatenated codes. Our manuscript was officially submitted on October 6, 2021.

\begin{acknowledgments}
This work was supported by the Army Research Office under Grant No. W911NF-18-1-0212. We thank V. Sivak and A. Eickbusch for critical reading of the manuscript, as well as I. Ciobotaru-Hriscu for pointing out that the $\hat d_j$ operators are approximate nullifiers.
\end{acknowledgments}

\appendix

\section{Lattice Theory}\label{sectApp:latticeTheory}
In this section we introduce in more details some lattice theory, first discussing lattices in Euclidian geometry and then the corresponding notions in symplectic geometry. We refer the reader to Ref.~\cite{Conway13c} for a more in-depth discussion of lattices in general.

\subsection{Lattices in Euclidian Geometry}
A (Bravais) lattice $\Lambda = \{\vlambda_j\}$ is an infinite set of points in $\mathbb R ^{n}$ that is closed under standard vector addition, $\vlambda_i + \vlambda_j \in \Lambda$ for all $\vlambda_i,\vlambda_j \in \Lambda$. To specify these points, we choose $n$ linearly independent points $\{\mathbf s_1,..., \mathbf s_n\}$ that form a basis for $\Lambda$, i.e. all lattice points can be expressed as $\vlambda =  \sum_{j=1}^n a_j\mathbf s_j$ for a set integers $\{a_j\}$. 

The fundamental parallelotope of the lattice is defined as the region within $\sum_{j=1}^n \theta_j \mathbf s_j $ for $\theta_j \in [0,1)$. Translating this fundamental parallelotope by each lattice point completely covers $\mathbb R^n$. The choice of basis generating $\Lambda$ is not unique, hence the shape of the fundamental parallelotope is not unique. However, its volume is the same for all basis choices.

It is convenient to stack the basis (row) vectors $\{\mathbf s_j\}$ into a generator matrix $S$, where the $j$th row of $S$ is given by $\mathbf s_j$, see \cref{eq:generatorMatrixS}. Since the basis vectors are linearly independent, the generator matrix $S$ has full rank $n$. From $S$, the set of lattice points can be succintly defined as in \cref{eq:LambdaDef}. The volume of the fundamental parallellotope is also easily computed using the generator matrix and is given by its determinant, $V_p = \det(S)$.

The Gram matrix of the lattice,
\begin{equation}
G = S S^T,
\end{equation}
is defined such that its $(i,j)$th entry is given by the inner product $\mathbf s_i \cdot \mathbf s_j$. From this definition, it is easy to see that $G = G^T$ is a symmetric matrix. The \emph{determinant} of the lattice $\Lambda$ is defined to be the square of the volume of the fundamental parallelotope, and is given by $\det(\Lambda) = V_p^2 = \det(S)^2 = \det(G)$.

The dual lattice $\Lambda^*$ is defined as the set of points which have integer inner product with all points in $\Lambda$. Since it is enough to impose an integer inner product with the basis vectors of $\Lambda$, the set of points defining $\Lambda^*$ can be expressed as
\begin{equation}
\Lambda^* = \left\{\vlambda^* \mid\, S\vlambda^* \in \mathbb Z^n \right\}.
\end{equation}
A generator matrix for the dual lattice is obtained from $S^* = (S^T)^{-1}$. The volume of the fundamental parallelotope of $\Lambda^*$ is inversly proportional to the volume of the fundamental parallelotope of $\Lambda$, $V_p^* = 1/V_p$. 

An \emph{integral} lattice is a lattice where $G$ only contains integers. Since $G$ corresponds to the pairwise inner product between the basis vectors of $\Lambda$, the definition of the dual lattice then imposes that $\Lambda \subseteq \Lambda^*$. We note that, in general, the dual lattice points do not have integer norm within themselves unless $\Lambda = \Lambda^*$, in which case the volume of the fundamental parallelotope must be one, $V_p = 1$.

There exists multiple choices of basis for a same lattice. A change of basis can be expressed in a unimodular matrix $R$, an invertible integer matrix with unit determinant. Under this change of basis, the new generator matrix is expressed as $S' = RS$. Note that the Gram matrix does depend on the basis choice since in general $G' = RG R^T \neq G$.

Generally, we refer to lattices related by an isometry with the same name, although the specific set of points can be different. For example, a rotated square lattice is still considered a square lattice. In particular, the Gram matrices are identical for two lattices related by an orthogonal transformation, $G' = S O O^T S^T = S S^T = G$. 
To summarize, by a lattice we mean the equivalence class of all generator matrices related by a basis change and an orthogonal transformation, $S' = R S O$.

The minimum of a lattice is defined as the smallest distance between any two of its points. Since the set of distances from a given lattice point is translation invariant, we can choose to define the minimum of the lattice from the minimum distance to the point at origin $\mathbf 0 \in \Lambda$,
\begin{equation}
\min(\Lambda) = \min_{\vlambda \in \Lambda,\vlambda \neq 0} |\lambda|^2.
\end{equation}\indent
Placing the center of a sphere at each lattice point, the sphere packing ratio $\Delta$ is defined as the maximal fraction of space that can be covered by these sphere without the spheres overlapping. Considering that the radius of each sphere cannot be extended beyond $\sqrt{\min(\Lambda)}/2$, we have 
\begin{equation}
\Delta = \frac{V_n[\min(\Lambda)]^{n/2}}{2^n\sqrt{\det(\Lambda)}},
\end{equation} 
where we have defined the volume of a sphere of unit radius in $\mathbb R^n$ as
\begin{equation}
V_n = \frac{\pi^{n/2}}{\Gamma(n/2 - 1)}.
\end{equation} 
The densest lattice packings (largest $\Delta$) are known for all dimensions $n \leq 8$ and for $n = 24$.

\subsection{Lattices in Symplectic Geometry}
Many notions defined above for Euclidian geometry in $\mathbb R^n$ have an analogous version in symplectic geometry, which we describe below. We consider symplectic vector spaces in $\mathbb R^{2m}$.

A symplectic \emph{matrix} is a matrix which, when viewed as a linear transformation on $\mathbb R^{2m}$, preserves the symplectic form. Mathematically, a matrix $M$ is symplectic if $M^T \Omega M = \Omega$. A matrix is symplectic if and only if its transpose is symplectic, $M^T \Omega M = \Omega \Leftrightarrow M \Omega M^T = \Omega$.

In analogy with the Gram matrix of the lattice $\Lambda$ where the $(i,j)$th entry is the inner product between the basis vector $\mathbf s_i$ and $\mathbf s_j$, we define the symplectic Gram matrix of a lattice $\Lambda$ as $A = S\Omega S^T$, where the $(i,j)$th entry is the symplectic form between the basis vectors $\mathbf s_i$ and $\mathbf s_j$, $A_{ij} = \omega(\mathbf s_i,\mathbf s_j) = \mathbf s_i \cdot \Omega \cdot \mathbf s_j$. Since the symplectic form is anti-symmetric, the symplectic Gram matrix is anti-symmetric, $A = -A^T$.

We also define the symplectic dual of a lattice as the set of points having integer symplectic form with the lattice points,
\begin{equation}
\Lambda_s^* = \left\{\vlambda^* \mid\, S\Omega\vlambda^* \in \mathbb Z^n \right\}.
\end{equation}
In this appendix, we denote the symplectic dual lattice and its generator matrix with an ``s'' index to distinguish it from the standard dual. A generator matrix for the dual lattice can be obtained from $S_s^* = A^{-1}S$. The standard and symplectic dual are related through $S^*_s = -S^* \Omega$. The volume of the fundamental parallelotope of the symplectic dual lattice is identical to the volume of the dual lattice, $|\det(S^*)| = |\det(S^*_s)|$. 

A lattice is called symplectically integral if its symplectic Gram matrix $A$ only contains integers. From the definition of the symplectic dual lattice, this implies that $\Lambda \subseteq \Lambda_s^*$.

A lattice is called symplectic if $\Omega$ (viewed as a linear transformation) acts as an isometry from $\Lambda$ to its dual $\Lambda^*$. Equivalently, the generator matrix of a symplectic lattice can be chosen to be a symplectic matrix when viewed as a linear transformation in $\mathbb R^{2m}$, $S\Omega S^T = \Omega$. A symplectic lattice has unit determinant.

\section{Gauge updates}\label{sectApp:gaugeUpdates}
In this section, we show how to update the gauges after different operations.

\emph{Translations}\textemdash We first derive \cref{eq:gaugeUpdateDisplacement}a), which sets the new gauge after a translation by half a lattice vector $\pmb \tau/2$, with $\pmb \tau \in \Lambda$. Computing how the eigenvalue of a translation by a lattice vector $\vlambda \in \Lambda$ changes, we get
\begin{equation}
\begin{aligned}
\hT(\vlambda) \hT(\pmb \tau/2)\ket{\psi} &= \expo{i\pi \vlambda^T \Omega \pmb \tau} \nu_{\pmb \mu}(\vlambda) \hT(\pmb \tau/2)\ket{\psi}.
\end{aligned}
\end{equation}
We therefore set the new gauge $\pmb \mu'$ such that
\begin{equation}
\nu_{\pmb \mu'}(\vlambda) = \expo{i\pi \vlambda^T \Omega \pmb \tau} \nu_{\pmb \mu}(\vlambda).
\end{equation}
Replacing the definition of the $\nu$ function \cref{eq:nuFunction}, we get that for all $\vlambda \in \Lambda$,
\begin{equation}
\begin{aligned}
\vlambda^TS^{-1}\vmu' &= \vlambda^T \Omega \pmb \tau + \vlambda^TS^{-1} \vmu,\\
\Rightarrow \vmu' &= S\Omega \pmb \tau + \vmu,
\end{aligned}
\end{equation}
with the equalities above understood modulo 2. \Cref{eq:gaugeUpdateDisplacement}b) is proven in a similar way.

\emph{Gaussian operations}\textemdash Next we show how to update the gauges after a quadratic operation $\hat Q(M)$. Using the commutation relation \cref{eq:commRelQofM}, we obtain that in the code space
\begin{equation}\label{eq:startingEqGaugeUpdateStab}
\hT(\vlambda) \hat Q(M)\ket{\psi} = \nu_{\vmu}(M^{-1}\vlambda) \hat Q(M) \ket{\psi},
\end{equation}
which then sets the equation that determines the new gauge, $\nu_{\vmu'}(\vlambda) = \nu_{\vmu}(M^{-1}\vlambda)$.
To make the following equations more concise, we define $\mathbf a = (S^T)^{-1}\vlambda$ with $\mathbf a \in \mathbb Z^{2m}$, as well as $N^T = SM^TS^{-1}$. Replacing the gauge function defined by \cref{eq:nuFunction}, we obtain
\begin{equation}
\begin{aligned}
\mathbf a^TN^T\left[A_\text{\lltriangle}N\mathbf a + \vmu'\right] &= \mathbf a^T \left[A_\text{\lltriangle}\mathbf a + \vmu\right],\\
\mathbf a^T\left[A_\text{\lltriangle} - N^TA_\text{\lltriangle}N\right]\mathbf a &= \mathbf a^T \left[N^T\vmu' - \vmu\right],\\
\end{aligned}
\end{equation}
which yields an equation that sets one element of $\vmu'$. To find the whole gauge, we choose a set of $2m$ vectors $\{\vlambda_j\} = \{\mathbf s_j\}$, which corresponds to choosing an ensemble of $\mathbf a$ as the canonical basis for $\mathbb Z^{2m}$. Using the fact that the diagonal elements of $A_\text{\lltriangle}$ are 0, $\mathbf a^T A_\text{\lltriangle} \mathbf a = 0$, we arrive at
\begin{equation}\label{eq:gaugeUpdateSymplectic}
\begin{aligned}
\vmu' ={}& (N^T)^{-1}(\vmu + \vec{\mathrm{diag}}[N^TA_\text{\lltriangle}N]) \,\,\mathrm{mod}\,\, 2,\\
\end{aligned}
\end{equation}

Next, we derive the equation to update the Pauli frame gauge $\vupsilon$ after a Gaussian operation $M$. We start with an equation similar to \cref{eq:startingEqGaugeUpdateStab} which yields
\begin{equation}
\begin{aligned}
\hat Q(M)\hT(\mathbf p_0)\ket{\psi_{+P}} &= \hT(M\mathbf p_0) \hat Q(M) \ket{\psi_{+P}}.
\end{aligned}
\end{equation}
After applying the gate, the state is again in the code space, and we define $\ket{\psi_{\iota \tilde P}} = \hat Q(M) \ket{\psi_{+P}}$, with $\iota \in \{\pm\}$. This yields an equation to solve for each Pauli $P$,
\begin{equation}
\begin{aligned}
(-1)^{\upsilon_p} ={}& (-1)^{\iota}\nu^{\tilde P}_{\vmu',\pmb \upsilon'}(M\mathbf p_0) ,\\
\upsilon_p ={}& \iota + \upsilon_{\tilde P}'+ \mathbf{\tilde p}_0^T\Omega M \mathbf p_0 + (M\mathbf p_0 - \mathbf{\tilde p_0})^TS^{-1}\\
&\times \left[A_\text{\lltriangle}(S^{-1})^T(M\mathbf p_0 - \mathbf{\tilde p_0}) + \vmu'\right].
\end{aligned}
\end{equation}
Conjugation of Pauli operators by Clifford operations can be written as a (signed) permutation matrix of the Paulis $\mathbf P = (X;Y;Z)$, $\mathbf{\tilde P}= M_L \mathbf P$ with $\det(M_L) = 1$. For example, the matrix associated with the logical $\bar S$ gate is
\begin{equation}
M_L = \begin{pmatrix}
0 & -1 & 0\\
1 & 0 & 0\\
0 & 0 & 1
\end{pmatrix}.
\end{equation}
Further defining a vector of ones, $\mathbf 1$, and $V \equiv (M_L L_0 M^T - L_0)S^{-1}$, we get the full gauge update equation
\begin{equation}
\begin{aligned}
\vupsilon' ={}& M_L\vupsilon + \frac{(M_L + I)}{2}\mathbf 1 + V \vmu'\\
& +\vec{\mathrm{diag}}\left[L_0 \Omega M L_0^T M_L + V A_\text{\lltriangle}V^T\right]\,\,\mathrm{mod}\,\, 2,
\end{aligned}
\end{equation}
where we have also defined $V \equiv (M_L L_0 M^T - L_0)S^{-1}$ and $\vec{\mathrm{diag}}$ is the operation of building a vector from the diagonal of a matrix.

\emph{Basis change}\textemdash We now show how to update the gauge after a change in the choice of basis, $S' = R S$. Choosing a vector $\vlambda \in \Lambda$, we apply $\hat T(\vlambda)$ to any code word to obtain
\begin{equation}
\begin{aligned}
\nu_{\vmu}(\vlambda) &= \nu_{\vmu'}(\vlambda),\\
\Rightarrow \mathbf b^T A_\text{\lltriangle} \mathbf b + \mathbf b^T \cdot \vmu &= \mathbf a^T A'_\text{\lltriangle} \mathbf a + \mathbf a\cdot \vmu',
\end{aligned}
\end{equation}
where we have defined $\vlambda = S'^T \mathbf a = S^T \mathbf b$ for some $\mathbf a,\mathbf b \in \mathbb Z^{2m}$. This yields an equation that sets one element of $\vmu'$. Replacing $\mathbf b = (S^T)^{-1} S'^T \mathbf a = R^T\mathbf a$ and choosing an ensemble of $\{\mathbf a_j\}$ such that they form the canonical basis for $\mathbb Z^{2m}$, we get an ensemble of $2m$ equations 
\begin{equation}\label{eq:gaugeUpdateSplitting}
\vmu' = R\vmu + \vec{\mathrm{diag}}[R A_\text{\lltriangle}R^T].
\end{equation}
We note that this equation does not require that the two lattices are identical, only that $\Lambda' \subseteq \Lambda$. As a result, it can also be used to perform a lattice splitting operation where, for qubit codes, $\det S' = 2\det S$.

When the two lattices are identical, $\Lambda' = \Lambda$, we can update the Pauli frame gauge using
\begin{equation}
\vupsilon' = \vupsilon + \Delta_L\vmu + \vec{\mathrm{diag}}[L_0 \Omega L_0'^T + \Delta_L A_\text{\lltriangle}\Delta_L^T],
\end{equation}
where we have defined $\Delta_L = (L_0' - L_0)S^{-1}$.

\emph{Gauge condition}\textemdash We now derive the condition to impose on the gauge $\vmu$ so that the eigenvalues of the logical Pauli operators are real. Imposing that $\nu_{\vmu}(2\mathbf p_0) = 1$ is equivalent to imposing that
\begin{equation}
(2\mathbf p_0^T) S^{-1}\left[A_\text{\lltriangle} (S^{-1})^T (2\mathbf p_0) + \vmu\right] \;\mathrm{mod}\; 2 = \mathbf 0.
\end{equation}
Writing $\mathbf p_0 = (S^*)^T \mathbf b$ for some integer vector $\mathbf b \in \mathbb Z^{2m}$ and replacing $S^* = A^{-1}S$ using \cref{eq:dualLatticeGenerator}, we obtain
\begin{equation}
4 \mathbf b^T A^{-1} A_\text{\lltriangle} \Omega \mathbf b + 2 \mathbf b^T A^{-1} \vmu \;\mathrm{mod}\; 2 = \mathbf 0.
\end{equation}
The matrix $2A^{-1}$ is integral and the first term always sums to $\mathbf 0$ modulo 2. Imposing the condition above for all $\mathbf b$ leads to \cref{eq:gaugeValidityCondition}.

\section{Gauge Updates in Code Switching}\label{sect:appGaugeCodeSwitching}
In this section we focus on the gauge update equation in code switchings.

\emph{Lattice Splitting}\textemdash The stabilizer gauge update can be realized using \cref{eq:gaugeUpdateSplitting}, so we focus on the Pauli frame gauge, $\vupsilon$. We can always express a representative of two-qubit Pauli operators in the $AB$ code as $\mathbf p^{(AB)} = L_{0,AB}^T \mathbf b$, where we have defined $L_{0,AB} = L_{0,A}\oplus L_{0,B}$ and $\mathbf b \in \mathbb Z^6$. The hierarchy \cref{eq:latticeSurgeryHierarchy} implies that all Pauli operators of the C code correspond to a product of Paulis of the AB code, such that we can also find Pauli representatives of the C code written as $\mathbf p^{(C)} = L_{0,AB}^T \mathbf b$. 
Taking two representatives $\mathbf p_1,\mathbf p_2$ for each Pauli operator of the C code such that they are all linearly independent, we write
\begin{equation}
\begin{pmatrix} \mathbf x_1 \\ \mathbf y_1 \\ \mathbf z_1 \\ \mathbf x_2 \\ \mathbf y_2 \\ \mathbf z_2
\end{pmatrix} = U L_{0,AB}.
\end{equation}
Computing the sign for each of these representives yields a set of equations,
\begin{equation}
\begin{aligned}
\mathbf b \cdot \vupsilon_{AB} &= \upsilon_{p,C} +\mathbf p_{0,C}^T\Omega L_{0,AB}^T \mathbf b \\
&+ (\mathbf b^TL_{0,AB} - \mathbf p_{0,C}^T)B(L_{0,AB}^T\mathbf b - \mathbf p_{0,C}),
\end{aligned}
\end{equation}
where we have defined $B = S_C^{-1}A_\text{\lltriangle}^{C}(S_C^T)^{-1}$. 
Solving the above, we set the Pauli frame gauges,
\begin{equation}\label{eq:gaugeUpdateSplittingPauli}
\vupsilon_{AB} = U^{-1}\vupsilon_{C}^{\oplus 2} + U^{-1}\vec{\mathrm{diag}}\left[L_{0,C}^{(2)}\Omega L_{0,AB}^T U^T + E B E^T\right],
\end{equation}
where we have defined 
\begin{equation}
L_{0,C}^{(2)} = \begin{pmatrix} L_{0,C} \\ L_{0,C}
\end{pmatrix},
\end{equation}
and $E = (U L_{0,AB} - L_{0,C}^{(2)})$.

\emph{Lattice Merging}\textemdash
To perform the merging operation, we first choose a merging vector $\vlambda_m \in \Lambda_C$ in the $C$ lattice stabilizers that is not in the $AB$ lattice, $\vlambda_m \notin \Lambda_A \oplus \Lambda_B$.
We then measure the eigenvalue of $\hat T(\vlambda_m)$, setting it to a definite value $\nu_m =\pm 1$.
This measurement can be done with circuits similar to those required for error correction, see \cref{sect:ErrorCorrection}.

After the measurement, we get a new lattice with generator matrix $S'$, which we obtain by replacing a row of $S_{AB}$ by $\vlambda_m$. The replaced vector should respect $\mathbf s_{AB}\cdot \vlambda_m \neq 0$, such that $S'$ is full rank. Without loss of generality, we choose to replace the last generator vector of the $B$ lattice, obtaining 
\begin{equation}
S' = \begin{pmatrix}
\mathbf s_{A,1}\oplus \mathbf 0 \\
 ... \\
\mathbf s_{A,2m_A} \oplus \mathbf 0 \\
\mathbf 0 \oplus \mathbf s_{B,1}  \\
 ... \\
\mathbf 0 \oplus \mathbf s_{B,2m_B - 1} \\
\vlambda_{m}
\end{pmatrix},\,\,
\pmb \mu = \begin{pmatrix}
\mu_{A,1} \\
... \\
\mu_{A,2m_A} \\
\mu_{B,1} \\
... \\
\mu_{B,2m_B - 1} \\
\nu_{m} 
\end{pmatrix}.
\end{equation}
By construction, $S'$ and $S_C$ generate the same lattice and are related by a basis change $R = S_C S'^{-1}$. Using \cref{eq:gaugeUpdateSplitting}, we can convert the gauge to match the desired generator matrix $S_C$.

We remark that there are multiple possible choices for the merging vector $\vlambda_m$, which correspond to different logical operators of the $AB$ code. As a result, the choice of $\vlambda_m$ also sets the projection operator from the two-qubit logical subspace of the $AB$ code to the single qubit logical subspace of the $C$ code. 

Writing the base Pauli representatives of the C code as $\mathbf p_{0,C} = \mathbf p_{A} \oplus \mathbf p_{B}$, the elements of the Pauli frame gauge are updated using
\begin{equation}
\upsilon_{p,C} = \varphi\left[\nu^{P_A}_{\vmu_A,\vupsilon_A}(\mathbf p_{A})\nu^{P_B}_{\vmu_B,\vupsilon_B}(\mathbf p_{B})\right],
\end{equation}
where we have defined $\varphi(1) = 0,\varphi(-1) = 1$.

\section{Multimode Controlled Translations}\label{sect:app:CT}
In this section, we describe how to realize a controlled displacement in multiple modes indexed by $j$.
We consider a single qubit coupled to multiple modes via a pairwise dispersive interaction, with a drive $\mathcal E_j(t)$ on each mode,
\begin{equation}
\hH(t) = \sum_j\left[\frac{\chi_j}{2}\had_j \ha_j \hat{ \sigma}_z + \mathcal E_j(t)\had_j + \mathcal E_j^*(t)\ha_j\right].
\end{equation}
The Hamiltonian generates displacements, but also an ancilla qubit phase and a qubit state-dependent rotation of the oscillator modes. In order to echo out the ancilla-dependent rotation due to the dispersive shift, we consider an evolution in $K$ steps with a qubit flip between each step in analogy with the scheme used in Ref.~\cite{Campagne-Ibarcq20a},
\begin{equation}
\hU = \Pi_{k=1}^{K} \hat \sigma_x \mathcal T \expo{-i\int_{k,i}^{t_{k,f}}d\tau \hH(\tau)},
\end{equation}
where we have defined the kth step ranging from $t_{k,i}$ to $t_{k,f}$ and the product is time-ordered.
The final qubit flip is omitted if $K$ is odd. We commute through the qubit flips $\hat \sigma_x$ such that, during the kth step, the sign of $\hat \sigma_z$ is multiplied by $z_k \in \{\pm1\}$ which we include in a continuous function $z(t)$. With this simplification, we can write the whole evolution in a single step, $\hU = \mathcal T\expo{-i\int_{0}^{T}d\tau \hH_z(\tau)}$, with
\begin{equation}\label{eq:zHamiltonian}
\begin{aligned}
\hH_z(t) &= \sum_j\left[\frac{\chi_j}{2}\had_j \ha_j \hat{ \sigma}_z z(t) + \mathcal E_j(t)\had_j + \mathcal E_j^*(t)\ha_j\right],\\
&= \frac{\vec \chi\cdot \vec{\hat n}}{2}\hat \sigma_z z(t) + \vec{\hat a}^\dag \cdot \vec{\mathcal E} + \vec{\mathcal E}^*\cdot \vec{\hat a}.
\end{aligned}
\end{equation}
Considering the form of the Hamiltonian, we take an ansatz for the resulting unitary
\begin{equation}
\begin{aligned}
\hU &= \expo{i\theta \frac{\hat \sigma_z}{2}}\expo{\vec{\hat a}^\dag\cdot (\vec \gamma + \vec \delta \hat \sigma_z) - (\vec \gamma^\dag + \vec \delta^\dag \hat \sigma_z)\cdot \vec{\hat a} }\expo{-i \vec \phi \cdot \vec{\hat n} \hat \sigma_z},\\
\end{aligned}
\end{equation}
where $\theta$ sets the ancilla qubit phase, $\vec \phi \in \mathbb R^m$ represents the qubit-dependent rotation of each mode and $\vec \gamma,\vec \delta \in \mathbb C^m$ represent the displacement and controlled-displacement of each mode, respectively. We extract a differential equation for each parameters using Schr\"oedinger's equation, $\dot{\hU} = -i \hH_z(t) \hU$. Neglecting terms leading to an irrelevant global phase, we obtain
\begin{subequations}
\begin{align}
\dot \theta &= - 2\mathrm{Re}[\vec{\mathcal E}^* \cdot \vec \delta] ,\\
\dot{\gamma}_j &=-i \frac{\chi_j}{2}z(t)\delta_j -i \mathcal E_j,\\
\dot{\delta}_j &=-i \frac{\chi_j}{2}z(t)\gamma_j ,\\
\dot{\phi}_j &= \frac{\chi_j}{2} z(t)
\end{align}
\end{subequations}
In order to echo out the qubit state-dependent rotation of the oscillators, we choose a $z(t)$ such that $\vec \phi(T) = 0$. The differential equations above can easily be solved, yielding the parameters 
\begin{subequations}
\begin{align}
\theta(t) &= -2 \int_0^t d\tau\, \mathrm{Re}[\vec\epsilon(\tau)^* \cdot \vec \delta(\tau)],\\
\gamma_j(t) &= -i\int_0^t d\tau \cos[\phi_j(\tau) - \phi_j(t)] \mathcal E_j(\tau),\\
\delta_j(t) &= \int_0^t d\tau \sin[\phi_j(\tau) - \phi_j(t)] \mathcal E_j(\tau),\\
\phi_j(t) &= \frac{\chi_j}{2}\int_0^t d\tau z(\tau),
\end{align}
\end{subequations}

Using the Baker-Campbell-Hausdorff (BCH) formula to separate the overall and controlled translations, we rewrite the resulting unitary as 
\begin{equation}\label{eq:controlledDisplacement3}
\begin{aligned}
\hU = \expo{i\theta'\frac{\hat \sigma_z}{2}}\times \hat T(\vec g)\times C\hat T(\vec d),
\end{aligned}
\end{equation} 
where we have defined $\vec g, \vec d \in \mathbb R^{2m}$ as $\vec g = \mathrm{Vec}[\vec \gamma(T)]$ and $\vec d = \mathrm{Vec}[2\vec \delta(T)]$ with $\mathrm{Vec}[\vec v] = \sqrt 2/l \times \mathrm{Re}[\vec v]\oplus \mathrm{Im}[\vec v]$. The (exact) BCH expansion term yields a correction to the phase
\begin{equation}
\theta' = \theta + 2\mathrm{Im}[\vec \gamma^\dag \cdot \vec \delta].
\end{equation} 
The desired controlled translation is therefore obtained by applying the drives $\{\mathcal E_j\}$, with a displacement and a qubit phase correction at the end. Alternatively, one can choose the drives such that $\vec \gamma(T) = \vec 0$ and $ \theta(T) = 0$. For example, to obtain a controlled translation $C\hat T(\vec b)$, we can split the evolution in two parts and choose the drives
\begin{equation}
\begin{aligned}
\mathcal E_j(t) ={}& \alpha_{d,j} \left[\delta(t) -2 \delta(t - T/2)\cos(\chi_j T/4)\right.\\
&\left. + \delta(t - T) \cos(\chi_j T/2)\right],\\
\alpha_{d,j} ={}& [C \vec b]_j \frac{l}{2 \sin(\chi_j T/2)},
\end{aligned}
\end{equation}
where $\delta(t)$ is the Dirac delta function. This drive can approximately be realized in a system where the displacements by $\alpha_{d,j}$ can be effected in a time scale much faster than $1/\chi_j$.

We now investigate the effect of ancilla qubit decay during a controlled displacement. More precisely, we study the effect of a decay event $\hat \sigma_-$ at time $t_\mathrm{err}\in [0,T]$. Assuming that the populations of the $g$ and $e$ states of the ancilla qubit are roughly equal during the evolution, the probability distribution for the time of errors is uniform.
Commuting the decay event through the unitary, we get
\begin{equation}
\hU_\mathrm{err} = \hat \sigma_{i\mathrm{err}} \mathcal T \expo{-i \int_0^T d\tau \hH_\mathrm{err}(\tau)},
\end{equation} 
where we have defined $i\mathrm{err} = - z(t_\mathrm{err})\in \{\pm 1\}$.
The Hamiltonian $\hH_\mathrm{err}(\tau)$ is identical to the Hamiltonian \cref{eq:zHamiltonian}, with the replacement $z(t) \rightarrow z_\mathrm{err}(t)$, which we define as $z_\mathrm{err} = + z(t)$ if $t < t_\mathrm{err}$ and $z_\mathrm{err} = - z(t)$ if $t \geq t_\mathrm{err}$. Following a similar approach as above and including the displacement correction by $\vec g$ at the end, we find that in the case of a decay error the effective operation is
\begin{equation}
C\hat T_\mathrm{err} = \hat \sigma_{i\mathrm{err}} \hat T[\vec e - \vec g] \Pi_j \hat R_j[\varphi_{j,\mathrm{err}}(T)],
\end{equation}
with parameters
\begin{subequations}
\begin{align}
\varphi_{j,\mathrm{err}}(t) &= -\frac{\chi_j z(t_\mathrm{err})}{2} \int_0^t d\tau  z_\mathrm{err}(\tau) ,\\
\gamma_{j,\mathrm{err}} &= -i \expo{-i \varphi_{j,\mathrm{err}}(T)}\int_0^T d\tau\, \mathcal E_j(\tau)\expo{i \varphi_{j,\mathrm{err}}(\tau)},
\end{align}
\end{subequations}
and $\vec e = \mathrm{Vec}[\vec \gamma_\mathrm{err}]$.
The qubit phase is irrelevant since the decay event destroys the ancilla coherence at time $t_\mathrm{err}$.
An error during a controlled translation therefore propagates to the modes as a rotation and a displacement. The rotation error is upper bounded by $|\varphi_{j,\mathrm{err}}| \leq \chi_j T/2 \times 1/\lceil K/2 \rceil$, such that it is advantageous to consider systems with small $\chi$, or alternatively a large number of echo pulses. While the displacement error can be large, it necessarily happens along a path parametrized by the time of the error, $t_\mathrm{err}$. Taking a pulse such that $\vec g = 0$ and in the limit that $\chi_j \rightarrow 0$, this displacement is exactly on the line spanned by the target generator $\mathbf s_j$, 
\begin{equation}
C\hat T_\mathrm{err} \approx \hat \sigma_{i\mathrm{err}} \hat T[\eta \mathbf s_j],
\end{equation}
with $\eta \in [0,1/2]$. We remark that $\hat T[\mathbf 0]$ is the worst case error in this situation since the gauge is updated based on the assumption that a translation $\hat T[\pm \mathbf s_j/2]$ occured.	

\section{Searching for Integral Lattices}\label{sect:IntegralLatticeSearch}
Given a basis $S$ for a lattice, we are looking for an orthogonal transformation $O$ such that $A' = S'\Omega S'^T$ is integral, with $S' = SO$. In general, a $2m$-dimensional rotational can be decomposed into $2m(2m-1)/2$ parameters, which we express as a series of Givens rotations,
\begin{equation}
\label{eq:OasGivensRotations}
O = \Pi_k G(i_k,j_k,\theta_k),
\end{equation}
where $i_k,j_k$ and $\theta_k$ are the axes and angle of the $k$th rotation, respectively.

To simplify the problem, we leverage the fact that $A'$ is invariant when $O$ is also symplectic. In particular, we can always transform the basis through an orthogonal and symplectic transformation such that one of the basis vectors points only in the $q_1$ direction. Without loss of generality, we choose to fix the global lattice orientation such that $\mathbf s_1 \propto (1,0,0,...)$, effectively removing one dimension from the problem and reducing the number of parameters to $(2m - 1)(m - 1)$. 

Finally, we choose a decomposition of $O$ such that it contains $m-1$ rotations contained in a single-mode (i.e.\ rotations that are also symplectic), and choose them to be the last $m-1$ rotations in \cref{eq:OasGivensRotations}. We can then neglect these rotations, which brings the number of effective parameters to $2(m - 1)^2$. In four dimensions, we obtain a two-parameter decomposition $O = G(2,3,\theta_1)G(2,4,\theta_2)$.

\subsection{Tesseract codes}
We search for codes of size $d$ based on a four-dimensional hypercube, i.e.\ lattice bases of the form $S = d^{1/4} G(p_1,x_2,\theta_1)G(p_1,p_2,\theta_2)$. Computing the symplectic Gram matrix for $S$, we obtain
\begin{equation}
A = \sqrt d \begin{pmatrix} 0 & \cos\theta_1\cos\theta_2 & \sin\theta_1\cos\theta_2 & \sin\theta_2\\
 & 0 & \sin\theta_2 & -\sin\theta_1\cos\theta_2\\
 & & 0 & \cos\theta_1\cos\theta_2\\
 & & & 0
\end{pmatrix},
\end{equation}
where we omitted the lower part of $A$ for clarity. For $A$ to be integral, we should therefore have
\begin{subequations}\label{eq:tesseractDimensionCondition}
\begin{align}
a &= \sqrt d \cos\theta_1\cos\theta_2,\\
b &= \sqrt d \sin\theta_1\cos\theta_2,\\
c &= \sqrt d \sin\theta_2,
\end{align}
\end{subequations}
for some $a,b,c\in \mathbb Z$. Squaring the equations above and replacing \cref{eq:tesseractDimensionCondition}(b,c) in (a), we obtain that $a,b,c$ must respect the Diophantine equation
\begin{equation}
d = a^2 + b^2 + c^2.
\end{equation}
According to Legendre's three-square theorem, there exists solutions for all code sizes except those that can be written as $d = 4^e(8f+7)$ for some non-negative integers $e,f$.

\subsection{D4 Codes}
Carrying out a similar procedure as above and starting from \cref{eq:SD4qunaughtState}, $S = d^{1/4} S_{D_4,\varnothing} G(p_1,x_2,\theta_1)G(p_1,p_2,\theta_2)$, we obtain that there exists solutions for $\theta_1,\theta_2$ when there exists a solution to the Diophantine equation 
\begin{equation}
4d = 3a^2 + 4ab + 4b^2 + c^2.
\end{equation}
There exists solution for this equation for all code sizes $d \leq 50$ except $d=14,30,46$.

\section{Probability of qubit decay}\label{app:qubitDecay}
In this section, we show that the probability of a qubit decay occuring during one sBs dissipation circuit is approximately given by $1 - \expo{-\gamma_\mathrm{anc} t_\mathrm{tot}/2}$, with $\gamma_\mathrm{anc}$ the ancilla decay rate and $t_\mathrm{tot}$ the time required to apply all three controlled translations the circuit of \cref{fig:StabilizationCircuits}a). Expressing the amplitude damping Kraus operators from \cref{eq:KrausOpsPhotonLoss} in the two-level case, we obtain
\begin{subequations}\label{eq:KrausOpsQubitDecay}
\begin{align}
\hat K_0(t) &= \ketbra{g} + \ketbra{e} \expo{-\gamma_\mathrm{anc} t/2},\\
\hat K_1(t) &= \left(\frac{1 - \expo{-\gamma_\mathrm{anc} t}}{\expo{-\gamma_\mathrm{anc} t}}\right)^{1/2}\hat \sigma_-(\ketbra{g} + \ketbra{e} \expo{-\gamma_\mathrm{anc} t/2}).
\end{align}
\end{subequations}
We consider an echo pulse during the controlled displacements as in \cref{sect:app:CT} and decompose the error channel in two steps of time $T/2$ each. The Kraus operator for the no-decay evolution after a time $T$ is therefore given by $\hat K_0(T) = \hat K_0(T/2) \hat \sigma_x \hat K_0(T/2) \hat \sigma_x = \expo{-\gamma_\mathrm{anc} T/4} \hat I$. The probability that this no-decay channel occurs is therefore given by $\mathbb P[\hat K_0(T)] = \bra{\psi}\hat K_0^\dag(T)\hat K_0(T)\ket{\psi} = \expo{-\gamma_\mathrm{anc} T/2}$. Neglecting the probabilility that two decay event occur, we take the probability that a single decay event occurs as $1 - \expo{-\gamma_\mathrm{anc} T/2}$. Extending this analysis to all three controlled translations in the sBs circuit, we obtain probabilities of $\{\expo{-\gamma_\mathrm{anc} t_\mathrm{tot}/2},1 - \expo{-\gamma_\mathrm{anc} t_\mathrm{tot}/2}\}$ for the no decay and single decay channels, respectively.

We remark that one additional benefit of the echo pulse is to symmetrize the effect of the ``no-jump'' part of the decay channel, which would otherwise increase the amplitude of the qubit $\ket{g}$ state at the end of the controlled translations.

\section{Code Switching in the D4 Code}\label{app:codeSwitchingD4}
\begin{figure}[!t]
    \centering
      \includegraphics[scale=1]{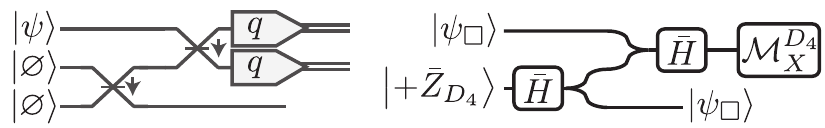}
    \caption{Teleportation-based error correction for a single-mode square code. The physical circuit on the left can be re-interpreted using as the logical circuit on the right.}
    \label{fig:TeleportationQEC}
\end{figure}

Consider two modes each encoding a logical qubit in the diamond code, with resulting generator matrix $S = S_{\Diamond}^{\oplus 2}$. A comparison between this lattice and the $D_4$ code reveals that all lattice points $\Lambda_\Diamond^{\oplus 2}$ are also in the set $\Lambda_{D_4}$. Differently put, we have
$\Lambda_{D_4}^* \subset (\Lambda_{\Diamond}^*)^{\oplus 2}$, such that every lattice point in the dual $D_4$ lattice corresponds to a point in the (denser) lattice obtained by combining two diamond GKP codes, and all code words of the $D_4$ code correspond to a code word of the two-qubit diamond code. Note that each Pauli operator of the $D_4$ code has multiple representatives, which are associated with different Pauli operators of the combined diamond codes. For example, representatives of $\bar Z_{D_4}$ with support on the first mode such as $\mathbf z = (1,0,0,0)$ or $(0,1,0,0)$ correspond to $\pm \bar Y_{\Diamond}\bar  I$, while representatives with support on the second mode correspond to $\pm \bar I \bar Y_{\Diamond}$.
In general, the signs in \cref{eq:D4toDiamond} depend on the gauges $\vmu$ and $\vupsilon$. \\ \indent
Splitting the $D_4$ code into two diamond codes does not require any operation and can be done in software.
However, in order to respect the gauge condition \cref{eq:gaugeValidityCondition} imposing that eigenvalues of the logical Pauli operators are real, the gauge should be set to $\vmu = (0,1,1,1)$ before ``performing'' the split. This gauge change can be realized with a translation $\hat T(\pmb \tau)$, with $\pmb \tau$ computed using \cref{eq:gaugeSettingDisplacement}. \\ \indent
In order to merge the two diamond codes toward the code space of the $D_4$ lattice, the operator $\bar Y_\Diamond\bar Y_\Diamond$ should be measured, for example using the methods in \cref{sect:ErrorCorrection}. There are four representives for $\bar Y_\Diamond$ in the diamond code, and there are 8 equivalent choices of representatives for $\bar Y_\Diamond\bar Y_\Diamond$ that lead to an equivalent merging towards the $D_4$ code (we do not count representatives that differ by a sign). Disregarding the continuous nature of the GKP code, the merging procedure then corresponds to non-deterministically projecting the space of two qubits onto the logical subspace of a repetition code. \\ \indent
A first example where code switching is useful is the teleportation-based error-correction circuit originally introduced in Ref.~\cite{Walshe20q} and shown in \cref{fig:TeleportationQEC}. On the left of the figure, we show the physical circuit, and on the right we show the equivalent logical circuit involving splitting and merging square codes into the $D_4$ code.
First, we initialize the two lower states in qunaught states, which equivalently correspond to a $\ket{+\bar Z_{D_4}}$ state or the tensor product of two (modified gauge) $\ket{+\bar Y_\Diamond}$ states. 
Then, the beam-splitter operation acts as a $D_4$ Hadamard gate which maps the state to a logical $\ket{+\bar X_{D_4}}$ state. Performing a code splitting operation then maps the single-qubit state of the $D_4$ code to the two-qubit Bell state $\propto \ket{+\bar Z_{\square},+\bar Z_{\square}} + \ket{-\bar Z_{\square},-\bar Z_{\square}}$, see \cref{eq:D4toDiamond}a), with the gauge naturally mapped to the required value for splitting by the beamsplitter.
Omitting the $\hat R(\pi/4)$ rotations of the Hadamard in \cref{eq:HadamardGateD4} allows to map the states from the diamond to the square qubit code. The upper part containing the second beam splitter is understood in a similar manner, with the merging realized through the final homodyne measurements. \\ \indent
A second example where code switching could be useful is the purification of magic states. For example, in the $D_4$ code, we could adopt an ``amputate and regrow'' strategy that leverages the fact that the nonlinear gates presented in \cref{sect:D4Gates} only affect one of the two modes. After the (cross-)Kerr gate, we can split the lattice into two diamond codes, yielding an entangled state of two diamond codes. 
In order to mitigate the effect of errors during the non-Clifford gate, one option is to measure the infected code and post-select on not measuring errors.
Finally, the lattice can be regrown by providing a fresh diamond GKP code word in the $\ket{+\bar X_{\Diamond}}$ state and measuring a $\bar Y_\Diamond\bar Y_\Diamond$ representative.


%

\end{document}